\documentclass[aps,prd,10pt,superscriptaddress,amsfonts,amssymb,amsmath,preprintnumbers,showpacs,notitlepage,a4paper,eqsecnum]{revtex4-1}

\usepackage{graphicx}
\usepackage{hyperref}

\begin{document}

\title{
Constraints on primordial black holes from Galactic gamma-ray background
}

\author{B. J. Carr}\email{B.J.Carr@qmul.ac.uk}
\affiliation{
Astronomy Unit, Queen Mary University of London,
Mile End Road, London E1 4NS, UK
}
\affiliation{
Research Center for the Early Universe (RESCEU),
Graduate School of Science, The University of Tokyo,
Tokyo 113-0033, Japan
}
\author{Kazunori Kohri}\email{kohri@post.kek.jp}
\affiliation{
IPNS, KEK and Sokendai, Tsukuba, Ibaraki 305-0881, Japan
}
\author{Yuuiti Sendouda}\email{sendouda@hirosaki-u.ac.jp}
\affiliation{
Graduate School of Science and Technology, Hirosaki University,
Hirosaki, Aomori 036-8561, Japan
}
\author{Jun'ichi Yokoyama}\email{yokoyama@resceu.s.u-tokyo.ac.jp}
\affiliation{
Research Center for the Early Universe (RESCEU),
Graduate School of Science, The University of Tokyo,
Tokyo 113-0033, Japan
}
\affiliation{
Department of Physics,
Graduate School of Science, The University of Tokyo,
Tokyo 113-0033, Japan
}
\affiliation{
Kavli Institute for the Physics and Mathematics of the Universe (Kavli IPMU),
The University of Tokyo,
Kashiwa, Chiba 277-8568, Japan
}

\date{April 18, 2016}

\begin{abstract}
The fraction of the Universe going into primordial black holes (PBHs) with initial mass $ M_* \approx 5 \times 10^{14}\,\mathrm g $\,, such that they are evaporating at the present epoch, is strongly constrained by observations of both the extragalactic and Galactic gamma-ray backgrounds.
However, while the dominant contribution to the extragalactic background comes from the time-integrated emission of PBHs with initial mass $ M_* $\,, the Galactic background is dominated by the instantaneous emission of those with initial mass slightly larger than $ M_* $ and current mass below $ M_* $\,.
Also, the instantaneous emission of PBHs smaller than $ 0.4\,M_* $ mostly comprises secondary particles produced by the decay of directly emitted quark and gluon jets.
These points were missed in the earlier analysis by Lehoucq \textit{et al.} using EGRET data.
For a monochromatic PBH mass function, with initial mass $ (1+\mu)\,M_* $ and $ \mu \ll 1 $, the current mass is $ (3 \mu)^{1/3}\,M_* $ and the Galactic background constrains the fraction of the Universe going into PBHs as a function of $ \mu $\,.
However, the initial mass function cannot be precisely monochromatic and even a tiny spread of mass around $ M_* $ would generate a current low-mass tail of PBHs below $ M_* $\,.
This tail would be the main contributor to the Galactic background, so we consider its form and the associated constraints for a variety of scenarios with both extended and nearly-monochromatic initial mass functions.
In particular, we consider a scenario in which the PBHs form from critical collapse and have a mass function which peaks well above $ M_* $\,.
In this case, the largest PBHs could provide the dark matter without the $ M_* $ ones exceeding the gamma-ray background limits.
\end{abstract}

\preprint{RESCEU-16/16}
\preprint{KEK-TH-1895}


\maketitle

\section{
Introduction
}

This paper develops a discussion which first appeared in our earlier paper \cite{Carr:2009jm}.
If primordial black holes (PBHs) of mass $ M_* \approx 5 \times 10^{14}\,\mathrm g $ such that they are evaporating at the present cosmologival epoch are clustered inside the Galactic halo, as expected, then their quantum evaporation should generate a Galactic $ \gamma $-ray background.
Since this would be anisotropic, it should be separable from the \textit{extragalactic} $ \gamma $-ray background, with the ratio of the anisotropic to isotropic intensities depending on the Galactic longitude and latitude.
This places important constraints on the number of PBHs, although their precise form depends upon such parameters as the Galactic core radius and the halo flattening.
Similar considerations apply if the dark matter is in the form of WIMPs, with a Galactic $ \gamma $-ray background being generated by their annihilations and decays \cite{Ackermann:2012rg,*Ackermann:2015zua}.

Many years ago Wright \cite{1996ApJ...459..487W} claimed that a Galactic background had been detected in EGRET observations between $ 30\,\mathrm{MeV} $ and $ 120\,\mathrm{MeV} $ \cite{Sreekumar:1997un} and attributed this to PBHs.
His detailed fit to the data, subtracting various other known components, required the PBH clustering factor to be $ (2\textnormal{--}12) \times 10^5\,h^{-1} $\,, comparable to that expected, and the local PBH explosion rate to be $ \mathcal R = 0.07\textnormal{--}0.42 \,\mathrm{pc}^{-3}\,\mathrm{yr}^{-1} $\,.
Here $ h $ is the Hubble parameter in units of $ 100\,\mathrm{km}\,\mathrm s^{-1}\,\mathrm{Mpc}^{-1} $ and the dependence on this arises because the mass $ M_* $ of the black holes evaporating now ($ t_0 \approx 13.8\,\mathrm{Gyr} $ \cite{Ade:2015xua}) scales as $ t_0^{1/3} \propto h^{-1/3} $\,, corresponding to an individual black hole luminosity $ L \propto M_*^{-2} \propto h^{2/3} $\,.
The contribution to the Galactic intensity is then $ I \propto \rho_\mathrm L\,M_*^{-3} \propto h $\,, where $ \rho_\mathrm L $ is the local PBH density, so the limit on $ \rho_\mathrm L $ and hence the clustering factor scales as $ h^{-1} $\,, while the limit on the local explosion rate, $ \mathcal R \propto \rho_\mathrm L\,t_0^{-1} $\,, is $ h $-independent.

A more recent analysis of EGRET data between $ 70\,\mathrm{MeV} $ and $ 150\,\mathrm{MeV} $\,, assuming a variety of PBH distributions, was given by Lehoucq \textit{et al.} \cite{Lehoucq:2009ge}.
In the isothermal model, which gives the most conservative limit, they found that the observed Galactic $ \gamma $-ray background required $ \mathcal R \leq 0.06 \,\mathrm{pc}^{-3}\,\mathrm{yr}^{-1} $\,.
They claimed that this corresponds to a limit on the cosmological PBH density of $ \Omega_\mathrm{PBH}(M_*) \leq 2.6 \times 10^{-9} $ in units of the critical density.
This implies that the fraction of the Universe's mass undergoing collapse at the PBH formation epoch is $ \beta(M_*) \leq 1.4 \times 10^{-26}\,\gamma^{-1/2} $\,, where $ \gamma $ is the size of the black hole relative to the particle horizon at formation, which is probably close to $ 1 $.
This is five times larger than the extragalactic background constraint derived in Ref.~\cite{Carr:2009jm}.
Lehoucq \textit{et al.} themselves claimed that it corresponds to $ \beta(M_*) < 1.9 \times 10^{-27}\,\gamma^{-1/2} $ but this is because they used a rather inaccurate formula relating $ \Omega_\mathrm{PBH} $ to $ \beta(M_*) $\,.
No dependence upon $ h $ was indicated since the Hubble parameter was assumed to be known ($ H_0=72\,\mathrm{km}\,\mathrm s^{-1}\,\mathrm{Mpc}^{-1} $).
The preferred value of the Hubble parameter is now lower ($ H_0 = 68\,\mathrm{km}\,\mathrm s^{-1}\,\mathrm{Mpc}^{-1} $) \cite{Ade:2015xua}.

In our previous paper \cite{Carr:2009jm}, we reassessed the Lehoucq \textit{et al.} analysis by including a more precise model for the PBH mass spectrum.
By deriving the relationship between the current PBH mass $ m $ and the initial mass $ M $\,, we showed that the dominant contribution to the Galactic background comes not from black holes with the mass $ M_* $ whose lifetime is the current age of the Universe but a somewhat larger mass of $ 1.08\,M_* $\,.
The point is that the Galactic background depends upon the \textit{current} PBH emission, whereas the extragalactic background depends upon the time-integrated emission.
The $ M_* $ black holes themselves no longer exist, so their emission could only be reaching us from the edge of the Universe rather than the edge of the Galaxy.
We also pointed out that the form of the Galactic background depends upon the low-mass tail of the PBH spectrum below $ M_* $\,, this naturally resulting from evaporations at the present epoch.
The implied limits on $ \beta(M) $ were rather different from those found by Lehoucq \textit{et al.}, both in their functional dependence on $ M $ and in their strength.

Several factors require a re-evaluation of our previous analysis.
First, it was based on the EGRET data and we now have more recent data from the Fermi satellite, extending to $ 100\,\mathrm{GeV} $\,.
At such high energies one must also consider the effects on the Galactic background of the secondary emission generated by quark and gluon jet decays above the QCD temperature \cite{MacGibbon:1991tj}.
Second, our own earlier analysis assumed that the initial mass function of the PBHs was \textit{almost} monochromatic.
This is because we were interested in constraints on the PBH abundance as a function of mass.
However, it cannot be \textit{exactly} monochromatic, else there would be no low-mass tail, so this involved a rather complicated calculation of how the bandwidth of the emission associated with the spread of the mass function compares to the bandwidth of the observations.

We cover the monochromatic case in this paper for completeness but stress that the monochromatic assumption is unlikely to apply in any realistic PBH formation scenario.
Even if the density fluctuations producing the PBHs are highly peaked on some scale (e.g.\ the horizon scale at their formation), the resulting PBH mass spectrum may be quite broad.
This applies, in particular, if the PBHs form from critical collapse since the spectrum then extends well below the horizon scale.
This problem was originally analysed by Yokoyama \cite{Yokoyama:1998xd} in the context of the extragalactic background and is here updated and applied to the Galactic background.
Coincidentally, the low-mass tails from evaporations and critical phenomena have similar form.
Even if the mass function were monochromatic, there is no reason why it should correspond to the mass $ M_* $\,.

The plan of the paper is as follows.
In Sec.~\ref{sec:BH} we recall the characteristics of black hole emission, deriving a precise relationship between the initial and current PBH mass, determining the mass at which secondary emission becomes important and comparing the characteristics of the primary and secondary emission.
In Sec.~\ref{sec:mf} we explain why the contribution of PBHs to the Galactic background is very sensitive to the initial mass function around $ M_* $ and we discuss the form of this function for a variety of scenarios, including one in which the PBHs form from primordial density perturbations via critical collapse.
In Sec.~\ref{sec:flux} we calculate the expected Galactic gamma-ray flux for the various scenarios.
In Sec.~\ref{sec:constraint} we infer the associated constraints for these scenarios on the fraction of the Universe going into PBHs over various mass ranges.
In Sec.~\ref{sec:conclusion} we draw some general conclusions and compare our limits with recent ones from the search for PBH bursts \cite{Abdo:2014apa,MacGibbon:2015mya}.

\section{\label{sec:BH}
Black hole gamma-ray emission
}

As first shown by Hawking \cite{Hawking:1974rv,Hawking:1974sw}, a black hole with mass $ M \equiv M_{14} \times 10^{14}\,\mathrm g $ emits thermal radiation with temperature
\begin{equation}
T_\mathrm{BH}
= \frac{1}{M}
\approx
  106\,M_{14}^{-1}\,\mathrm{MeV}\,,
\label{eq:temp}
\end{equation}
where throughout this paper we use natural units with $ \hbar = c = k_\mathrm B = 8 \pi G = 1 $ in any formulae but appropriate physical units when giving numerical values.
This assumes that the hole has no charge or angular momentum, which is reasonable since charge and angular momentum will also be lost through quantum emission on a shorter timescale than the mass \cite{Page:1976df,Page:1976ki,Page:1977um}.
The mass loss rate can be expressed as
\begin{equation}
\frac{\mathrm dM_{14}}{\mathrm dt}
= -5.34 \times 10^{-17}\,f(M)\,M_{14}^{-2}\,\mathrm s^{-1}\,,
\label{eq:massloss}
\end{equation}
where $ f(M) $ measures the number of emitted particle species and is normalised to unity for the holes with $ M \gg 10^{17}\,\mathrm g $ which emit only (effectively) massless particles (photons and neutrinos).
The contribution of each relativistic degree of freedom to $ f(M) $ depends on the spin $ s $ \cite{MacGibbon:1991tj}:
\begin{equation}
\begin{aligned}
&
f_{s=0} = 0.267\,,
\quad
f_{s=1} = 0.060\,,
\quad
f_{s=3/2} = 0.020\,,
\quad
f_{s=2} = 0.007\,, \\
&
f_{s=1/2} = 0.147~(\textnormal{neutral})\,,
\quad
f_{s=1/2} = 0.142~(\textnormal{charge}~\pm e)\,.
\end{aligned}
\label{eq:spin}
\end{equation}
The average energies of the emitted particles are $ 4.22\,T_\mathrm{BH} $ for $ s = 1/2 $ neutral, $ 4.18\,T_\mathrm{BH} $ for $ s = 1/2 $ charged, and $ 5.71\,T_\mathrm{BH} $ for $ s = 1 $, respectively.
The peak energies of the flux and power are within $ 7\,\% $ of these values \cite{MacGibbon:1990zk}.
Therefore holes in the mass range $ 10^{15}\,\mathrm g < M < 10^{17}\,\mathrm g $ emit electrons but not muons, while those in the range $ 10^{14}\,\mathrm g < M < 10^{15}\,\mathrm g $ also emit muons, which subsequently decay into electrons and neutrinos.
At this point the value of $ f $ is $ 2 \times 0.06 + 8 \times 0.142 + 6 \times 0.147 = 2.14 $, allowing for all spin states of relevant particles and antiparticles.

\begin{figure}[htb]
\includegraphics[scale=0.65]{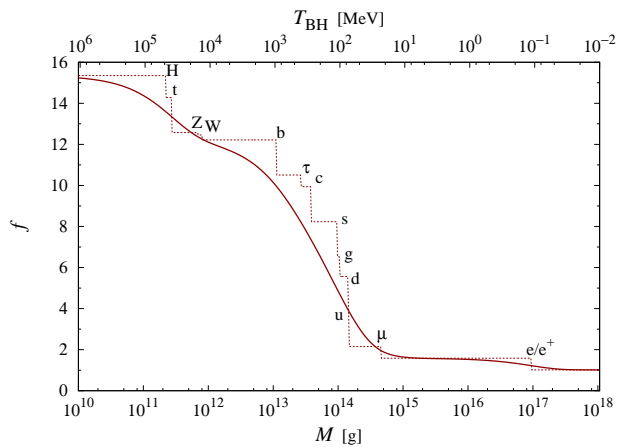}
\caption{\label{fig:dof}
Dependence of $ f(M) $ on $ M $\,.
Dotted red line shows step-function approximation at mass thresholds for quarks (up, down, strange, charm, bottom, top), gluons, $ W $/$ Z $ bosons and the Higgs particle.
Solid red curve shows MacGibbon's approximation formula \cite{MacGibbon:1991tj}, updated to include $ W $/$ Z $ boson, top quark and Higgs.
The most notable feature is the increase by factor $ \alpha = 4 $ at $ 2 \times10^{14}\,\mathrm g $\,.
}
\end{figure}

A black hole begins to emit pions ($ m_\pi \approx 140\,\mathrm{MeV} $) once $ M $ falls below about $ 5 \times 10^{14}\,\mathrm g $ and then other hadron species as $ M $ continues to fall.
However, hadrons are composite particles made up of quarks held together by gluons, so one would expect only these fundamental particles to be emitted for temperatures exceeding the QCD confinement scale, $ \Lambda_\mathrm{QCD} = 250\textnormal{--}300\,\mathrm{MeV} $\,.
Taking the peak emission energy to be $ 4\,T_\mathrm{BH} $ for quarks ($ s = 1/2 $) and $ 6\,T_\mathrm{BH} $ for gluons ($ s = 1 $) , this corresponds to a mass below
\begin{equation}
M_\mathrm q
= (1.4\textnormal{--}2.5) \times 10^{14}\,\mathrm g
\approx
  2 \times 10^{14}\,\mathrm g\,.
\label{eq:Mq}
\end{equation}
Since there are $ 12 $ quark degrees of freedom per flavour and $ 16 $ gluon degrees of freedom, one would expect the emission rate (i.e.\ the value of $ f $) to increase suddenly once the QCD temperature is reached.
If one includes just $ u $\,, $ d $ and $ s $ quarks and gluons, Eq.~\eqref{eq:spin} implies that their contribution to $ f $ is $ 3 \times 12 \times 0.14 + 16 \times 0.06 \approx 6 $, compared to the pre-QCD value of about $ 2 $.
Thus the value of $ f $ roughly quadruples, although there will be a further increase in $ f $ at somewhat higher temperatures due to the emission of the heavier quarks.
After their emission, quarks and gluons fragment into further quarks and gluons until they cluster into the observable hadrons when they have travelled a distance $ \Lambda_\mathrm{QCD}^{-1} \sim 10^{-13}\,\mathrm{cm} $\,.
We describe the products of quark and gluon decays as ``secondary'' emission.
The dependence of $ f(M) $ on $ M $ is indicated in Fig.~\ref{fig:dof}.
One could also add another step at around $ 10^{12}\,\mathrm g $ due to the emission of $ W $ and $ Z $ bosons, the top quark and the Higgs boson, leading to a maximum value $ f \approx 15 $ in the standard model.
However, this last step is smaller (a factor of $ 2 $ instead of $ 4 $) and it is not relevant to later considerations, so we neglect it below.
An equivalent figure has been derived by Ukwatta \textit{et al.} \cite{Ukwatta:2015iba}.

Integrating the mass loss rate over time \cite{MacGibbon:1991tj} gives a lifetime
\begin{equation}
\tau
\approx
  2.7 \times 10^{14}\,f(M)^{-1}\,M_{14}^3\,\mathrm s\,.
\label{eq:tau}
\end{equation}
This can be inverted to give the mass of a PBH evaporating at time $ \tau $ after the big bang.
Since the current age of the Universe is $ 13.8\,\mathrm{Gyr} $ \cite{Ade:2015xua}, the mass of a PBH completing its evaporation at the present epoch is
\begin{equation}
M_*
\approx
  5.1 \times 10^{14}\,
  \left(\frac{f_*}{1.9}\right)^{1/3}\,\mathrm g\,,
\end{equation}
where $ f_* $ is the value of $ f $ at the temperature $ T_\mathrm{BH}(M_*) \approx 21\,\mathrm{MeV} $ implied by Eq.~\eqref{eq:temp}.
We note that the mass $ M_\mathrm q $ is smaller than this by a factor of $ 0.4 $.
The above analysis is not exact because the value of $ f(M) $ in Eq.~\eqref{eq:tau} should really be the weighted average over the lifetime of the black hole.
Another recent calculation gives $ 5.0 $ rather than $ 5.1 $ \cite{MacGibbon:2007yq}.
We present a more accurate calculation below.

\subsection{
More accurate relationship between $ M_* $ and $ m $
}

We now obtain a more precise expression for $ M_* $ taking hadron emission into account.
The mass-loss rate is
\begin{equation}
\frac{\mathrm dM(t)}{\mathrm dt}
= -\frac{\phi[M(t)]}{3\,M(t)^2}\,,
\end{equation}
where $ \phi(M) $ represents the number of emitted particle degrees of freedom for a PBH with mass $ M $\,.
This is hereafter assumed to have the simplified form
\begin{equation}
\phi(M)
=
\begin{cases}
\phi_*
& (M_\mathrm q \leq M \lesssim M_*) \\
\alpha\,\phi_*
& (M \leq M_\mathrm q)\,,
\end{cases}
\end{equation}
where $ \alpha \approx 4 $ to sufficient precision.
The function $ \phi(M) $ is equivalent to the function $ f(M) $ but the latter is normalized to $ 1 $ at high $ M $\,.
The PBH mass at time $ t $ is then
\begin{equation}
M(t)^3
= M^3 - \int_{t_\mathrm f}^t\!\mathrm dt\,\phi[M(t)]\,,
\end{equation}
where in this section $ M \equiv M(t_\mathrm f) $ is the mass at the formation epoch $ t_\mathrm f $\,.
The time $ \tau_\mathrm q $ at which a PBH with initial mass $ M $ falls to the mass $ M_\mathrm q $ is
\begin{equation}
\tau_\mathrm q(M)
\approx
  \frac{M^3 - M_\mathrm q^3}{\phi_*}
\equiv
  t_0\,\left(\frac{M^3 - M_\mathrm q^3}{\bar M_*^3}\right)\,,
\end{equation}
where $ t_0 = 13.8\,\mathrm{Gyr} $ and
\begin{equation}
\bar M_*
\equiv
  (\phi_*\,t_0)^{1/3}
\approx
  5.07 \times 10^{14}\,\mathrm g\,\left(\frac{f_*}{1.9}\right)^{1/3}
\end{equation}
is the mass of a PBH currently evaporating if one neglects secondary emission once $ M(t) $ falls below $ M_\mathrm q $\,.
Note that $ \tau_\mathrm q(M) \leq t_0 $ implies
\begin{equation}
M
\leq
  M_\mathrm c
\equiv
  \bar M_*\,\left[1 + \left(M_\mathrm q/\bar M_*\right)^3\right]^{1/3}
= (1.01\textnormal{--}1.04)\,\bar M_*
\approx
  1.02\,\bar M_*\,,
\label{eq:Mc}
\end{equation}
using the range of $ M_\mathrm q $ given by Eq.~\eqref{eq:Mq} and then some intermediate value at the last step.
So only PBHs slightly larger than $ \bar M_* $ generate secondary emission by the present epoch.

\begin{figure}[htb]
\includegraphics[scale=0.65]{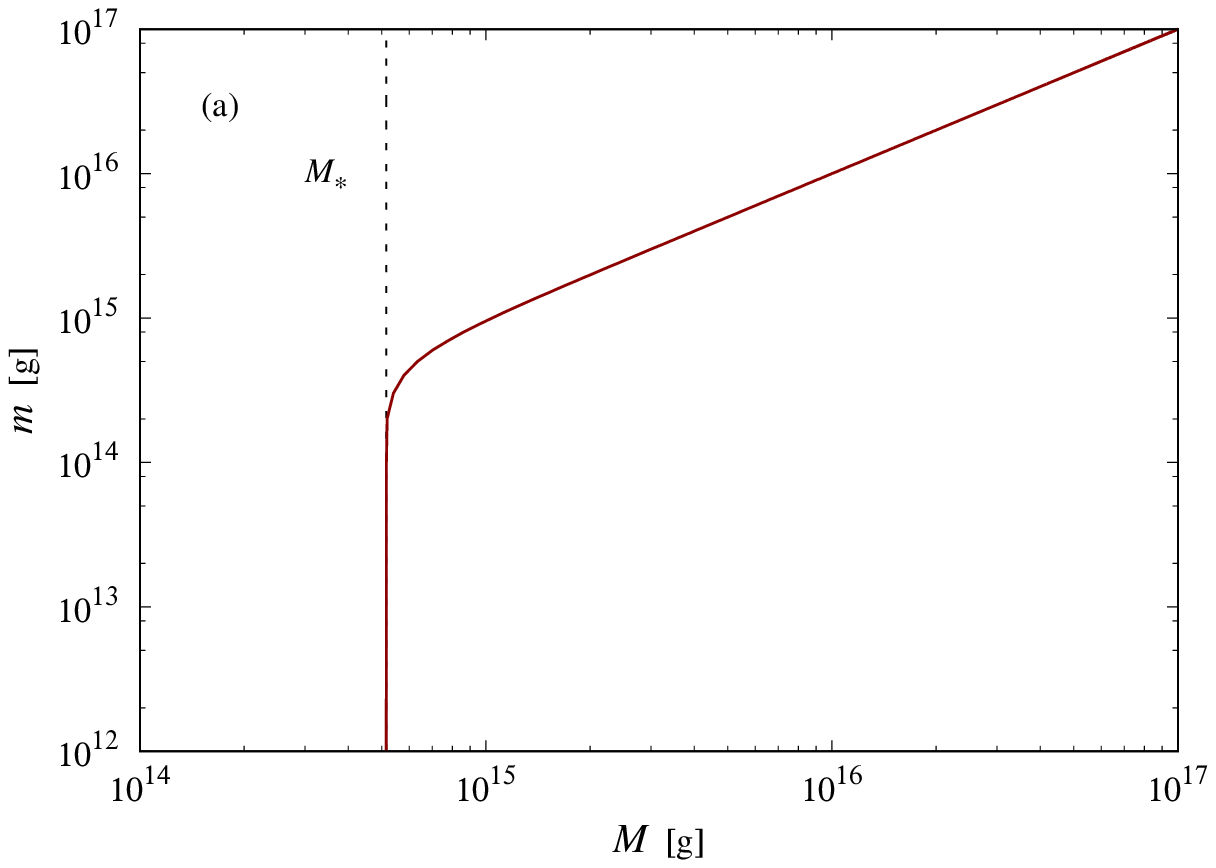}
\includegraphics[scale=0.65]{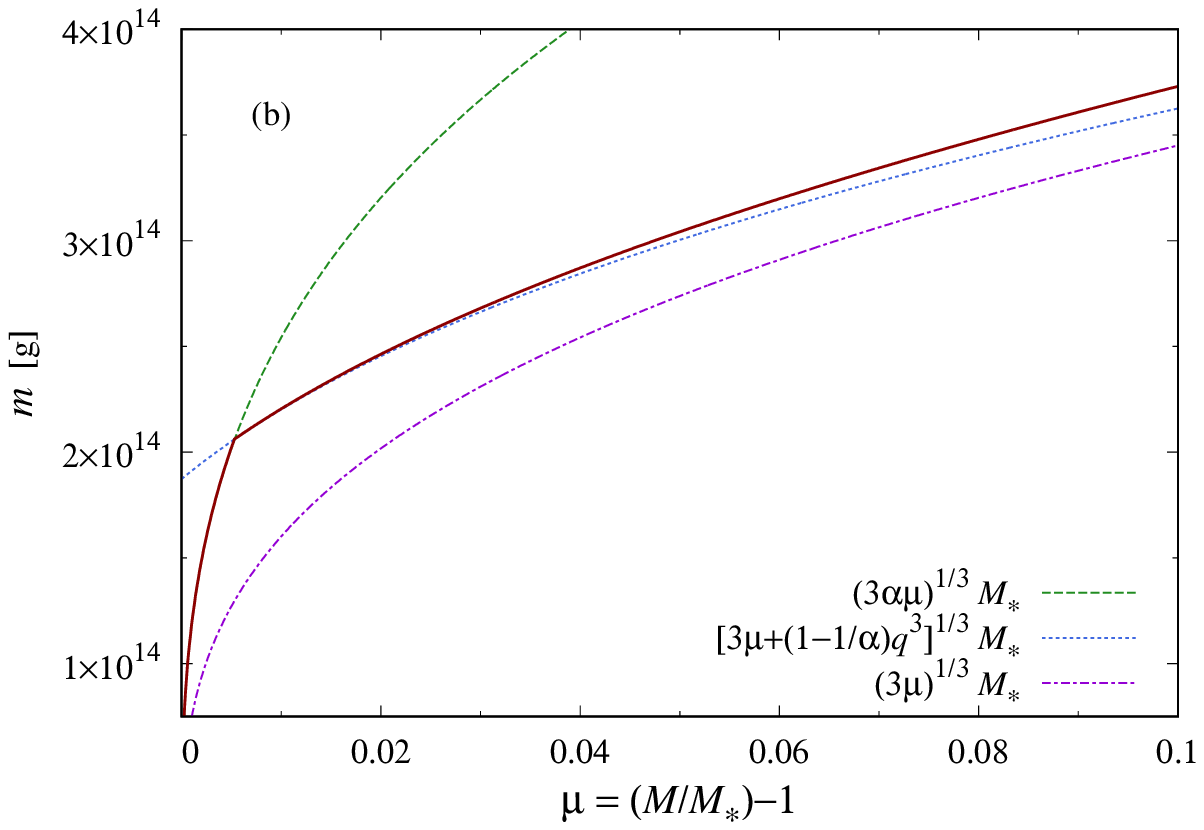}
\caption{\label{fig:m}
(a): $ m $ versus $ M $\,.
(b): Comparison of exact $ m(\mu) $ relation (red line) with various approximations.
}
\end{figure}

For $ M \geq M_\mathrm c $\,, we have $ \tau_\mathrm q(M) \geq t_0 $ and the current mass $ m \equiv M(t_0) $ is given by
\begin{equation}
m^3
= M^3 - \bar M_*^3
\quad
(M \geq M_\mathrm c)\,.
\end{equation}
For $ M \leq M_\mathrm c $\,, we approximate the mass at $ t \in [\tau_\mathrm q(M),t_0] $ by
\begin{equation}
\begin{aligned}
M(t)^3
&
= M^3
  - \left(\int_{t_\mathrm f}^{\tau_\mathrm q(M)} + \int_{\tau_\mathrm q(M)}^t\right)
    \mathrm dt\,\phi[M(t)] \\
&
\approx
  \alpha\,\left[
   M^3
   - (1-\alpha^{-1})\,M_\mathrm q^3
   - \bar M_*^3\,\frac{t}{t_0}
  \right]\,,
\end{aligned}
\end{equation}
so the current mass is given by
\begin{equation}
m^3
= \alpha\,\left[
   M^3
   - (1-\alpha^{-1})\,M_\mathrm q^3
   - \bar M_*^3
  \right]
\quad
(M \leq M_\mathrm c)\,.
\end{equation}
PBHs completing their evaporation today have $ m = 0 $ and therefore an initial mass
\begin{equation}
M_*
= \left[\bar M_*^3 + (1-\alpha^{-1})\,M_\mathrm q^3\right]^{1/3}\,.
\end{equation}
Defining $ q \equiv M_\mathrm q/M_* = 0.3\textnormal{--}0.5 \approx 0.4 $, we obtain
\begin{equation}
M_*
= \frac{\bar M_*}{[1 - (1-\alpha^{-1})\,q^3]^{1/3}}
\approx
  \left[1 + \frac{1}{3} (1-\alpha^{-1})\,q^3\right]\,\bar M_*
= (1.007\textnormal{--}1.031)\,\bar M_*
\approx
  1.017\,\bar M_*
\approx
  5.15 \times 10^{14}\,\mathrm{g}\,,
\end{equation}
where the small correction $ 0.017 $ differs from the correction $ 0.020 $ in Eq.~\eqref{eq:Mc} by the factor $ 1 - \alpha^{-1} \approx 3/4 $.
The current mass can then be expressed as
\begin{equation}
m
=
\begin{cases}
\left[M^3 - M_*^3 + (1 - \alpha^{-1})\,q^3\,M_*^3\right]^{1/3}
& (M \geq M_\mathrm c) \\
\alpha^{1/3}\,(M^3 - M_*^3)^{1/3}
& (M_* \leq M \leq M_\mathrm c)\,,
\end{cases}
\label{eq:precisem}
\end{equation}
where
\begin{equation}
M_\mathrm c
= (1+q^3/\alpha)^{1/3}\,M_*
= (1.002\textnormal{--}1.010)\,M_*
\approx
  1.005\,M_*
\approx
  5.17 \times 10^{14}\,\mathrm g
\end{equation}
is the initial mass corresponding to a current mass $ M_\mathrm q $\,.
This relationship is indicated in Fig.~\ref{fig:m}(a).
The function $ m(M) $ is continuous at $ M_\mathrm c $ but its derivative is discontinuous.

It is also convenient to write the mass of PBHs somewhat larger than $ M_* $ in the form
\begin{equation}
M
= M_*\,(1+\mu)\,,
\end{equation}
where $ \mu $ is dimensionless and generally small.
One can then write the $ m(M) $ relationship \eqref{eq:precisem} as
\begin{equation}
m
=
\begin{cases}
\left[(\mu+1)^3 -1 + (1-\alpha^{-1})\,q^3\right]^{1/3}\,M_*
& (\mu \geq \mu_\mathrm c) \\
(3\,\alpha\,\mu)^{1/3}\,(1 + \mu + \mu^2/3)^{1/3}\,M_*
& (0 \leq \mu \leq \mu_\mathrm c)\,,
\end{cases}
\label{eq:mmu}
\end{equation}
where
\begin{equation}
\mu_\mathrm c
\approx
  q^3/(3\,\alpha)
= 0.005\,(\alpha/4)^{-1}\,(q/0.4)^3
\end{equation}
is the value of $ \mu $ corresponding to $ M_\mathrm c $\,.
The $ m(\mu) $ relationship can be approximated in various regimes by
\begin{equation}
m
=
\begin{cases}
\mu\,M_*
& (\mu \gg 1) \\
(3\,\mu)^{1/3}\,M_*
& (\mu_\mathrm d \leq \mu \ll 1) \\
\left[3\,\mu + q^3(1-\alpha^{-1})\right]^{1/3}\,M_*
& (\mu_\mathrm c \leq \mu \leq \mu_\mathrm d) \\
(3\,\alpha\,\mu)^{1/3}\,M_*
& (0 \leq \mu \leq \mu_\mathrm c)\,,
\end{cases}
\label{eq:mmu_approx}
\end{equation}
where
\begin{equation}
\mu_\mathrm d
\approx
  q^3/3
= 0.02\,(q/0.4)^3
\end{equation}
corresponds to the mass above which the second expression applies (i.e.\ it is accurate for $ \mu > \mu_\mathrm d $).
This will be useful when calculating the Galactic gamma-ray background and is accurate to about $ 10\,\% $ over the relevant range of $ \mu $\,.
The validity of these approximations is indicated in Fig.~\ref{fig:m}(b).

\subsection{
Instantaneous primary and secondary emission
}

\begin{figure}[htb]
\includegraphics[scale=0.65]{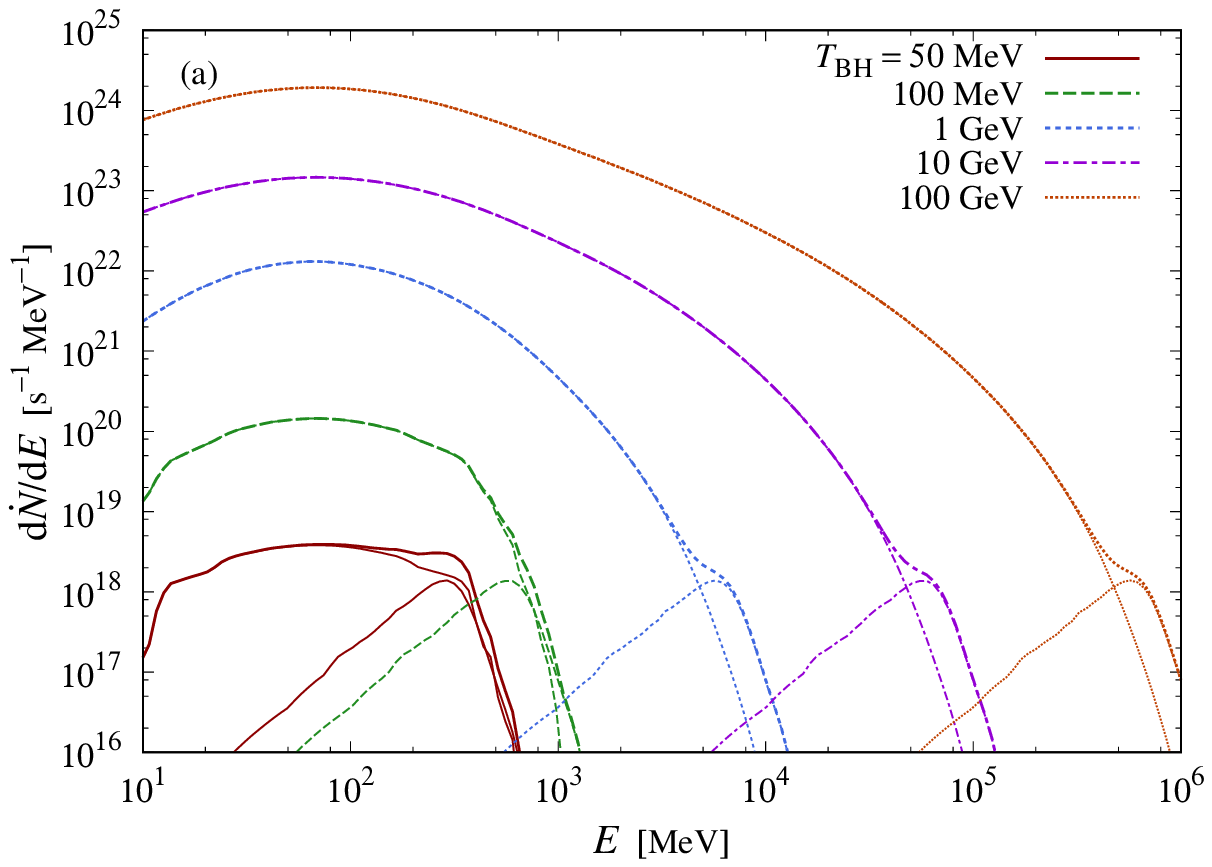}
\includegraphics[scale=0.65]{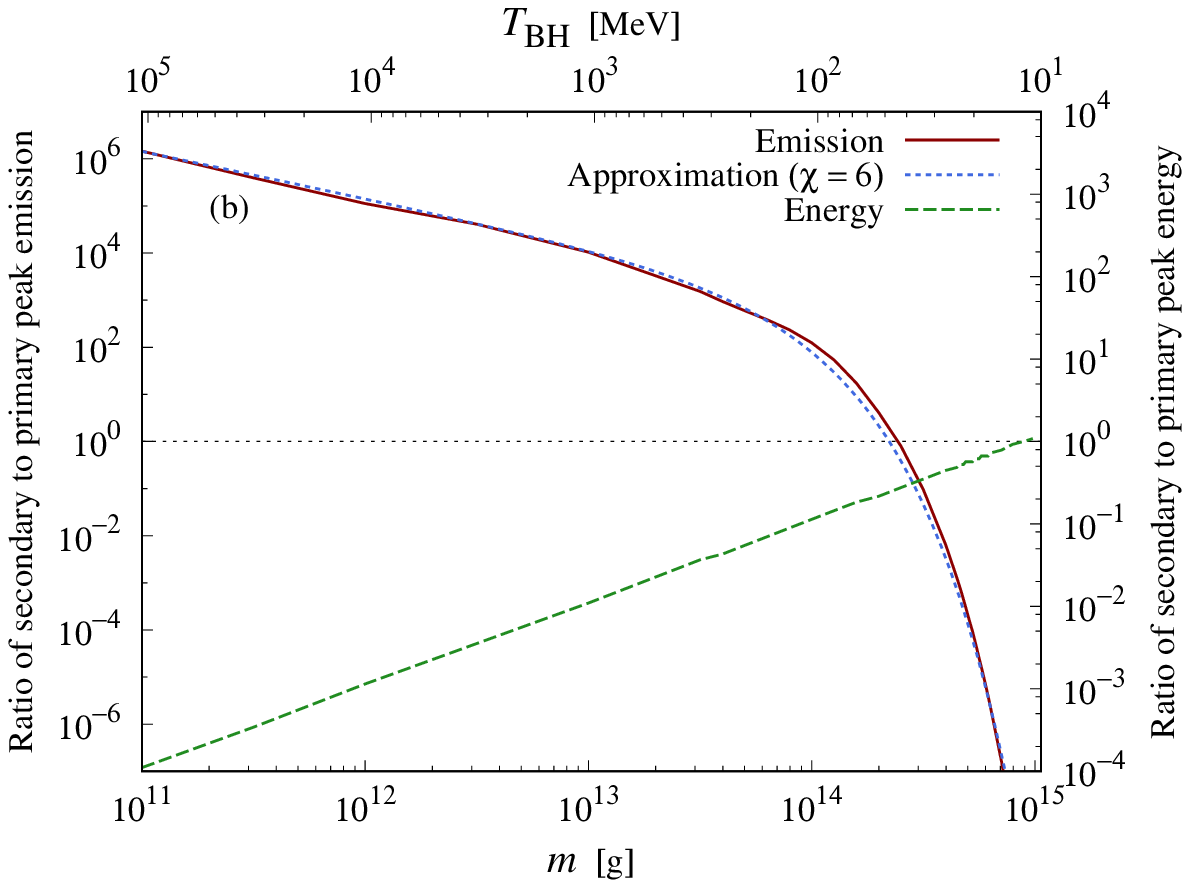}
\caption{\label{fig:ratios}
(a) Instantaneous emission rates for black holes of various temperatures, with primary component at bottom.
The $ 50\,\mathrm{MeV} $ emission rate (solid red) corresponds to $ T_\mathrm{BH} \approx M_\mathrm q^{-1} $\,.
(b) Ratios of secondary to primary peak energies (dashed green) and fluxes (solid red) for the instantaneous emission.
The right figure confirms the secondary peak is proportional to the temperature for $ m < M_\mathrm q $ (as the primary peak is always constant).
In (a) it is noted that the widths of secondary emission are roughly proportional to the temperature.
}
\end{figure}

In calculating the Galactic gamma-ray background, we need the instantaneous emission as a function of the mass $ M $\,.
This evolves from the initial mass $ M_\mathrm i $ to the current mass $ m $\,.
Only black holes with $ M \geq M_* $ are relevant since smaller ones do not contribute.
As we use units with $ 8 \pi G = 1 $, the temperature of a black hole of mass $ M $ is $ T_\mathrm{BH} = 1/M $\,.
The instantaneous emission rate for primary photons of energy $ E $ can be written as
\begin{equation}
\frac{\mathrm d\dot N^\mathrm P}{\mathrm dE}(M,E)
= \frac{1}{2\pi^2}\,\frac{E^2\,\sigma(M,E)}{\mathrm e^{M E}-1}
\propto
\begin{cases}
E^3\,M^3
&(E < M^{-1}) \\
E^2\,M^2\,\mathrm e^{-M E}
& (E > M^{-1})\,,
\end{cases}
\label{eq:rate_pri}
\end{equation}
where $ \sigma(E,M) $ is the absorption cross-section for photons ($ s = 1 $), given by \cite{Page:1976df}
\begin{equation}
\sigma(E,M)
\propto
\begin{cases}
E^2\,M^4 & (E < M^{-1} ) \\
M^2 & (E > M^{-1})\,.
\end{cases}
\label{eq:sigma}
\end{equation}
The form of the spectrum is illustrated by the lower curves in Fig.~\ref{fig:ratios}(a) for various values of $ M $\,.
It peaks at $ \bar E^\mathrm P \approx 5.8\,T_\mathrm{BH} \approx 600\,M_{14}^{-1}\,\mathrm{MeV} $ with a value
\begin{equation}
\frac{\mathrm d\dot N^\mathrm P}{\mathrm dE}(E = \bar E^\mathrm P)
\approx
  1.4 \times 10^{18}\,\mathrm s^{-1}\,\mathrm{MeV}^{-1}\,.
\label{eq:ratepeak}
\end{equation}
The average energy of the primary photons is $ \bar E^\mathrm P \approx 5.7\,T_\mathrm{BH} $\,, while that of the primary quarks which generate the secondary photons is $ 4.2\,T_\mathrm{BH} $\,.

Once secondary emission becomes important, as is always the case for $ M_\mathrm i < M_\mathrm q $\,, the analysis of MacGibbon and Webber \cite{MacGibbon:1990zk} shows that the instantaneous emission rate can be expressed as an integral over the jet energy $ Q $:
\begin{equation}
\frac{\mathrm d\dot N^\mathrm S}{\mathrm dE}
= E^{-1}
  \int_E^\infty\!
  \frac{Q^2\,T_\mathrm{BH}^{-2}\,(1-E/Q)^{2s-1}\,\Theta(E - k\,m_{\pi})}
       {\mathrm e^{Q/T_\mathrm{BH}} \pm 1}\,
  \mathrm dQ
\propto
\begin{cases}
G(E,M)
& (E < M_\mathrm q^{-1}) \\
E^{-1}\,M^{-1}
& (M^{-1} > E > M_\mathrm q^{-1}) \\
E^2\,M^2\,\mathrm e^{-E M}
& (E > M^{-1})\,,
\end{cases}
\label{eq:rate_sec}
\end{equation}
where the function $ G(E,M) $ reflects the form of the jet fragmentation peak, $ k $ is a constant of $ \mathcal O(1) $\,, $ M_\mathrm q \approx 0.2 \Lambda_\mathrm{QCD} \approx 53\,\mathrm{MeV} $\,, and $ + $ and $ - $ signs apply for quark and gluon jets, respectively.
The form of the spectrum is illustrated by the upper curves in Fig.~\ref{fig:ratios}(a).
This has a rather broad peak at half the pion mass, corresponding to an average photon energy of $ \bar E^\mathrm S \approx m_{\pi^0}/2 \approx 68\,\mathrm{MeV} $\,, and its form reflects the low-energy fragmentation function.
More precisely, the secondary peak flux is
\begin{equation}
\begin{aligned}
\frac{\mathrm d\dot N^\mathrm S}{\mathrm dE}(E = \bar E^\mathrm S)
&
\approx
  2 \sum_{i=q,g} \mathcal B_{i\to\pi^0}(\bar E_i)\,\frac{\bar E_i}{m_{\pi^0}}\,
  \frac{\mathrm d\dot N^\mathrm P_i}{\mathrm dE_i}(E_i=\bar E_i) \\
&
\approx
  8.4 \times 10^{18}\,\mathrm s^{-1}\,\mathrm{MeV}^{-1}\,
  \left(\frac{M}{M_*}\right)^{-1}\,
  \sum_{i=q,g} \mathcal B_{i\to\pi^0}(\bar E_i)\,,
\end{aligned}
\end{equation}
where $ \bar E_i = 4.2\,T_\mathrm{BH} $ and the last term is $ \mathcal O(1) $\,.
This expression only applies for $ M < M_\mathrm q $ since the secondary emission drops off exponentially for $ M > M_\mathrm q $ because of the Wien factor.
The function $ G(E,M) $ can be empirically represented for the interval $ 10\,\mathrm{MeV} < E < M_\mathrm q^{-1} $ as
\begin{equation}
G(E,M)
\sim
  E\,M^{-1}\,M_\mathrm q^2\,\mathrm e^{-\chi M/M_\mathrm q}\,,
\label{eq:frag}
\end{equation}
where $ \chi \approx 6 $.
Figure~\ref{fig:ratios}(a) is compatible with the equivalent figure in Ukwatta \textit{et al.} \cite{Ukwatta:2015iba} but they are interested in the higher energies associated with the final black hole burst.

We will be interested in the ratios of the secondary to primary peak energies and fluxes at the present epoch ($ M = m $).
The energy ratio is
\begin{equation}
\frac{\bar E^\mathrm S}{\bar E^\mathrm P}
\approx
  (68\,\mathrm{MeV})/(600\,m_{14}^{-1}\,\mathrm{MeV})
\approx
  0.6\,(m/M_*)\,,
\label{eq:energyratio}
\end{equation}
while the flux ratio is
\begin{equation}
\left.\left(\frac{\mathrm d\dot N^\mathrm S}{\mathrm dE}\right)_{\bar E^\mathrm S}\right/
\left(\frac{\mathrm d\dot N^\mathrm P}{\mathrm dE}\right)_{\bar E^\mathrm P}
\approx
  1.4\,\left(\frac{m}{M_*}\right)^{-1}\,\mathrm e^{-\chi\,m/M_\mathrm q}\,.
\end{equation}
These ratios are plotted as a function of $ m $ in Fig.~\ref{fig:ratios}(b), along with the exact PBH emission spectra, which are computed numerically.
This shows that the analytic dependence on $ m $ and $ E $ fits the expected form quite well.

\subsection{\label{sec:BH_tail}
Time-integrated primary and secondary emission
}

The time-integrated spectrum has been studied in detail by MacGibbon \cite{MacGibbon:1991tj} and more recently Linton \textit{et al.} \cite{Linton:2006yu} but we now derive the qualitative features using simple analytical arguments.
Throughout this section, $ M $ is the evolving mass and $ M_\mathrm i $ is the initial mass.
There are three different cases, depending on the value of $ M_\mathrm i $\,.

For $ M_\mathrm i > M_\mathrm c $\,, only primary emission is important and we can integrate Eq.~\eqref{eq:rate_pri} over time (i.e.\ $ M $) for fixed $ E $:
\begin{equation}
\frac{\mathrm dN^\mathrm P}{\mathrm dE}
= \int^m_{M_\mathrm i}\!\frac{\mathrm d\dot N^\mathrm P}{\mathrm dE}\,
  \frac{\mathrm dt}{\mathrm dM}\,\mathrm dM
\propto
  E^2\,
  \int_m^{M_\mathrm i}\!\frac{\sigma (E,M)\,M^2}{\mathrm e^{E M}-1}\,
  \mathrm dM\,,
\end{equation}
where $ m > M_\mathrm q $\,, we have used the relation $ \mathrm dM/\mathrm dt \propto M^{-2} $ and $ \sigma(E,M) $ is given by Eq.~\eqref{eq:sigma}.
In this case, nearly all the emission occurs at the present epoch, so we neglect redshift effects.
For $ E < M_\mathrm i^{-1} $\,, the mass integral just involves the Rayleigh--Jeans part of the spectrum and is dominated by the upper limit $ M_\mathrm i $\,.
For $ M_\mathrm i^{-1} < E < m^{-1} $\,, the exponential term cuts the integral off above a mass $ M \sim E^{-1} $\,.
For $ E > m^{-1} $\,, the integral is dominated by the lower limit $ m $ and falls off exponentially.
The time-integrated spectrum of photons from a PBH with $ M_\mathrm i > M_\mathrm c $ can therefore be expressed as
\begin{equation}
\frac{\mathrm dN^\mathrm P}{\mathrm dE}
\propto
\begin{cases}
E^3\,M_\mathrm i^6
& (E < M_\mathrm i^{-1}) \\
E^{-3}
& (M_\mathrm i^{-1}< E < m^{-1}) \\
E\,m^4\,\mathrm e^{-E m}
& (E > m^{-1})\,.
\end{cases}
\label{eq:int_pri}
\end{equation}
This peaks at $ E \sim M_\mathrm i^{-1} $ with a value $ \sim M_\mathrm i^3 $ but there is a high-energy tail, falling as $ E^{-3} $ until the cut-off at $ m^{-1} $\,.

For the narrow mass band $ M_* < M_\mathrm i < M_\mathrm c $\,, Eq.~\eqref{eq:int_pri} still gives the primary emission but there is also secondary emission once $ M $ falls below $ M_\mathrm q $ and this gives
\begin{equation}
\frac{\mathrm dN^\mathrm S}{\mathrm dE}
= \int^m_{M_\mathrm q}\!\frac{\mathrm d\dot N^\mathrm S}{\mathrm dE}\,
  \frac{\mathrm dt}{\mathrm dM}\,\mathrm dM
\propto
\begin{cases}
E\,M_\mathrm q^4
& (E < M_\mathrm q^{-1}) \\
E^{-3}
& (M_\mathrm q^{-1} < E < m^{-1}) \\
E\,m^4\,\mathrm e^{-E m}
& (E > m^{-1})\,.
\end{cases}
\label{eq:int_sec}
\end{equation}
This has the same form as Eq.~\eqref{eq:int_pri} except in the low-energy regime, where quark fragmentation dominates and we assume $ \chi\,M_\mathrm i \ll M_\mathrm q $ and $ \epsilon \ll 1 $.
The mass integral is determined by the form of the fragmentation function for $ E < M_\mathrm q^{-1} $\,, by the $ M \sim E^{-1} $ contribution for $ M_\mathrm q^{-1} < E < m^{-1} $\,, and by the exponential tail for $ E > m^{-1} $\,.
Since $ m \to 0 $ as $ M_\mathrm i \to M_* $\,, the exponential cut-off disappears in this limit.

\begin{figure}[htb]
\includegraphics[scale=0.65]{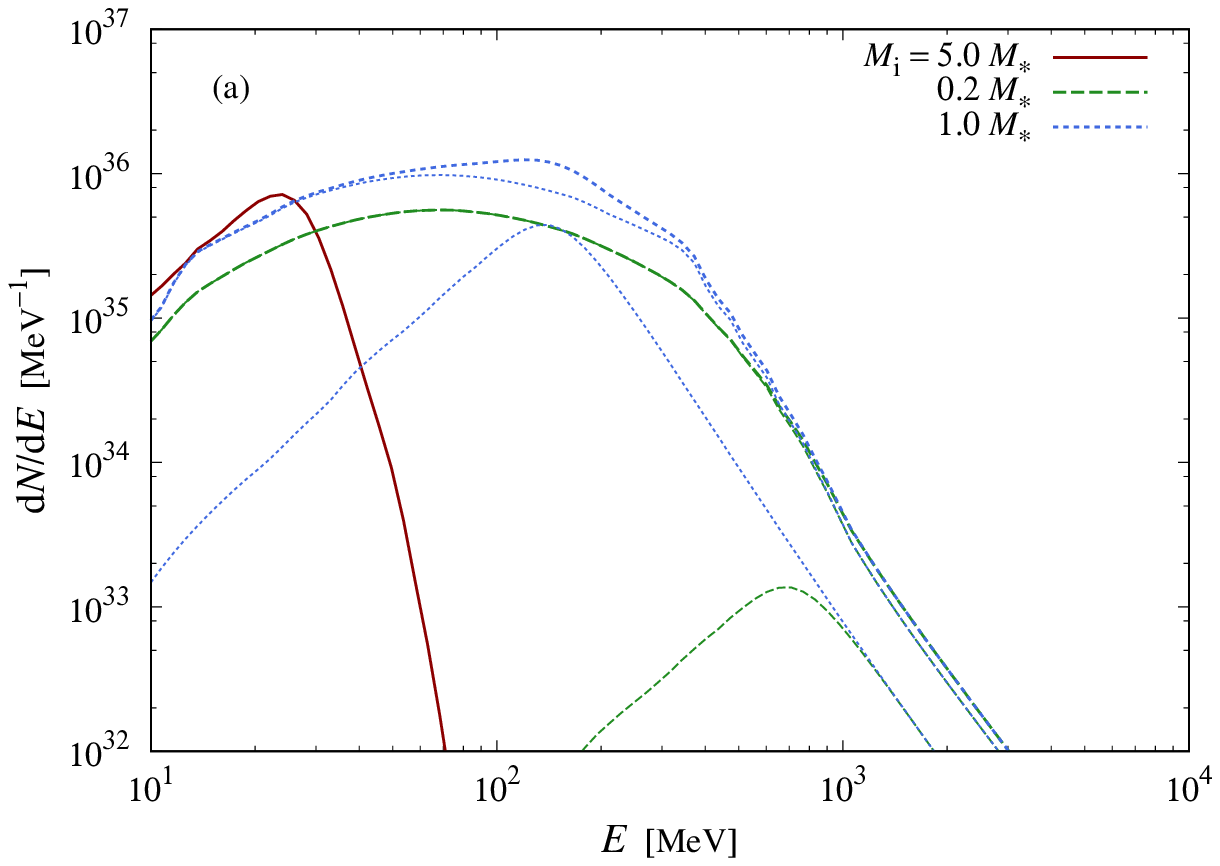}
\includegraphics[scale=0.65]{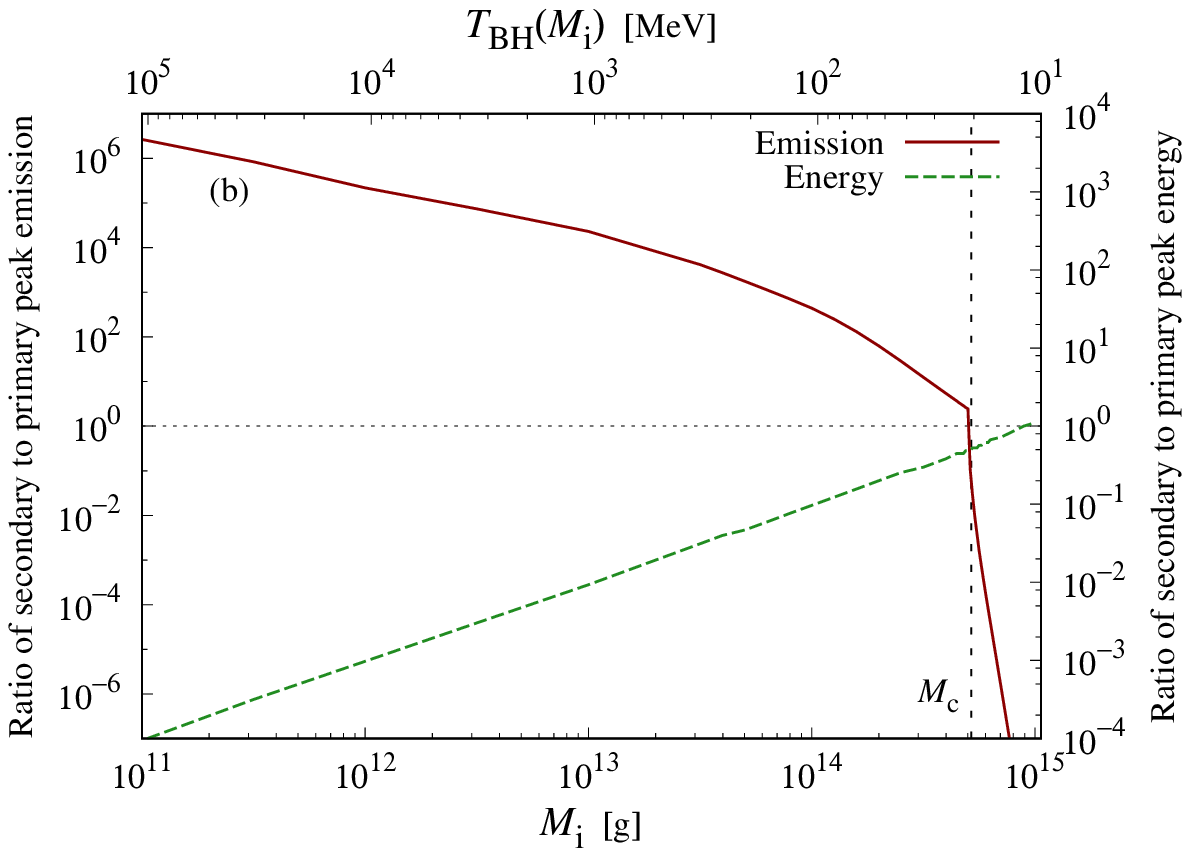}
\caption{\label{fig:ratios_integrated}
(a) Time-integrated spectra for different values of $ M_\mathrm i $\,.
(b) Ratios of secondary to primary peak energies (solid red) and fluxes (dashed green) for the time-integrated emission as a function of $ M_\mathrm i $\,.
}
\end{figure}

For $ M_* > M_\mathrm i > M_\mathrm q $\,, one again has secondary emission once $ M $ falls below $ M_\mathrm q $ but $ m = 0 $ and effectively all the emission occurs at the redshift $ z_\mathrm{evap}(M_\mathrm i) $ of evaporation.
This can be accounted for by replacing $ E $ by the \textit{present} photon energy $ E_0 = E\,(1+z_\mathrm{evap})^{-1} $ but we do not include this dependence explicitly here.
Figure~\ref{fig:ratios_integrated}(a) shows that secondary emission dominates at all energies except around $ E \sim M_\mathrm i^{-1} $\,.
We therefore have
\begin{equation}
\frac{\mathrm dN^\mathrm S}{\mathrm dE}
= \int^0_{M_\mathrm q}\!\frac{\mathrm d\dot N^\mathrm S}{\mathrm dE}\,
  \frac{\mathrm dt}{\mathrm dM}\,\mathrm dM
\propto
\begin{cases}
E\,M_\mathrm q^4
& (E < M_\mathrm q^{-1}) \\
E^{-3}
& (E > M_\mathrm q^{-1})\,.
\end{cases}
\end{equation}
For $ E \leq M_\mathrm q^{-1} $\,, the mass integral is dominated by the limit $ M_\mathrm q $\,, while for $ E > M_\mathrm q^{-1} $ it is dominated by $ M \sim E^{-1} $\,.
This is the same as Eq.~\eqref{eq:int_sec} but without the exponential cut-off.

For $ M_\mathrm i < M_\mathrm q $\,, secondary emission always dominates and it is generated at the redshift $ z_\mathrm{evap}(M_\mathrm i) $\,, so we have
\begin{equation}
\frac{\mathrm dN^\mathrm S}{\mathrm dE}
= \int^0_{M_\mathrm i}\!\frac{\mathrm d\dot N^\mathrm S}{\mathrm dE}\,
  \frac{\mathrm dt}{\mathrm dM}\,\mathrm dM
\propto
\begin{cases}
E\,M_\mathrm i^2\,M_\mathrm q^2
& (E < M_\mathrm q^{-1}) \\
E^{-1}\,M_\mathrm i^2
& (M_\mathrm q^{-1} < E < M_\mathrm i^{-1}) \\
E^{-3}
& (E > M_\mathrm i^{-1})\,.
\end{cases}
\label{eq:int_sec2}
\end{equation}
The first expression reflects the form of the low-energy fragmentation function and applies for $ \chi\,M_\mathrm i \ll M_\mathrm q $ and $ \epsilon \ll 1 $.
For $ E < M_\mathrm i^{-1} $\,, the mass integral is dominated by the limit $ M_\mathrm i $\,, while for $ E > M_\mathrm i^{-1} $ it is dominated by the mass $ M \sim E^{-1} $\,.
One again has a high-energy $ E^{-3} $ tail but there is no exponential cut-off because $ m = 0 $.

The forms of Eqs.~\eqref{eq:int_pri} to \eqref{eq:int_sec2} are shown in Fig.~\ref{fig:ratios_integrated}(a).
The ratio of the secondary and primary time-integrated peak energies is still given roughly by Eq.~\eqref{eq:energyratio}, as illustrated in Fig.~\ref{fig:ratios_integrated}(b).
For many effects, one is interested in the time-integrated fraction of a black hole's initial mass going into secondaries.
For $ M_\mathrm i < M_* $\,, the fraction is
\begin{equation}
f_\mathrm S
\approx
\begin{cases}
1 - (M_\mathrm i/M_\mathrm q)^5
& (M_\mathrm i < M_\mathrm q) \\
M_\mathrm q/M_\mathrm i
& (M_\mathrm q < M_\mathrm i< M_*)\,,
\end{cases}
\end{equation}
where the first expression accounts for the fact that even PBHs with $ M_\mathrm i < M_\mathrm q $ have some primary emission because of the Rayleigh--Jeans part of the emission.
The fraction falls off sharply above $ M_* $ because $ m $ falls below $ M_\mathrm q $ only when $ \mu $ is less than the tiny value $ \mu_\mathrm c \approx 0.005 $.
Since the mass radiated into secondary particles by the present epoch is $ M_\mathrm q - m $\,, the fraction of the initial mass going into secondaries over the narrow range between $ M_* $ and $ M_\mathrm c $ is
\begin{equation}
f_\mathrm S
\approx
  0.4\,(1-M_\mathrm i/M_\mathrm c)
\quad
(M_* < M_\mathrm i < M_\mathrm c\,, 0 < \mu < 0.005)
\label{eq:f_S}
\end{equation}
and this goes to zero at $ M_\mathrm i = M_\mathrm c $\,.
However, even for PBHs with $ M_\mathrm i > M_\mathrm c $\,, there will still be some secondary emission, albeit exponentially reduced by the Wien factor.
We therefore write the secondary fraction as
\begin{equation}
f_\mathrm S
\approx
  0.4\,\exp(-\chi\,M_\mathrm i/M_\mathrm c)
\quad
(M_\mathrm i > M_\mathrm c\,, \mu > 0.005)
\end{equation}
for some constant $ \chi $\,.
This formula extends the linear expression \eqref{eq:f_S} beyond $ M_\mathrm c $ for $ m \ll M_\mathrm q $\,.

\section{\label{sec:mf}
PBH mass function
}

In determining the Galactic $ \gamma $-ray background, one is mainly interested in the effects of PBHs with initial mass around $ M_* $\,.
Smaller ones have already evaporated and much larger ones are too cool to contribute appreciably.
In fact, we will find that the dominant contribution comes from PBHs with initial mass slightly larger than $ M_* $\,, since these are the ones which have not quite completed their evaporation.
As discussed in Sec.~\ref{sec:mf_tail}, these generate a current mass function with a low-mass tail below $ M_* $\,, whose form can be predicted very precisely.
While the extragalactic background is dominated by the time-integrated emission of PBHs with $ M \approx M_* $\,, the Galactic background is dominated by the instantaneous emission from this low-mass tail.

In our previous paper \cite{Carr:2009jm} we assumed that the PBHs have an effectively monochromatic mass function at formation.
This assumption was useful in discussing cosmological constraints on the fraction of the Universe $ \beta(M) $ going into PBHs with a specific mass.
However, the low-mass tail would not exist if the initial mass function were \textit{precisely} monochromatic at $ M_* $\,, so one needs at least some spread of mass.
One might naturally expect a spread $ \Delta M $ comparable to $ M_* $ but even a tiny spread (with $ \Delta M \ll M_* $) would suffice to generate an extended low-mass tail of PBHs at the current epoch.
We discuss these nearly-monochromatic scenarios in Sec.~\ref{sec:mf_mono}.
We also consider a variant of these scenarios in which the mass spectrum is narrow but not centred at exactly $ M_* $\,.
Of course, \textit{a priori} a narrow mass function is unlikely to contain or be close to the mass $ M_* $ but it is still interesting to calculate the associated Galactic background.

In practice, a nearly-monochromatic mass function at \textit{any} $ M $ may be implausible for realistic formation processes.
Even if the density fluctuations producing the PBHs peak on some scale, the resulting PBH mass spectrum may still be quite broad.
In some scenarios (such as PBH formation from cosmic strings) the spectrum may cover a wide range of masses and have no peak at all.
These considerations motivate us to consider in Sec.~\ref{sec:mf_ext} scenarios in which the mass function is extended.
A particular realization of this scenario, discussed in Sec.~\ref{sec:mf_cc}, arises if the PBHs form from primordial density perturbations as a result of critical collapse.
In this case, the mass function extends well below the peak and can be predicted rather precisely.
Coincidentally, its form is close to that of the low-mass tail.

\subsection{\label{sec:mf_tail}
Low mass tail
}

We first discuss the low-mass tail effect and its connection with the high-energy tail described in Sec.~\ref{sec:BH_tail}.
For simplicity, we take the \textit{formation} mass function to have the power-law form
\begin{equation}
\frac{\mathrm dn}{\mathrm dM}
= \left(\frac{\mathrm dn}{\mathrm dM}\right)_*\,
  \left(\frac{M}{M_*}\right)^\nu
\label{eq:mf_pl}
\end{equation}
in some mass range containing $ M_* $\,, where the exponent $ \nu $ is arbitrary.
The expression~\eqref{eq:massloss} for the evaporation rate then implies that the \textit{current} mass function is
\begin{equation}
\frac{\mathrm dn}{\mathrm dm}
= \left(\frac{m}{M_*}\right)^2\,
  \left[\frac{1}{1+\mu(m)}\right]^2\,
  \left(\frac{\mathrm dn}{\mathrm dM}\right)
\approx
  \left(\frac{m}{M_*}\right)^2\,
  \left(\frac{\mathrm dn}{\mathrm dM}\right)_*
\quad
(M_\mathrm q \leq m \ll M_*)\,,
\label{eq:currmf_pl}
\end{equation}
both mass functions being comoving.
The first expression is exact, with $ \mu(m) $ being implicitly determined by Eq.~\eqref{eq:mmu}, while the second expression applies for $ \mu \ll 1 $.
In the latter case, $ m \approx (3 \mu)^{1/3}\,M_* $ for $ \mu > \mu_\mathrm d \approx 0.02 $ from Eq.~\eqref{eq:mmu_approx} and the integrated number density of holes with mass below $ m $ can be approximated by
\begin{equation}
n(m)
\approx
  \frac{1}{3}\,\left(\frac{m}{M_*}\right)^3\,n_*
\quad
(M_\mathrm q \leq m \ll M_*)\,,
\label{eq:n_curr}
\end{equation}
where $ n_* \equiv M_*\,(\mathrm dn/\mathrm dM)_* $ is the original comoving number density of PBHs with mass around $ M_* $\,.
For $ m \leq M_\mathrm q $\,, an extra factor $ \alpha^{-1} \approx 1/4 $ appears on the right-hand-side of Eq.~\eqref{eq:currmf_pl} and $ m \approx (3 \alpha \mu)^{1/3}\,M_* $\,.
Therefore we can approximate the current mass function by
\begin{equation}
\frac{\mathrm dn}{\mathrm dm}
= \left[
    \frac{1}{\alpha}\,
    \left(\frac{m}{M_*}\right)^2\,
    \left(\frac{\mathrm dn}{\mathrm dM}\right)_*\,,
    \left(\frac{m}{M_*}\right)^2\,
    \left(\frac{\mathrm dn}{\mathrm dM}\right)_*\,,
    \left(\frac{\mathrm dn}{\mathrm dM}\right)
  \right]
\quad\textnormal{for}\quad
[m < M_\mathrm q\,, M_\mathrm q < m < M_*\,, m > M_*]\,.
\end{equation}
This is the same as the formation mass function well above $ M_* $ (i.e.\ $ \mathrm dn/\mathrm dm \approx \mathrm dn/\mathrm dM $ for $ \mu \gg 1 $), reflecting the fact that $ m \approx M $ in this regime.
More precisely, Eq.~\eqref{eq:mmu} implies
\begin{equation}
m/M
\approx
  [1-0.95\,(1+\mu)^{-3}]^{1/3}
\quad
(\mu > \mu_\mathrm c)\,,
\end{equation}
where the factor $ 0.95 $ corresponds to $ 1-(1-\alpha^{-1})\,q^3 $\,.
For example, the ratio is $ 0.95 $ for $ \mu = 2 $ and $ 0.99 $ for $ \mu = 3 $.
For $ m \ll M_* $\,, we have $ \mathrm dn/\mathrm dm \propto m^2 $ and we describe this as the ``low mass tail''.
For intermediate values of $ m $ ($ \mu \sim 1 $), Eqs.~\eqref{eq:mmu} and \eqref{eq:currmf_pl} imply that the local slope of the mass function is given by
\begin{equation}
\frac {\mathrm dn}{\mathrm dm}
\propto
  m^\beta\,,
\quad
\beta
= 2-(2-\nu)\,\frac{\ln[(m/M_*)^3 + 0.95]}{3\,\ln(m/M_*)}
\approx
\begin{cases}
2
& (m \ll M_*) \\
\nu
& (m \gg M_*)\,.
\end{cases}
\label{eq:slope_mf}
\end{equation}

\begin{figure}[htb]
\includegraphics[scale=0.65]{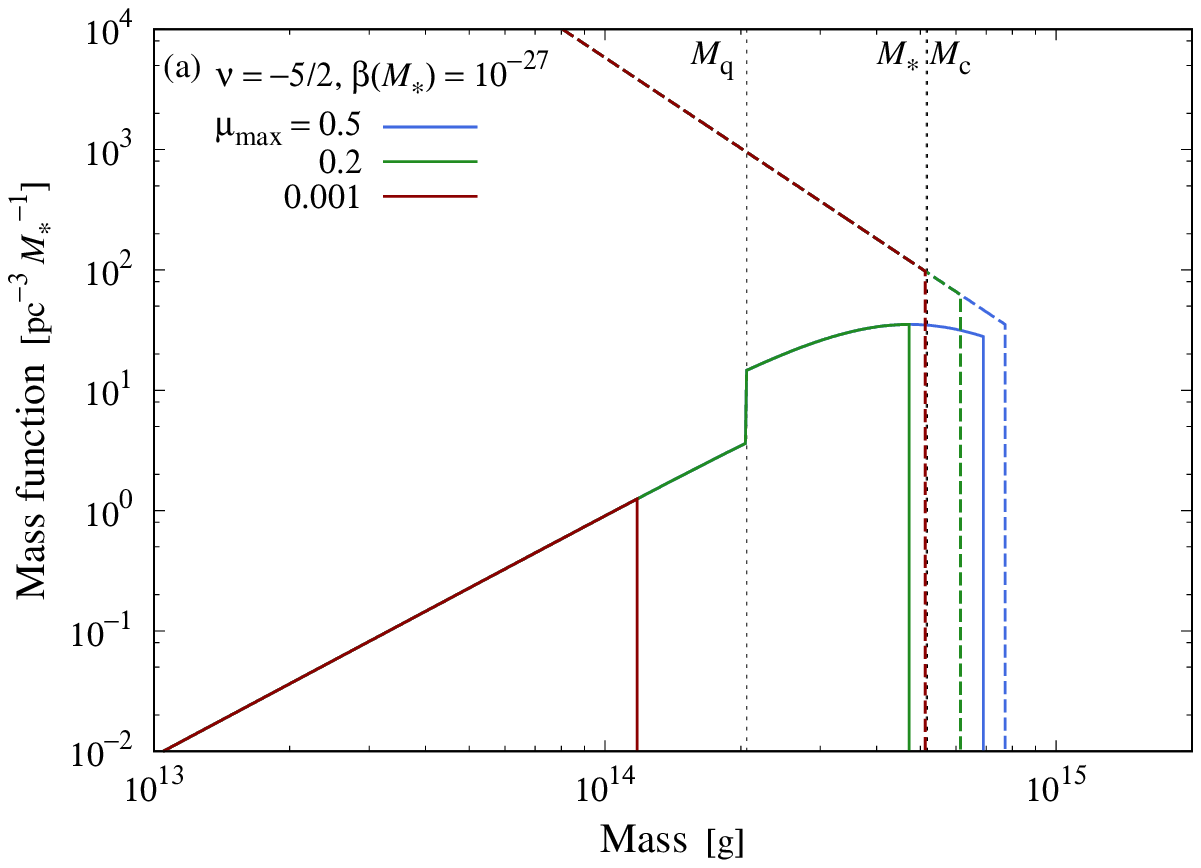}
\includegraphics[scale=0.65]{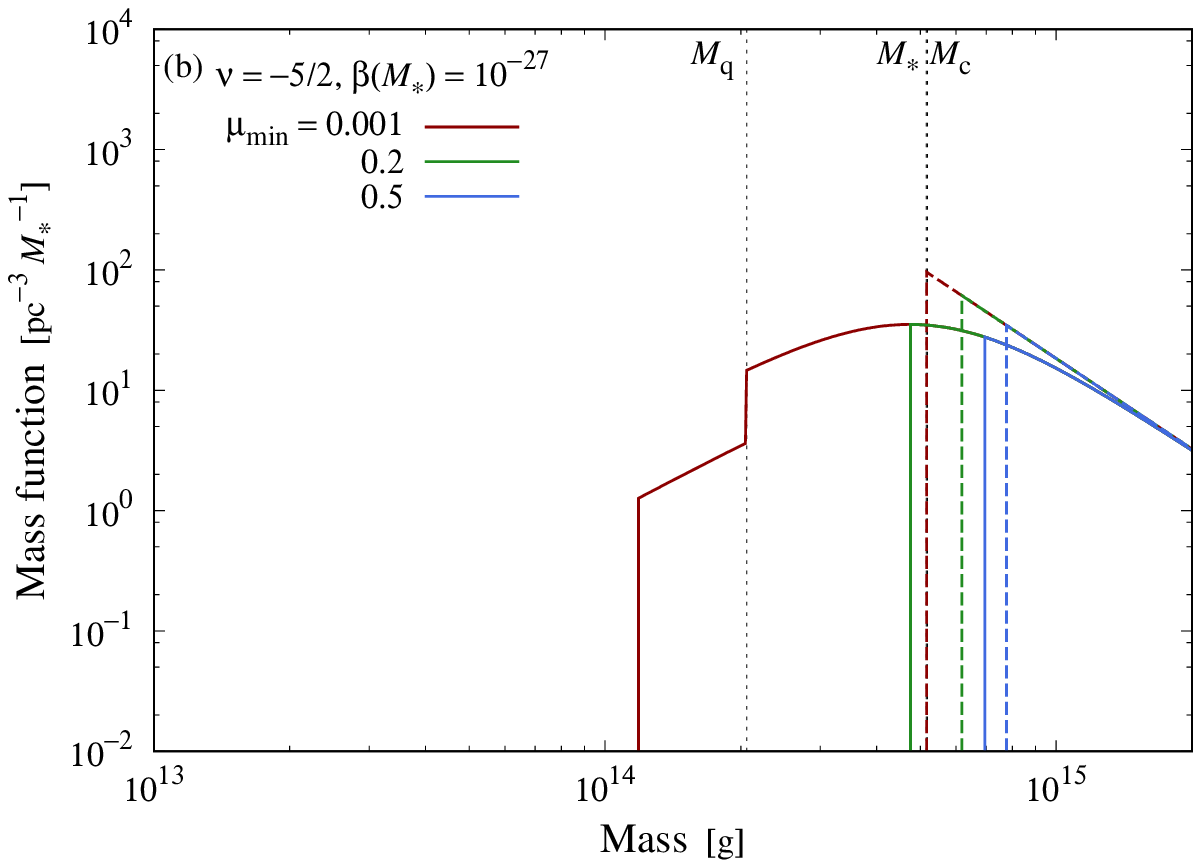}
\caption{\label{fig:mf}
This shows the relationship between the PBH mass functions at formation (dotted) and currently (solid) for extended mass functions with a fine-tuned upper (a) or lower (b) cut-off.
}
\end{figure}

We stress that the low mass tail is only present if the formation mass function contains or gets sufficiently close to $ M_* $\,.
One has three possible situations: (1) an extended mass function which reaches $ M_* $ from below and extends slightly above it; (2) an extended mass function which nearly reaches $ M_* $ from above but not quite; (3) an extended mass function which contains $ M_* $ and goes well below and well above it.
The first two cases correspond to a fine-tuning of the upper or lower cut-off and the degree of fine-tuning determines how much of the tail is present.
For example, if the initial mass function extends from below $ M_* $ to $ (1+\mu)\,M_* $ with $ \mu \ll 1 $, then the bottom of the tail appears between $ 0 $ and $ (3\mu)^{1/3}\,M_* $; the whole tail up to $ M_* $ appears for $ \mu \sim 1 $.
On the other hand, if the initial mass function extends down to $ (1+\mu)\,M_* $\,, then only the top of the tail above $ (3\mu)^{1/3}\,M_* $ appears and there will be very little tail if $ \mu \sim 1 $.
Such fine-tuning would not be expected \textit{a priori} but we discuss this for completeness.
These two situations are represented in Figs.~\ref{fig:mf}(a) and (b) for a spectrum with $ \nu = -5/2 $.
If the initial mass function extends well below and well above $ M_* $\,, as in the more likely case (3), then the entire tail is present.

Since the photon production rate of an individual hole is $ \dot N \propto m^{-1} $\,, the instantaneous flux from the tail population is $ I \propto n(m)\,m^{-1} \propto m^2\,\propto E^{-2} $\,, where we have used Eq.~\eqref{eq:n_curr}.
This relates to the high energy $ E^{-3} $ tail of the time-integrated emission from PBHs with $ M \leq M_* $\,, given by Eqs.~\eqref{eq:int_pri} to \eqref{eq:int_sec2}.
However, the connection between the low mass tail and high energy tail requires some clarification.
All PBHs generate an $ E^{-3} $ energy tail \textit{eventually} but only those with $ M \approx M_* $ produce one at the present epoch and those with mass only slightly above $ M_* $ do not produce the \textit{entire} energy tail because they have still not completed their evaporation (i.e.\ the highest energy part is missing).
But it is precisely these unevaporated remnants which provide the low-mass tail, so the energy and mass tails are complementary.

\subsection{\label{sec:mf_mono}
Nearly monochromatic initial mass function
}

\begin{figure}[htb]
\includegraphics[scale=0.65]{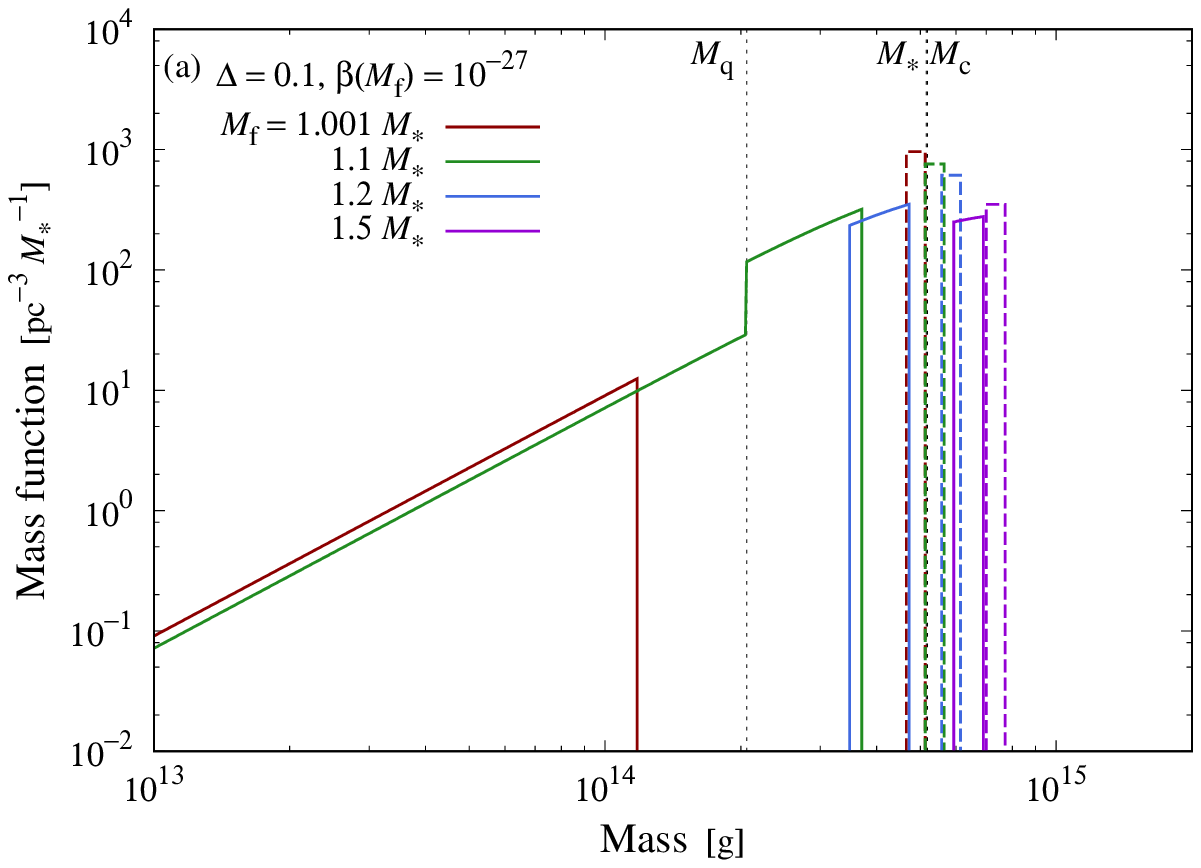}
\includegraphics[scale=0.65]{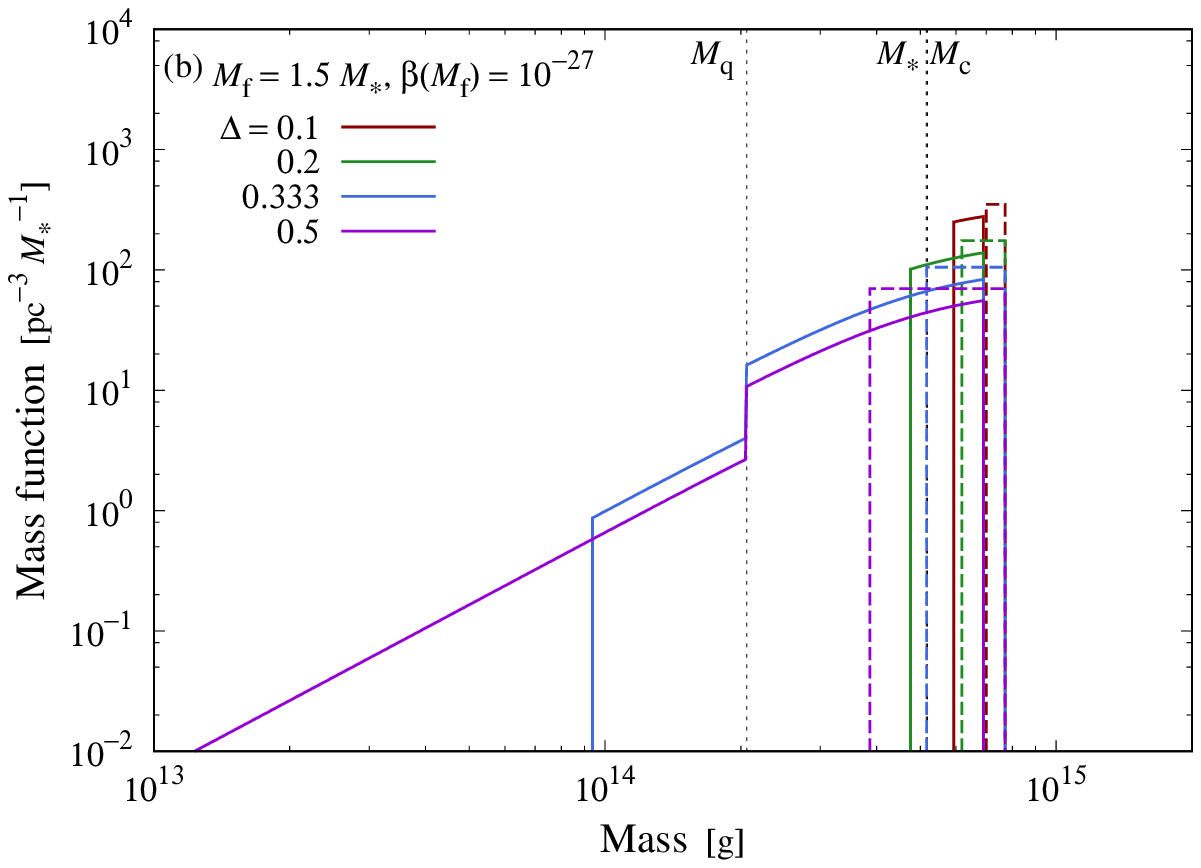}
\caption{\label{fig:mf_mono}
(a) Nearly-monochromatic mass functions with different mass scales $ M_\mathrm f $ but the same fractional width $ \Delta = 0.1 $.
The dashed lines represent the initial mass functions and the solid lines the current ones, the color coding the value of $ M_\mathrm f $\,.
(b) Nearly-monochromatic mass functions with $ M_\mathrm f = 1.5 M_*$ but different values of $ \Delta $\,.
In each case $ n $ is normalised so that $ \beta(M_\mathrm f) = 10^{-27} $, the maximum value consistent with the extragalactic $ \gamma $-ray background.
}
\end{figure}

One may model a nearly monochromatic initial mass function using a top-hat distribution
\begin{equation}
\frac{\mathrm dn}{\mathrm dM}
=
\begin{cases}
\frac{n}{\Delta M_\mathrm f}
& ((1-\Delta)M_\mathrm f < M < M_\mathrm f) \\
0
& (M < (1-\Delta)\,M_\mathrm f\,, M_\mathrm f < M)
\end{cases}
\label{eq:mf_tophat}
\end{equation}
with fractional width $ 0 < \Delta < 1 $.
Here $ n $ is the total comoving number density.
If the PBHs form in a radiation-dominated era, this is related to $ \beta $\,, the fraction of the Universe collapsing at their formation epoch, by
\begin{equation}
\frac{n\,M_\mathrm f}{\rho_\mathrm{CMB}}
= \beta(M_\mathrm f)\,(1+z_\mathrm f)
\propto
  \beta(M_\mathrm f)\,M_\mathrm f^{-1/2}
\quad\Rightarrow\quad
n
\approx
  96.5\,\mathrm{pc}^{-3}\,
  \left(\frac{\beta(M_\mathrm f)}{10^{-27}}\right)\,
  \left(\frac{M_\mathrm f}{M_*}\right)^{-3/2}\,.
\label{eq:nbeta}
\end{equation}
For $ M_\mathrm q < m < \mathcal O(10)\,M_* $ (i.e.\ $ M_\mathrm c < M < \mathcal O(10)\,M_* $), the number of emission degrees of freedom is approximately $ \phi_* $ all the way until $ t_0 $\,, so Eqs.~\eqref{eq:currmf_pl} and \eqref{eq:mmu} imply that the current mass function is
\begin{equation}
\frac{\mathrm dn}{\mathrm dm}
\approx
  \left(1+\frac{\phi_*\,t_0}{m^3}\right)^{-2/3}\,
  \frac{n}{\Delta\,M_\mathrm f}
\quad
(M_\mathrm q < m < \mathcal O(10)\,M_*)\,.
\end{equation}
This just scales as $ m^2 $ for $ m \ll M_* $\,, so the monochromatic initial mass function is spread out into the expected low-mass tail.
Fig.~\ref{fig:mf_mono} depicts the evolution of typical mass functions of interest.

In Fig.~\ref{fig:mf_mono}(a), the initial (dashed) and current (solid) mass functions are plotted for $ M_\mathrm f = 1.001\,M_*, 1.1\,M_*\,, 1.2\,M_*\,, 1.5\,M_* $ with fixed fractional width $ \Delta = 0.1 $.
For $ M_\mathrm f = 1.1\,M_* $ (green), the initial spectrum extends down to $ 0.99\,M_* $\,, so there is an extensive $ m^2 $ tail at the present epoch with a drop at $ M_\mathrm q $\,.
For $ M_\mathrm f = 1.2\,M_* $ (blue), the spectrum extends down to $ 1.08\,M_* $\,, so the tail is quite short and does not reach the drop at $ M_\mathrm q $\,.
For $ M_\mathrm f = 1.5\,M_* $ (magenta), the spectrum only extends down to $ 1.35\,M_* $\,, so the mass function does not evolve significantly and there is no appreciable tail at all.
For $ M_\mathrm f = 1.01\,M_*$ (red), the initial spectrum does not extend up to $ M_\mathrm c $\,, so only the bottom end of the low-mass tail appears.
Plotted in Fig.~\ref{fig:mf_mono}(b) are mass functions with different values of $ \Delta $ for $ M_\mathrm f = 1.5\,M_* $\,.
For $ \Delta = 0.333 $, $ (1-\Delta)\,M_\mathrm f $ is slightly above $ M_\mathrm c $\,, so only the top end of the low-mass tail appears.
Note that $ M_\mathrm c $ is so close to $ M_* $ that these scales cannot be distinguished in the figures.

\subsection{\label{sec:mf_ext}
Extended initial mass function
}

We now consider the consequences of the PBHs having an initial mass function which extends well above $ M_* $\,, thereby generating the entire low-mass tail below $ M_* $\,.
We assume this has the form given by Eq.~\eqref{eq:mf_pl} over some mass range $ M_\mathrm{min} < M < M_\mathrm{max} $\,, where the exponent $ \nu $ is arbitrary and may be positive or negative.
In the present context, we assume $ M_\mathrm{min} = M_* $\,, since PBHs smaller than $ M_* $ are irrelevant to the Galactic gamma-ray background, and we use the notation $ M_\mathrm{max} = M_\mathrm f $\,.
If the PBHs form from initial scale-invarant density fluctuations, $ \beta(M) $ is independent of $ M $ and $ \nu = -5/2 $ for PBHs forming in a radiation-dominated era \cite{Carr:1975qj} but one does not expect this in general.
Later we discuss a fairly generic scenario in which $ \nu $ is close to $ 2 $.
From Eq.~\eqref{eq:slope_mf}, the current mass function is
\begin{equation}
\frac{\mathrm dn}{\mathrm dm}
\approx
  \left(\frac{\mathrm dn}{\mathrm dM}\right)_*
  \times
\begin{cases}
\left(\frac{m}{M_*}\right)^2
& (m < M_*) \\
\left(\frac{m}{M_*}\right)^\nu
& (m > M_*)\,.
\end{cases}
\end{equation}
For the purpose of numerical computations, we normalize $ (\mathrm dn/\mathrm dM)_* $ to $ 100\,\mathrm{pc}^{-3}\,M_*^{-1} $\,.
Then $ n(M_*) \sim M_*\,(\mathrm dn/\mathrm dM)_* \sim 100\,\mathrm{pc}^{-3} $\,, which corresponds to $ \beta(M_*) \sim 10^{-27} $ from Eq.~\eqref{eq:nbeta}, and this is the maximum value allowed by the extragalactic $ \gamma $-ray background limit.

Independent of the $ \gamma $-ray background constraint at $ M_* $\,, there will be another constraint if $ \nu > -2 $ since black holes with mass around $ M $ will have a density parameter $ \Omega_\mathrm{PBH}(M) \propto M^{2+\nu} $ which increases with $ M $\,.
Therefore the mass function will be constrained by dark matter observations at the largest value of $ M $\,.
Indeed, as emphasized by Yokoyama \cite{Yokoyama:1998xd}, one could envisage a scenario in which $ M_\mathrm f $ is sufficiently large that the PBHs could provide the dark matter even if the mass function is strongly constrained at $ M_* $\,.
More precisely, the density parameter of the PBHs with mass around $ M_* $ would be
\begin{equation}
\Omega_\mathrm{PBH}(M_*)
= \Omega_\mathrm{PBH}(M_\mathrm f)\,(M_*/M_\mathrm f)^{2+\nu}
= f\,\Omega_\mathrm{CDM}\,(M_\mathrm f/10^{15}\,\mathrm g)^{-2-\nu}\,,
\end{equation}
where $ f $ is the fraction of the Cold Dark Matter density in the PBHs.
For PBHs to explain both the extragalactic background, $ \Omega_\mathrm{PBH}(M_*)\approx 10^{-9} $, and the dark matter, $ \Omega_\mathrm{PBH}(M_\mathrm f) = \Omega_\mathrm{CDM} \approx 0.2 $, one would require
\begin{equation}
M_\mathrm f
\approx
  10^{(39+15\nu)/(2+\nu)}\,\mathrm g
\approx
\begin{cases}
10^{17}\mathrm g
& (\nu =2) \\
10^{20}\mathrm g
& (\nu =0) \\
10^{24}\mathrm g
& (\nu =-1)\,.
\end{cases}
\label{eq:cdm}
\end{equation}
If $ \nu < -2 $, $ \Omega_\mathrm{PBH}(M) $ decreases with increasing $ M $\,, so there is no dark matter constraint.

\subsection{\label{sec:mf_cc}
Critical mass function
}

It is well known that black hole formation is associated with critical phenomena and this applies, in particular, to PBHs which form from initial density fluctuations \cite{Choptuik:1992jv,Abrahams:1993wa,Evans:1994pj,Koike:1995jm,Niemeyer:1997mt}.
In this section, we will show that this implies that the PBH mass spectrum should have a simple power-low form which extends well below the peak mass.
This scenario has been considered before \cite{Yokoyama:1998xd} but we generalize it to a wider range of situations.

Let us assume a monochromatic power spectrum for the density fluctuations on some mass scale and identify the amplitude of the density fluctuation when that scale crosses the horizon, $ \delta $\,, as the control parameter.
Then the black hole mass is given by
\begin{equation}
M
= K\,(\delta - \delta_\mathrm c)^c\,,
\label{eq:M_cc}
\end{equation}
where $ \delta_\mathrm c $ is the critical magnitude of the density fluctuation required for PBH formation ($ 0.4 $ in a radiation-dominated era), the exponent has a universal value $ c \approx 0.35 $ \cite{Evans:1994pj,Koike:1995jm,Niemeyer:1997mt} and $ K $ is a mass scale of order the horizon mass $ M_\mathrm H $\,.
Although the scaling relation \eqref{eq:M_cc} is expected to be valid only in the immediate neighborhood of $ \delta_\mathrm c $\,, most black holes are expected to form from fluctuations with this value because the probability distribution function (PDF) declines exponentially beyond $ \delta = \delta_\mathrm c $\,.
Hence it is sensible to calculate the expected mass function of PBHs using the formula \eqref{eq:M_cc}.
This allows us to estimate the mass function fairly independently of the specific form of the PDF of primordial density or curvature fluctuations.

Since the PDF of $ \delta $\,, $ P(\delta) $\,, is a steeply declining function around $ \delta_\mathrm c $\,, we write it as
\begin{equation}
P(\delta)\,\mathrm d\delta
= \mathrm e^{-f(\delta)}\,\mathrm d\delta\,,
\label{eq:pdf}
\end{equation}
where $ f(\delta) $ is a well-behaved function around $ \delta \simeq \delta_\mathrm c $\,.
If $ \delta $ has a Gaussian distribution, as assumed in most literature \cite{Carr:1975qj,Carr:1993aq,*Carr:1994ar,Green:1997sz}, $ f(\delta) $ is given by
\begin{equation}
f(\delta)
= \ln\left(\sqrt{2 \pi}\,\sigma\right) + \frac{\delta^2}{2\,\sigma^2}\,,
\label{eq:Gauss}
\end{equation}
with $ \sigma $ being the dispersion of $ \delta $\,.
Then the probability, $ \beta(M_\mathrm H) $\,, that the relevant horizon mass-scale has a fluctuation above the threshold required for black hole formation is given by
\begin{equation}
\beta(M_\mathrm H)
= \int_{\delta_\mathrm c}\!P(\delta)\,\mathrm d\delta
= \int_{\delta_\mathrm c}\!\mathrm e^{-f(\delta)}\,\mathrm d\delta\,.
\end{equation}
This is the volume fraction of the regions collapsing to PBHs at time $ t_\mathrm f $\,.
Since $ P(\delta) $ is a steeply decreasing function, the integral is sensitive to only its lower bound $ \delta_\mathrm c $\,.
Furthermore we can Taylor expand $ f(\delta) $ as
\begin{equation}
f(\delta)
= f(\delta_\mathrm c)
  + f'(\delta_\mathrm c)\,(\delta - \delta_\mathrm c)
  + \cdots
\equiv
  f_\mathrm c
  + s\,(\delta-\delta_\mathrm c)
  + \cdots\,,
\label{eq:Taylor}
\end{equation}
to find
\begin{equation}
\beta(M_\mathrm H)
\approx
  \frac{1}{s}\,\mathrm e^{-f_\mathrm c}\,.
\end{equation}
Here $ s = \delta_\mathrm c/\sigma^2 $ and the approximation is applicable if $ |f''(\delta_\mathrm c)| \ll s^2 $\,.
For a Gaussian distribution, this corresponds to $ \sigma \ll \delta_\mathrm c $\,.
Putting $ \delta_\mathrm c = 1/3 $ and using \eqref{eq:Gauss}, we recover Carr's formula \cite{Carr:1975qj},
\begin{equation}
\beta
\simeq
  \sigma\,\exp[-1/(18\,\sigma^2)]\,.
\end{equation}
If $ P(\delta) $ is non-Gaussian, we must analyze on a case-by-case basis, but so long as $ \beta(M_\mathrm H) $ is small enough to satisfy the cosmological constraints and we use specific non-Gaussian distributions reported in the literature \cite{Bullock:1996at,Yokoyama:1998pt,Ivanov:1997ia}, we may justify \eqref{eq:Taylor}.

Let us present two specific examples.
One is the lognormal matter distribution
\begin{equation}
P_\mathrm{LN}(\delta)
= \frac{1}{(1+\delta)\,\sqrt{2\pi\,\ln(1+\sigma^2)}}\,
  \exp\left[
   -\frac{\left\{\ln\left[(1+\delta)\,\sqrt{1+\sigma^2}\right]\right\}^2}
         {2\,\ln\left(1+\sigma^2\right)}
  \right]\,,
\end{equation}
which is more plausible than Gaussian because it gives null probability for negative density and is known to fit the nonlinear density distribution of large-scale structures well \cite{1991MNRAS.248....1C,Bernardeau:1994aq}.
For $ \sigma^2 \ll 1 $, we find
\begin{align}
f_\mathrm{LN}(\delta_\mathrm c)
&
= \frac{\left\{\ln\left[(1+\delta_\mathrm c)\,\sqrt{1+\sigma^2}\right]\right\}^2}
  {2\,\ln\left(1+\sigma^2\right)}
  + \ln\left(1+\delta_\mathrm c\right)
  + \frac{1}{2}\,\ln\left[2\pi\,\ln(1+\sigma^2)\right]\,, \\
s_\mathrm{LN}
&
= f_\mathrm{LN}'(\delta_\mathrm c)
= \frac{1}{1+\delta_\mathrm c}\,\left[
   \frac{\ln\left[(1+\delta_\mathrm c)\,\sqrt{1+\sigma^2}\right]}
        {\ln\left(1+\sigma^2\right)}+1
  \right]
\simeq
  \frac{\ln\left(1+\delta_\mathrm c\right)}
       {(1+\delta_\mathrm c)\,\sigma^2}\,, \\
f_\mathrm{LN}''(\delta_\mathrm c)
&
\simeq
  -\frac{\ln(1+\delta_\mathrm c)-1}{(1+\delta_\mathrm{c})^2\,\sigma^2}\,.
\end{align}
Hence $ |f_\mathrm{LN}''(\delta_\mathrm c)| \ll s_\mathrm{LN}^2 $ corresponds to $ \sigma^2 \ll \delta_\mathrm c $\,, which is satisfied unless PBHs are overproduced.

Next we consider a more phenomenological description of non-Gaussianity where the density contrast is expressed in terms of a random Gaussian variable $ \delta_\mathrm G $ as
\begin{equation}
\delta
= \delta_\mathrm G + \tilde f_\mathrm{NL}\,\delta_\mathrm G^2
\label{eq:localNG}
\end{equation}
at each point \cite{Komatsu:2001rj}.
Such a model has been widely used in studies of the CMB and large-scale structure.
The local nonlinearity parameter $ f_\mathrm{NL} $ has been stringently constrained and can be at most $ \mathcal O(1) $ on the scales probed by the Planck satellite \cite{Ade:2015ava}.
It has been argued that $ \tilde f_\mathrm{NL} $ of order of unity significantly affects the PBH abundance, leading to significant constraints \cite{Young:2013oia}.
However, this is misleading because $ f_\mathrm{NL} = \mathcal O(1) $ on large scales corresponds to non-Gaussian corrections of order $ 10^{-5} $, while $ \tilde f_\mathrm{NL} = \mathcal O(1) $ on the PBH scale means corrections of $ 0.01\textnormal{--}0.1 $ because appreciable PBH formation requires $ \sqrt{\langle\delta_\mathrm G^2\rangle} \gtrsim 0.05 $.
Hence the same values of $ f_\mathrm{NL} $ and $ \tilde f_\mathrm{NL} $ can have totally different implications.
In fact, it is not sensible to identify $ f_\mathrm{NL} $ on large scale with $ \tilde f_\mathrm{NL} $\,.
Keeping these remarks in mind, Eq.~\eqref{eq:localNG} implies
\begin{equation}
P_\mathrm{loc}(\delta)
= \frac{1}{\sqrt{2\pi}\,\xi\,\sigma}
  \exp\left[
   -\frac{1}{2\,\sigma^2}\,
   \left(\frac{\xi-1}{2\,\tilde f_\mathrm{NL}}\right)^2
  \right]
\quad\textnormal{with}\quad
\xi
\equiv
  \sqrt{1+4\,\tilde f_\mathrm{NL}\,\delta}\,,
\end{equation}
so
\begin{align}
f_\mathrm{loc}(\delta_\mathrm c)
&
= \frac{1}{2\,\sigma^2}\,
  \left(\frac{\xi_\mathrm c-1}{2\,\tilde f_\mathrm{NL}}\right)^2
  - \frac{1}{2}\,\ln\left(2\pi\,\xi_\mathrm c^2\,\sigma^2\right)\,, \\
s_\mathrm{loc}
&
= f_\mathrm{loc}'(\delta_\mathrm c)
= \frac{1}{2\,\tilde f_\mathrm{NL}\,\sigma^2}\,\left(1-\frac{1}{\xi_\mathrm c}\right)
  - \frac{2\,\tilde f_\mathrm{NL}}{\xi _\mathrm c^2}\,, \\
f_\mathrm{loc}''(\delta_\mathrm c)
&
= \frac{1}{\sigma^2\,\xi_\mathrm c^3}
  + \frac{8\,\tilde f_\mathrm{NL}^2}{\xi_\mathrm c^4}\,.
\end{align}
The condition $ |f_\mathrm{loc}''(\delta_\mathrm c)| \ll s_\mathrm{loc}^2 $ is equivalent to $ \sigma^2 \ll \delta_\mathrm c $ for $ 4\,\tilde f_\mathrm{NL}\,\delta_\mathrm c \ll 1 $ and $ \sigma^2 \ll 2\,\tilde f_\mathrm{NL}^{1/2}\,\delta_\mathrm c^{3/2} \ll 1 $ for $ 4\,\tilde f_\mathrm{NL}\,\delta_\mathrm c \gg 1 $.
Both are satisfied for realistic abundance of PBHs.
Hence this simple approximation is widely applicable.

Using \eqref{eq:M_cc}, \eqref{eq:pdf} can be expressed in terms of $ M $ as
\begin{equation}
P(\delta)\,\mathrm d\delta
= P\left[\delta_\mathrm c+\left(\frac{M}{K}\right)^{1/c}\right]\,
  \left(\frac{M}{K}\right)^{1/c-1}\,
  \frac{\mathrm dM}{c\,K}\,,
\end{equation}
where the right-hand-side can be interpreted as the formation probability of a PBH with mass $ M $ in each horizon volume as the fluctuation peak enters the Hubble radius at $ t_\mathrm f $\,.
The comoving mass function of PBHs at formation is therefore
\begin{equation}
\frac{\mathrm dn}{\mathrm dM}
= \frac{1}{V_\mathrm f\,c\,K}\,
  \left(\frac{M}{K}\right)^{1/c-1}\,
  P\left[\delta_\mathrm c+\left(\frac{M}{K}\right)^{1/c}\right]\,,
\end{equation}
where $ V_\mathrm f $ is the \textit{comoving} horizon volume at $ t_\mathrm f $\,, related to the horizon mass $ M_\mathrm f $ at that time by
\begin{equation}
V_\mathrm f
= \frac{4\pi}{3}\,\left(\frac{M_\mathrm f}{4 \pi\,a(t_\mathrm f)}\right)^3\,.
\end{equation}
We therefore find
\begin{equation}
\frac{\mathrm dn}{\mathrm dM}
\simeq
  \frac{1}{V_\mathrm f\,c\,K}\,
  \left(\frac{M}{K}\right)^{1/c-1}\,
  \exp\left[-f_\mathrm c - s\,\left(\frac{M}{K}\right)^{1/c}\right]\,,
\label{eq:initmf_cc}
\end{equation}
which peaks at
\begin{equation}
M_\mathrm{peak}
= \gamma\,M_\mathrm f
\quad\textnormal{with}\quad
\gamma
\equiv
  \left(\frac{1-c}{s}\right)^c\,,
\quad
M_\mathrm f
= K\,.
\end{equation}
We can express \eqref{eq:initmf_cc} in terms of $ \gamma $ and $ M_\mathrm f $ as
\begin{equation}
\frac{\mathrm dn}{\mathrm dM}
\simeq
  \frac{\beta(M_\mathrm f)}{V_\mathrm f\,\gamma\,M_\mathrm f}\,
  \left(\frac{1-c}{c}\right)\,
  \left(\frac{M}{\gamma\,M_\mathrm f}\right)^{1/c-1}\,
  \exp\left[-(1-c)\,\left(\frac{M}{\gamma\,M_\mathrm f}\right)^{1/c}\right]\,.
\label{eq:initmf_cc2}
\end{equation}
Thus the possible non-Gaussianity of the density fluctuation PDF affects only $ \beta(M_\mathrm f) $ and the peak mass; the overall shape of the mass function is independent of the PDF apart from these two parameters.

\begin{figure}[htb]
\includegraphics[scale=0.65]{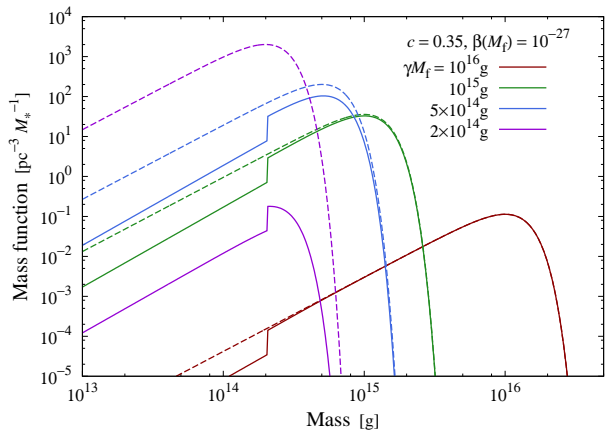}
\caption{\label{fig:mfs_cc}
Initial mass function (broken) and current mass function (solid) for critical collapse with $ \gamma\,M_\mathrm f = 10^{16}\,\mathrm g $ (red), $ 10^{15}\,\mathrm g $ (green), $ 5 \times 10^{14}\,\mathrm g $ (blue), and $ 2 \times 10^{14}\,\mathrm g $ (magenta).
For $ \gamma\,M_\mathrm f \gtrsim 10^{15}\,\mathrm g $\,, the broken and solid curves differ only below $ M_\mathrm q = 0.4\,M_* \approx 2 \times 10^{14}\,\mathrm g $\,, but the difference becomes more significant as $ \gamma\,M_\mathrm f $ decreases.
All curves have the value $ \beta(M_*) = 10^{-27} $ associated with the extragalactic $ \gamma $-ray background limit, corresponding to $ (\mathrm dn/\mathrm dM)_* \approx 100\,\mathrm{pc}^{-3}\,M_*^{-1} $\,.
}
\end{figure}

Equation~\eqref{eq:initmf_cc2} gives the mass function at the formation epoch.
In order to calculate the current (comoving) mass function, we must incorporate mass loss due to the Hawking radiation since this is important for holes with $ M \simeq M_* $\,.
The PBH mass at the formation time, $ M $\,, is related to the current mass, $ m $ through $ M^3 = M_*^3 + m^3 $\,.
The more exact expression is given by \eqref{eq:precisem}.
We can therefore write the current comoving mass spectrum as
\begin{equation}
\frac{\mathrm dn}{\mathrm dm}
\simeq
  \frac{\beta(M_\mathrm f)}{V_\mathrm f\,\gamma\,M_\mathrm f}\,
  \left(\frac{1-c}{c}\right)\,
  \left(\frac{M(m)}{\gamma\,M_\mathrm f}\right)^{1/c-1}\,
  \exp\left[
   -(1-c)\,\left(\frac{M(m)}{\gamma\,M_\mathrm f}\right)^{1/c}
  \right]
  \times
\begin{cases}
\frac{m^2}{M(m)^2}
& (m \geq M_\mathrm q) \\
\frac{m^2}{\alpha\,M(m)^2}
& (m \leq M_\mathrm q)
\end{cases}\,,
\label{eq:currmf_cc}
\end{equation}
where we have distinguished between the $ m > M_\mathrm q $ and $ m < M_\mathrm q $ cases.
This function is plotted in Fig.~\ref{fig:mfs_cc} for $ \beta(M_\mathrm f) = 10^{-27} $ and different values of $ M_\mathrm f $\,.
Since $ c \approx 0.35 $, it scales as $ M^{-0.15}\,m^2 $\,.
So coincidentally the effects of critical collapse on the initial mass function and of evaporations on the current mass function give slopes close to $ M^2 $ and $ m^2 $\,, respectively, for $ M \ll \gamma\,M_\mathrm f $\,.
Note that there would be another drop in the curves below $ 10^{12}\,\mathrm g $ if one allowed for the increase in $ f $ due to the emission of extra particles but this is not shown.
One can also obtain a constraint on the mass $ M_\mathrm f $ in order to avoid overproducing dark matter.
If one applies Eq.~\eqref{eq:cdm} with $ \nu \approx 1.85 $, this gives $ M_\mathrm f < 10^{17}\,\mathrm g $\,.

\section{\label{sec:flux}
Galactic gamma-ray flux
}

If the PBHs evaporating at the present epoch are clustered inside galactic halos (as expected), then there will be a Galactic background generated by PBHs with $ M \geq M_* $\,.
As discussed later, this is dominated by the low mass tail of PBHs with initial mass just above $ M_* $ but with an intensity increased by the local density enhancement.
While this mass tail makes a negligible contribution to the time-integrated extragalactic background, it is crucial for the Galactic background.
There would also be a Galactic contribution from PBHs which were slightly \textit{smaller} than $ M_* $ but sufficiently distant for their emitted particles to have only just reached us; since the light-travel time across the Galaxy is $ t_\mathrm{gal} \sim 10^5\,\mathrm{yr} $\,, this corresponds to PBHs initially smaller than $ M_* $ by $ (t_\mathrm{gal}/3t_0)\,M_* \sim 10^{-5}\,M_* $\,, so this contribution is very small.

In this section, an overbar denotes the cosmological average of a quantity.
For instance, $ \mathrm d\bar n/\mathrm dm $ is the average current mass function, which we discussed in the last section.
It is related to the average PBH mass density by
\begin{equation}
\bar\rho_\mathrm{PBH}
= \int_0^\infty\!\mathrm dm\,m\,\frac{\mathrm d\bar n}{\mathrm dm}\,,
\end{equation}
where the mass function will be zero below some lower limit and above some upper limit.
For simplicity, we assume that the local mass function of PBHs in the Galaxy is the same as the average except for the enhancement $ \rho_\mathrm{PBH}(\boldsymbol R)/\bar\rho_\mathrm{PBH} $\,, where $ \rho_\mathrm{PBH}(\boldsymbol R) $ is the local mass density of PBHs and $ \boldsymbol R $ represents the position in the Galaxy.
Then the local PBH mass function at $ \boldsymbol R $ is
\begin{equation}
\frac{\mathrm dn}{\mathrm dm}(\boldsymbol R)
= \frac{\mathrm d\bar n}{\mathrm dm}\,
  \frac{\rho_\mathrm{PBH}(\boldsymbol R)}{\bar\rho_\mathrm{PBH}}\,.
\end{equation}
The local photon emission rate from PBHs per unit energy per unit volume is
\begin{equation}
\mathcal E(E,\boldsymbol R)
= \int_0^\infty\!\mathrm dm\,
  \frac{\mathrm dn}{\mathrm dm}(\boldsymbol R)\,
  \frac{\mathrm d\dot N}{\mathrm dE}(m,E)
= \frac{\rho_\mathrm{PBH}(\boldsymbol R)}{\bar\rho_\mathrm{PBH}}\,
  \bar{\mathcal E}(E)\,,
\end{equation}
where expressions for $ \mathrm d\dot N/\mathrm dE $ were derived in Sec.~\ref{sec:BH} and $ \bar{\mathcal E} $ is the average emission rate, defined by
\begin{equation}
\bar{\mathcal E}(E)
= \int_0^\infty\!\mathrm dm\,
  \frac{\mathrm d\bar n}{\mathrm dm}\,
  \frac{\mathrm d\dot N}{\mathrm dE}(m,E)\,.
\label{eq:em_avg}
\end{equation}
Both $ \mathcal E $ and $ \bar{\mathcal E} $ have dimensions $ \mathrm s^{-1}\,\mathrm{MeV}^{-1}\,\mathrm{cm}^{-3} $\,.
We evaluate the average emissivity for various scenarios below.

We regard the position $ \boldsymbol R $ as a function of the line of sight vector $ \boldsymbol n $ and the distance $ r $ from the Sun (or us).
Consider photons emitted from PBHs at a distance $ r $ within a volume $ \mathrm dV = r^2\,\mathrm d\Omega\,\mathrm dr $\,, where $ \mathrm d\Omega $ denotes the solid angle of the volume relative to us.
A fraction $ 1/(4 \pi r^2) $ of these photons come into a unit area of our detector, so the photon number flux per unit solid angle from PBHs at a distance between $ r $ and $ r + \mathrm dr $ is
\begin{equation}
\mathrm d\Phi(\boldsymbol n,E)
= \frac{1}{4\pi}\,\mathcal E(E,\boldsymbol R(\boldsymbol n,r))\,\mathrm dr\,.
\end{equation}
Integrating over the radial distance gives the total flux
\begin{equation}
\Phi(\boldsymbol n,E)
= \frac{1}{4\pi}\,\int_0^\infty\!\mathrm dr\,
  \mathcal E(E,\boldsymbol R(\boldsymbol n,r))
= \frac{\bar{\mathcal E}(E)}{4\pi}\,\int_0^\infty\!\mathrm dr\,
  \frac{\rho_\mathrm{PBH}(\boldsymbol R(\boldsymbol n,r))}{\bar\rho_\mathrm{PBH}}
\label{eq:flux}
\end{equation}
with dimensions $ \mathrm s^{-1}\,\mathrm{MeV}^{-1}\,\mathrm{sr}^{-1}\,\mathrm{cm}^{-2} $\,.
The associated intensity is
\begin{equation}
I(E)
\equiv
  E\,\Phi(E)
\end{equation}
with dimensions $ \mathrm s^{-1}\,\mathrm{sr}^{-1}\,\mathrm{cm}^{-2} $\,.

\subsection{
Integration along line of sight
}

In order to carry out the integration along line of sight in Eq.~\eqref{eq:flux}, we assume a spherically symmetric PBH distribution about the centre of the Galaxy, $ \rho_\mathrm{PBH}(\boldsymbol R) = \rho_\mathrm{PBH}(R) $ with $ R $ denoting the Galactocentric distance.
Then we must compute the expression for $ \Phi(\boldsymbol n,E) $ given by Eq.~\eqref{eq:flux} by writing $ R $ in terms of the line-of-sight distance $ r $ and Galactic coordinates $ (b,l) $\,.
This gives
\begin{equation}
R(\boldsymbol n,r)
= \sqrt{r^2-2 r\,R_\odot\,\cos b\,\cos l+R_\odot^2}\,,
\end{equation}
with $ R_\odot = 9\,\mathrm{kpc} $ being the distance of the Sun from the Galactic centre.

\begin{figure}[htb]
\includegraphics[scale=0.65]{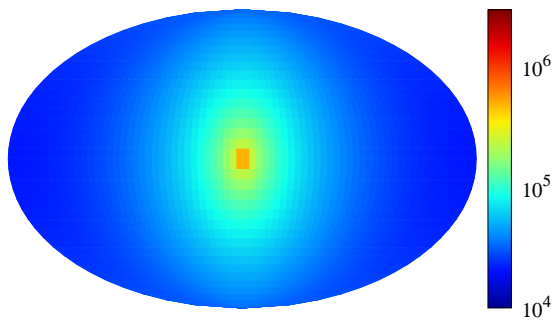}
\caption{\label{fig:gf}
Line-of-sight Galactic density enhancement $ g(\boldsymbol n) $ for an NFW profile with $ f = 1 $.
}
\end{figure}

For illustrative purposes we define a Galactic line-of-sight enhancement factor as
\begin{equation}
g(\boldsymbol n)
= \frac{1}{r_\mathrm{gal}}\,
  \int_0^{r_\mathrm{gal}}\!\mathrm dr\,
  \frac{\rho_\mathrm{PBH}(R(\boldsymbol n,r))}{\bar\rho_\mathrm{PBH}}\,,
\end{equation}
with $ r_\mathrm{gal} = 100\,\mathrm{kpc} $ being our distance from the edge of the dark matter halo (assumed direction-independent as an approximation).
We adopt the Navarro--Frenk--White (NFW) profile \cite{Navarro:1995iw},
\begin{equation}
\rho_\mathrm{PBH}(R)
= \frac{f\,\rho_\mathrm s}
       {(R/R_\mathrm s)^\gamma\,[1+(R/R_\mathrm s)^\alpha]^{(\beta-\alpha)/\alpha}}\,,
\end{equation}
with a set of best-fitting parameters taken from Diemand \textit{et al.} \cite{Diemand:2008in}:
$ \gamma = 1.24, \alpha = 1, \beta = 4-\gamma = 2.86, R_\mathrm s = 28.1\,\mathrm{kpc}\,, \rho_\mathrm s = 3.50 \times 10^{-3}\,M_\odot\,\mathrm{pc}^{-3} $\,.
Then we can compute $ g(\boldsymbol n) $ numerically and this is shown in Fig.~\ref{fig:gf}.

We shall require that the $ \gamma $-ray background from PBHs be below the Diffuse Galactic Emission (DGE).
For a spherically symmetric PBH distribution, the $ \gamma $-ray flux from PBHs averaged over the region of interest, $ |b| \geq 20^\circ $ to avoid contamination from the Galactic disc, is
\begin{equation}
\Phi(|b|\geq20^\circ;E)
= \frac{\int_{|b|\geq20^\circ} \Phi(\boldsymbol n,E)\,\mathrm d\Omega}
       {\int_{|b|\geq20^\circ} \mathrm d\Omega}
= \mathcal G\,r_\mathrm{gal}\,
  \frac{\bar{\mathcal E}(E)}{4\pi}\,,
\end{equation}
where $ \mathrm d\Omega = \cos b\,\mathrm db\,\mathrm dl $ and $ \mathcal G $ denotes the average value of $ g(\boldsymbol n) $ over the region of interest:
\begin{equation}
\mathcal G(|b|\geq20^\circ)
= \frac{\int_{|b|\geq20^\circ} g(\boldsymbol n)\,\mathrm d\Omega}
       {\int_{|b|\geq20^\circ} \mathrm d\Omega}\,.
\end{equation}
A numerical computation gives $ \mathcal G(|b|\geq20^\circ) \approx 4.3 \times 10^4 $ if $ r_\mathrm{gal} = 100\,\mathrm{kpc} $\,.

\subsection{
Emissivity for monochromatic mass function
}

We assume that the initial mass function is given by Eq.~\eqref{eq:mf_tophat} and that $ M_\mathrm f $ exceeds $ M_* $\,, else there is no emission at the present epoch.
The current emissivity is given by Eq.~\eqref{eq:em_avg} with an upper limit at $ m(M_\mathrm f) $ and a lower limit at $ m((1-\Delta)\,M_\mathrm f) $\,.
Here $ m(M) \approx M $ for $ \mu \gg 1 $, $ (3 \mu)^{1/3}\,M_* $ for $ 0.02 < \mu \ll 1 $ and $ (3 \alpha \mu)^{1/3}\,M_* $ for $ \mu < 0.005 $, with the more exact expression \eqref{eq:mmu} covering the transitions between these ranges.
We will usually be interested in the situation where the mass band encompasses or is close to $ M_* $\,, since this is relevant to the Galactic background.

Analysing the emissivity associated with the various monochromatic scenarios is quite complicated, so we start off with some general qualitative remarks.
For $ (1-\Delta)\,M_\mathrm f \gg M_\mathrm c $\,, the mass function does not evolve significantly, so it remains very narrow and there is only primary emission.
As $ (1-\Delta)\,M_\mathrm f $ falls towards $ M_* $\,, the current mass function broadens at the low end and begins to acquire a tail with $ \mathrm d\bar n/\mathrm dm \propto m^2 $ just below $ M_* $\,.
Secondary emission becomes important once the tail extends below $ M_\mathrm c $ and it extends all the way down to $ 0 $ for $ (1-\Delta)\,M_\mathrm f < M_* $\,.
So long as $ M_\mathrm f $ remains above $ M_\mathrm c $\,, there will also be some primary emission but both the upper and lower ends of the mass function cross $ M_\mathrm c $ nearly together if $ \Delta $ is small, so emission will usually be dominated by the primary or secondary component.
In any case, secondary emission will dominate once $ M_\mathrm f $ also falls below $ M_\mathrm c $\,.
An important qualitative point is that the low-mass tail will produce a high-energy emissivity tail ($ \bar{\mathcal E} \propto E^{-3} $) providing the mass scale $ E^{-1} $ lies between the limits in the mass integral \eqref{eq:em_avg}.
Otherwise the emissivity will fall off exponentially, so a crucial transition occurs when $ (1-\Delta)\,M_\mathrm f $ gets close to $ M_* $ since this marks the onset of the low-mass tail.
Another important feature is that the dependence of the emissivity on the parameter $ \Delta $ often drops out and always in scenarios where $ \Delta $ can go to zero.

We now discuss some of the possible cases in more detail.
Although we calculate all the emissivities numerically, our purpose is to understand the results analytically.
We use units with $ \hbar = c = k_\mathrm B = 8 \pi G = 1 $ but there is an implicit factor of $ \hbar^{-1} $ in the emissivity expressions below and they all scale with the PBH number density $ \bar n $\,, so the units are $ \mathrm s^{-1}\,\mathrm{MeV}^{-1}\,\mathrm{cm}^{-3} $\,.
Note that the expressions always involve factors of $ E\,m $\,, this combination being dimensionless with our chosen units.

\textit{Case A}.
If $ (1-\Delta)\,M_\mathrm f \gg M_* $\,, the current mass function preserves its original (narrow) form and the only secondary emission comes from the Wien tail, which we neglect here.
Equations~\eqref{eq:rate_pri}, \eqref{eq:mf_tophat} and \eqref{eq:em_avg} then imply
\begin{equation}
\begin{aligned}
\bar{\mathcal E}^\mathrm P(E)
&
\approx
  \int_{(1-\Delta) M_\mathrm f}^{M_\mathrm f}\!\mathrm dm\,
  \frac{\bar n}{\Delta\,M_\mathrm f}\,
  \frac{\mathrm d\dot N^\mathrm P}{\mathrm dE}(m,E) \\
&
\propto
  \bar n
  \times
\begin{cases}
E^3\,M_\mathrm f^3\,
[1 - 3 \Delta/2 + \mathcal O(\Delta^2)]
& (E < M_\mathrm f^{-1}) \\
E^2\,M_\mathrm f^2\,\mathrm e^{-E M_\mathrm f}\,
[1 + (E\,M_\mathrm f-2)\,\Delta/2 + \mathcal O(\Delta^2)]
& (E > M_\mathrm f^{-1})\,.
\end{cases}
\end{aligned}
\label{eq:em_mono}
\end{equation}
For $ E < M_\mathrm f^{-1} $\,, the factor of $ \Delta $ in the denominator is cancelled by the difference in the integral limits if only $ \Delta \ll 1 $.
For $ E \gg M_\mathrm f^{-1} $\,, a more severe condition $ \Delta \ll (E\,M_\mathrm f)^{-1}\,(\ll 1) $ has to be imposed for the cancellation.
So $ \Delta $ cancels except in the high energy regime $ E > \Delta^{-1}\,M_\mathrm f^{-1} $\,.
The emissivity is effectively black-body radiation from holes with mass $ M_\mathrm f $\,, so one has a peak at $ E^\mathrm P \approx 6\,M_\mathrm f ^{-1}$ with an $ M_\mathrm f $-independent value
\begin{equation}
\bar{\mathcal E}^\mathrm P(E^\mathrm P)
\approx
  \bar n\,\frac{\mathrm d\dot N^\mathrm P}{\mathrm dE}(M_\mathrm f,E^\mathrm P)
\approx
  1 \times 10^{18}\,\bar n\,\mathrm s^{-1}\,\mathrm{MeV}^{-1}\,,
\label{eq:em_mono2}
\end{equation}
using Eq.~\eqref{eq:ratepeak}.
The important point is that Eq.~\eqref{eq:em_mono} has an exponential upper cut-off for $ E > M_\mathrm f^{-1} $; there is no power-law high-energy tail because $ E^{-1} $ is less than the lower mass limit.

\textit{Case B}.
If $ M_\mathrm c < M_\mathrm d < (1-\Delta)\,M_\mathrm f < M_\mathrm f < 1.25 M_* $\,, which requires a high degree of tuning since the entire mass range is close to $ M_* $\,, the mass function evolves significantly but there is no secondary emission.
The mass limits become
\begin{equation}
m_\mathrm f
\equiv
  m(M_\mathrm f)
\approx
  (3 \mu_\mathrm f)^{1/3}\,M_*\,,
\quad
m((1-\Delta)\,M_\mathrm f)
\approx
  (1 - \Delta/\mu_\mathrm f)^{1/3}\,m_\mathrm f\,.
\end{equation}
The emissivity is therefore
\begin{equation}
\begin{aligned}
\bar{\mathcal E}^\mathrm P(E)
&
\approx
  \int_{(1-\Delta/ \mu_\mathrm f)^{1/3}m_\mathrm f}^{m_\mathrm f}\!\mathrm dm\,
  \frac{\bar n}{\Delta\,M_\mathrm f}\,
  \left(\frac{m}{M_*}\right)^2\,
  \frac{\mathrm d\dot N^\mathrm P}{\mathrm dE}(m,E) \\
&
\propto
  \bar n
  \times
\begin{cases}
E^3\,M_*^3\,\mu_\mathrm f\,
[1-\Delta/(2 \mu_\mathrm f)+\mathcal O(\Delta^2)]
& (E < m_\mathrm f^{-1}) \\
E^2\,M_*^2\,\mu_\mathrm f^{2/3}\,\mathrm e^{-m_\mathrm f E}\,
[1 + (E\,m_\mathrm f-2)\,\Delta/(6 \mu_\mathrm f) + \mathcal O(\Delta^2)]
& (E > m_\mathrm f^{-1})\,,
\end{cases}
\end{aligned}
\end{equation}
where we have used $ M_\mathrm f \approx M_* $ and $ m_\mathrm f \approx (3 \mu_\mathrm f)^{1/3}\,M_* $ at the last step.
For $ E < m_\mathrm f^{-1} $\,, the factor of $ \Delta $ in the denominator is cancelled if only $ \Delta \ll \mu_\mathrm f $\,.
For $ E \gg m_\mathrm f^{-1} $\,, a more severe condition $ \Delta \ll \mu_\mathrm f/(E\,m_\mathrm f)\,(\ll \mu_\mathrm f) $ is necessary.
So $ \Delta $ cancels except in the high energy regime $ E > (m_\mathrm f\,\Delta/\mu_\mathrm f)^{-1} $\,.
The emissivity peaks at $ E \approx 6\,m_\mathrm f^{-1} $ with the value given by Eq.~\eqref{eq:em_mono2}.
There is no power-law high-energy tail because the mass-scale $ E^{-1} $ lies below the integration range.
For $ M_\mathrm f > M_* $\,, there would be an additional primary component given by Eq.~\eqref{eq:em_mono}, so one would effectively have a combination of cases A and B.

\textit{Case C}.
If $ (1-\Delta)\,M_\mathrm f < M_* < M_\mathrm c < M_\mathrm f < 1.25\,M_* $\,, which requires $ \Delta > 0.005 $, the low-mass tail is complete (i.e.\ extends down to $ 0 $) and secondary emission is important below $ M_\mathrm q $\,.
Using Eq.~\eqref{eq:rate_sec}, the secondary emissivity is
\begin{equation}
\bar{\mathcal E}^\mathrm S(E)
\approx
  \int_0^{M_\mathrm q}\!\mathrm dm\,
  \frac{\bar n}{\Delta\,\alpha M_\mathrm f}\,
  \left(\frac{m}{M_*}\right)^2\,
  \frac{\mathrm d\dot N^\mathrm S}{\mathrm dE}(m,E)
\propto
\frac{\bar n}{\Delta\,\alpha}
  \times
\begin{cases}
q^4\,E\,M_*^2\,M_\mathrm f^{-1}
& (E < M_\mathrm q^{-1}) \\
E^{-3}\,M_*^{-2}\,M_\mathrm f^{-1}
& (E > M_\mathrm q^{-1})\,.
\end{cases}
\end{equation}
In this case, there is an $ E^{-3} $ high-energy tail for $ E > m_\mathrm f^{-1} $ because $ E^{-1} $ lies below the integral upper limit.
There is also now a dependence on $ \Delta $ but this cannot go to $ 0 $ because $ \Delta > 0.005 $.
The emissivity peaks at $ E^\mathrm S \approx 6\,M_\mathrm q^{-1} $ with a value
\begin{equation}
\bar{\mathcal E}^\mathrm S(E^\mathrm S)
\sim
  \frac{\bar n\,q^2\,M_*}{3 \alpha\,\Delta\,M_\mathrm f}
\sim
  10^{17}\,\mathrm s^{-1}\,\mathrm{MeV}^{-1}\,\mathrm{pc}^{-3}\,
  \left(\frac{\Delta}{0.1}\right)^{-1}\,
  \left(\frac{M_*}{M_\mathrm f}\right)
  \left( \frac{\bar n}{\mathrm{pc}^{-3}}\right)\,.
\end{equation}
In this case, there is also a primary component from holes between $ M_\mathrm q $ and $ m_\mathrm f $ with emissivity
\begin{equation}
\bar{\mathcal E}^\mathrm P(E)
\approx
  \int_{M_\mathrm q}^{m_\mathrm f}\!\mathrm dm\,
  \frac{\bar n}{\Delta\,M_\mathrm f}\,
  \left(\frac{m}{M_*}\right)^2\,
  \frac{\mathrm d\dot N^\mathrm P}{\mathrm dE}(m,E)
\propto
\frac{\bar n}{\Delta}
  \times
\begin{cases}
E^3\,m_\mathrm f^6\,M_\mathrm f^{-1}\,M_*^{-2}
& (E < m_\mathrm f^{-1}) \\
E^{-3}\,M_*^{-2}\,M_\mathrm f^{-1}
& (E > m_\mathrm f^{-1})\,.
\end{cases}
\end{equation}
There is a high-energy $ E^{-3} $ tail since $ E^{-1} $ lies between $ M_\mathrm q $ and $ m_\mathrm f $\,.
This has the same form as the high-energy tail of the secondary emission and it also scales as $ \Delta^{-1} $ but it has a different low-energy form.

\textit{Case D}.
If $ (1-\Delta)\,M_\mathrm f < M_* < M_\mathrm f < M_\mathrm c $\,, which requires fine-tuning of $ M_* $ and $ \mu_\mathrm f < 0.005 $, there is only secondary emission and the mass tail goes between $ 0 $ and $ m_\mathrm f $\,, so the emissivity is
\begin{equation}
\bar{\mathcal E}^\mathrm S(E)
\approx
  \int_0^{m_\mathrm f}\!\mathrm dm\,
  \frac{\bar n}{\Delta\,\alpha\,M_\mathrm f}\,
  \left(\frac{m}{M_*}\right)^2\,
  \frac{\mathrm d\dot N^\mathrm S}{\mathrm dE}(m,E)
\propto
  \frac{\bar n}{\Delta\,\alpha}
  \times
\begin{cases}
q^2\,E\,m_\mathrm f^2\,M_*^{-1}
& (E < M_\mathrm q^{-1}) \\
E^{-1}\,m_\mathrm f^2\,M_*^{-3}
& (M_\mathrm q^{-1} < E < m_\mathrm f^{-1}) \\
E^{-3}\,M_*^{-3}
& (E > m_\mathrm f^{-1})\,,
\end{cases}
\end{equation}
where we have put $ M_\mathrm f \approx M_* $ at the last step.
Because $ E^{-1} $ is below $ m_\mathrm f $\,, one again has an $ E^{-3} $ high-energy tail and a $ \Delta $-dependence.
This peaks at $ E^\mathrm S \approx 6\,M_\mathrm q^{-1} $ with a value
\begin{equation}
\bar{\mathcal E}^\mathrm S(E^\mathrm S)
\sim
  \frac{\bar n\,q\,m_\mathrm f^2}{\Delta\,M_*^2}
\sim
  10^{18}\,\mathrm s^{-1}\,\mathrm{MeV}^{-1}\,\mathrm{pc}^{-3}\,
  \left(\frac{\Delta}{0.1}\right)^{-1}\,
  \left(\frac{\bar n}{\mathrm{pc}^{-3}}\right)\,
  \left(\frac{\mu_\mathrm f}{0.005}\right)^{2/3}\,.
\end{equation}

\begin{figure}[htb]
\includegraphics[scale=0.65]{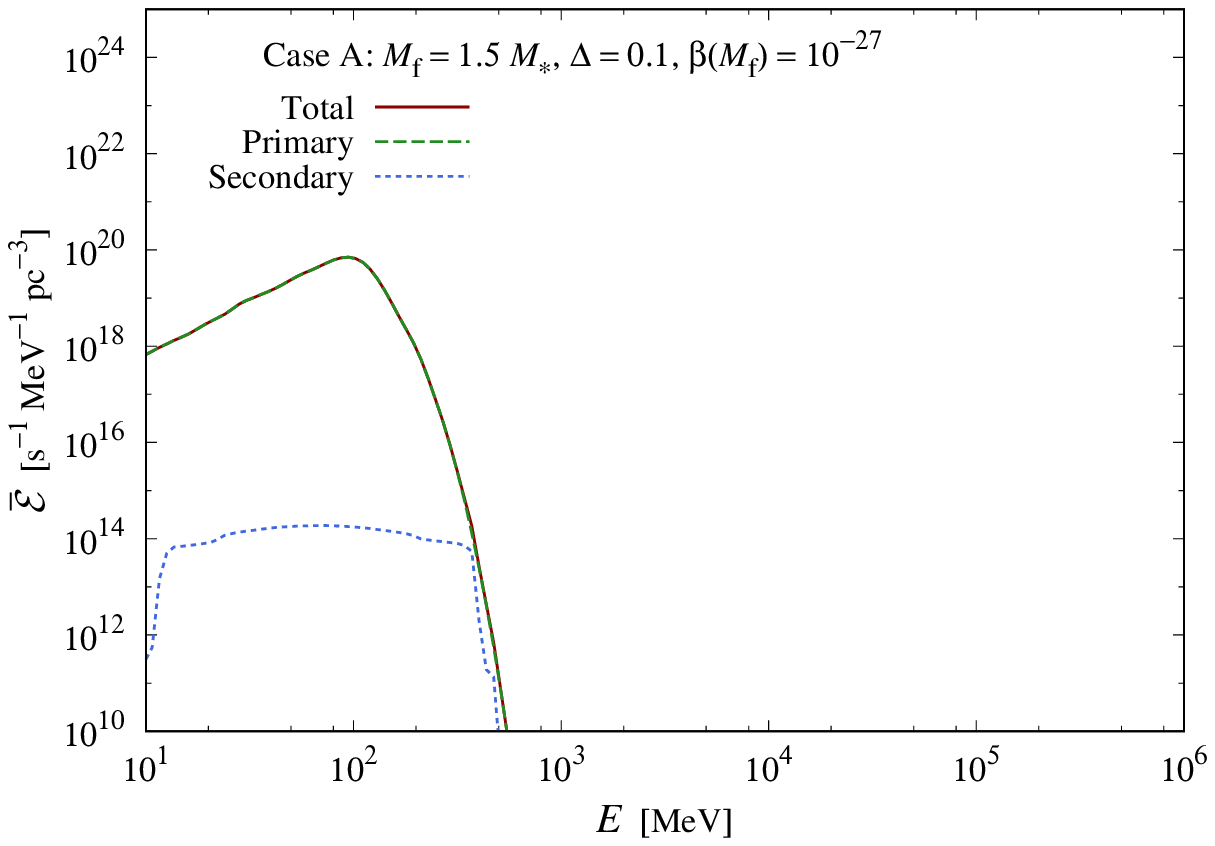}
\includegraphics[scale=0.65]{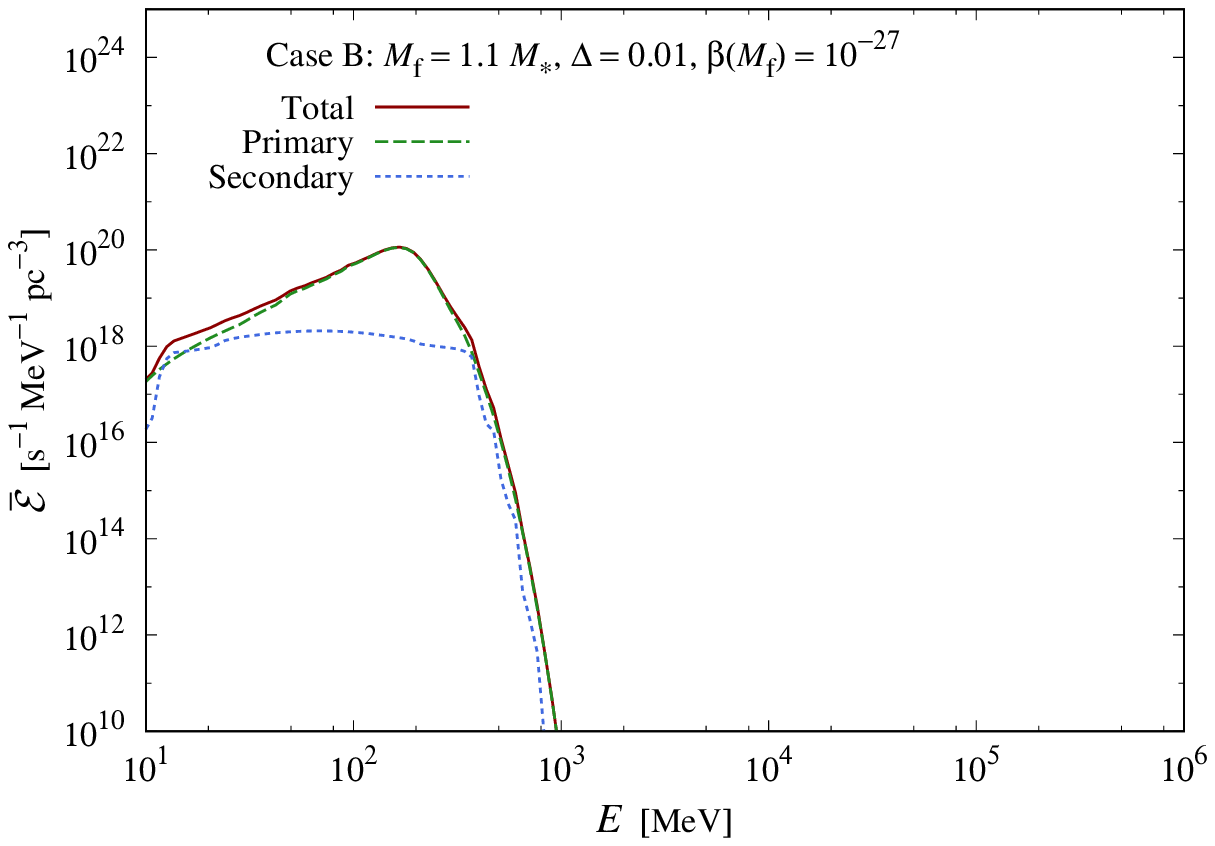} \\
\includegraphics[scale=0.65]{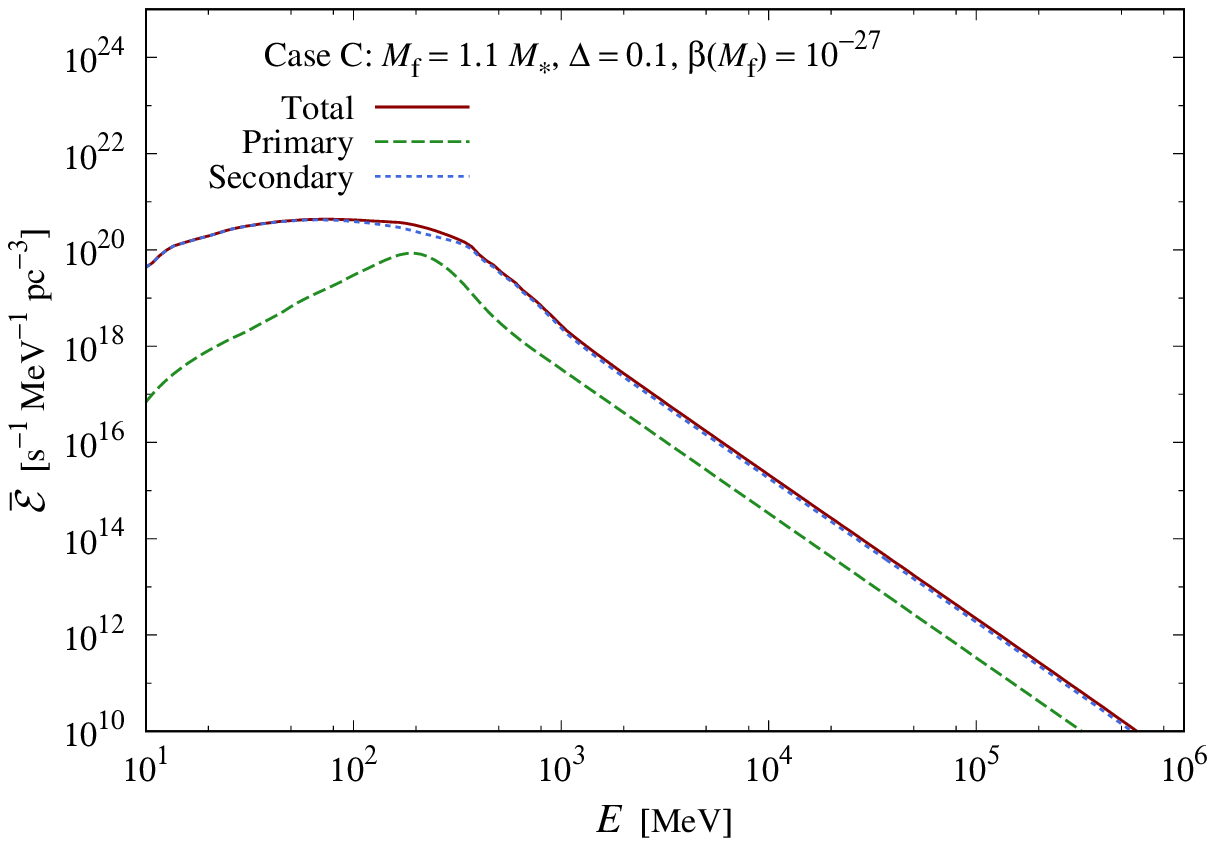}
\includegraphics[scale=0.65]{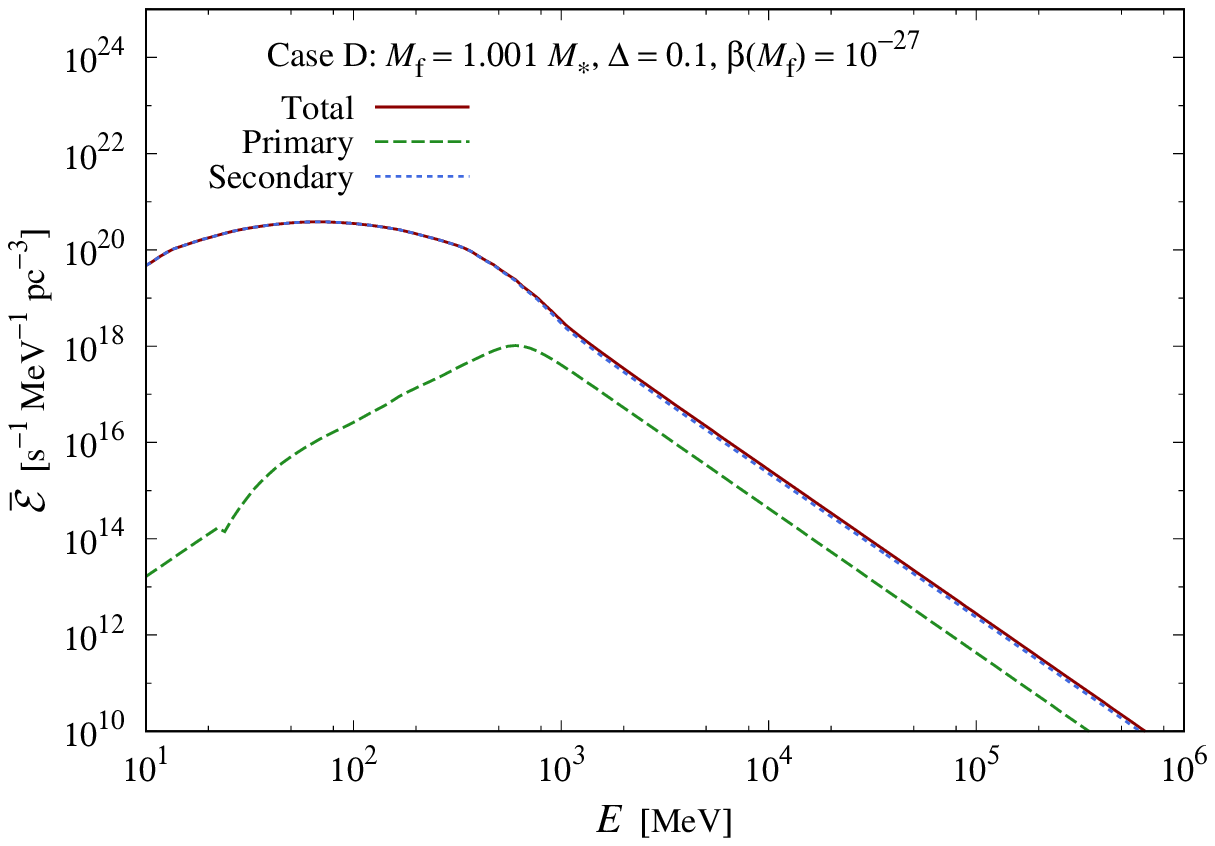} \\
\includegraphics[scale=0.65]{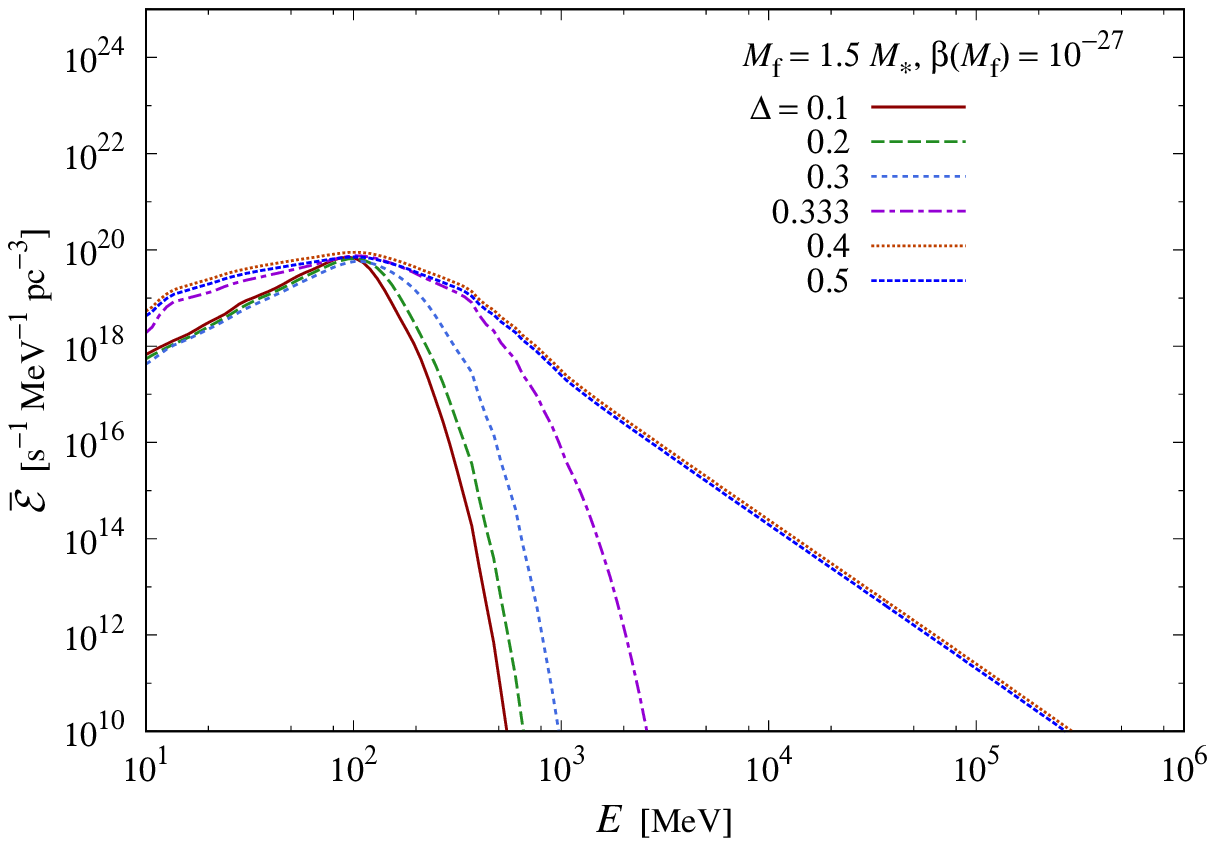}
\includegraphics[scale=0.65]{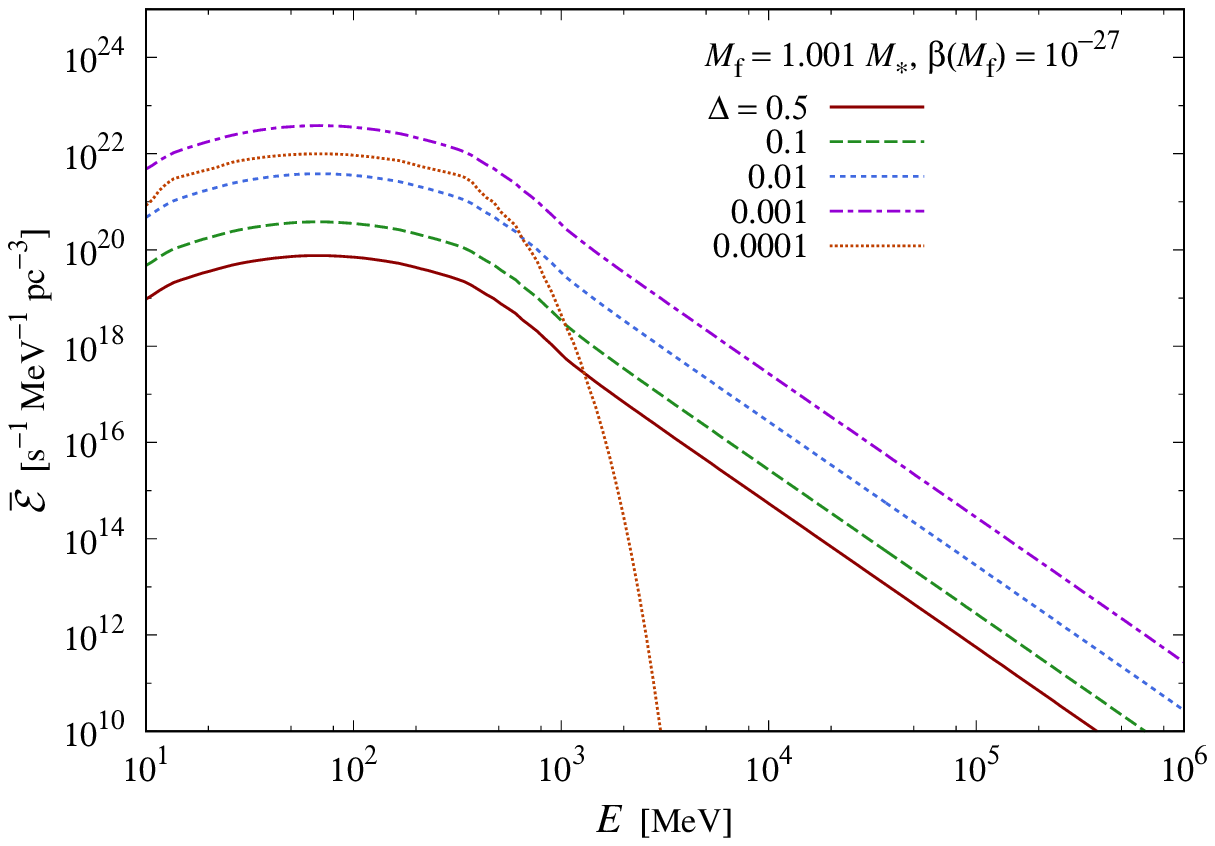}
\caption{\label{fig:em_mono}
Top: components of $ \gamma $-ray emissivity for $ M_\mathrm f = 1.5\,M_* $ with $ \Delta = 0.1 $ (left) and $ M_\mathrm f = 1.1\,M_* $ with $ \Delta = 0.3 $ (right).
The secondary components are suppressed in both cases since $ (1-\Delta)\,M_\mathrm f > M_\mathrm c $\,.
The peak energies for the primary components are $ E^\mathrm P \approx 6 M_\mathrm f^{-1} $ (left) and $ E^\mathrm P \approx 6 m_\mathrm f^{-1} $ (right), respectively.
Middle: components of $ \gamma $-ray emissivity for $ M_\mathrm f = 1.1\,M_* $ (left) and $ M_\mathrm f = 1.001\,M_* $ with $ \Delta = 0.1 $.
The secondary emission occurs since $ (1-\Delta)\,M_\mathrm f < M_\mathrm c $\,.
Bottom: total $ \gamma $-ray emissivities for $ M_\mathrm f = 1.5\,M_* $ (left) and $ M_\mathrm f = 1.001\,M_* $ (right) with various values of $ \Delta $\,.
As seen in the right figure, the secondary emissivity is proportional to $ \Delta^{-1} $ as long as the entire mass tail is formed, i.e.\ $ (1-\Delta)\,M_\mathrm f < M_* $\,, and is exponentially suppressed once $ (1-\Delta)\,M_\mathrm f $ exceeds $ M_* $\,.
All figures use the normalisation $ \beta(M_\mathrm f) = 10^{-27} $.
}
\end{figure}

To be specific, we calculate the $ \gamma $-ray emissivity numerically for some of the mass functions shown in Fig.~\ref{fig:mf_mono} and the results are shown in Fig.~\ref{fig:em_mono}.
In the top-left figure, $ M_\mathrm f = 1.5\,M_* $ and $ \Delta = 0.1 $ (case A);
since $ (1-\Delta)\,M_\mathrm f \gg M_\mathrm c $\,, the primary component dominates and this falls exponentially for $ E > 6\,M_\mathrm f^{-1} $\,.
In the top-right figure, $ M_\mathrm f = 1.1\,M_* $ and $ \Delta = 0.01 $ (case B); this is similar to case~A but the peak shifts to a higher energy $ E \approx 6\,m_\mathrm f^{-1} $ since $ m_\mathrm f \ll M_\mathrm f $\,.
In the middle-left figure, $ M_\mathrm f = 1.1\,M_* $ and $ \Delta = 0.1 $ (case C); since $ (1-\Delta)\,M_\mathrm f < M_\mathrm c $\,, there is secondary emission and this is comparable to the primary emission, both falling as $ E^{-3} $ for $ E > 100\,\mathrm{MeV} $\,.
In the middle-right figure, $ M_\mathrm f = 1.001\,M_* $ and $ \Delta = 0.1 $ (case D); this is similar to case~C, but the secondary emissivity is suppressed as $ \mu_\mathrm f^{2/3} $\,.
The crucial feature of these figures is that the high-energy emissivity scales as $ E^{-3} $ if there is low mass tail but falls exponentially otherwise.
Also the emissivity has no dependence on $ \Delta $ in the latter case but it scales as $ \Delta^{-1} $ in the former case.
A subtle effect on the primary emission is the broadening of the peak towards higher energy and this shows up as $ \Delta $ increases.
The bottom-left figure shows the effect of varying the width $ \Delta $ for $ M_\mathrm f = 1.5\,M_* $; the form changes as $ \Delta $ approaches $ 0.333 $, below which there is an exponential cut-off for secondary emission (case A).
The bottom-right figure shows the same for $ M_\mathrm f = 1.001\,M_* $; the form only changes for $ \Delta = 0.0001 $, when there is a power-law fall-off and secondary emission (case C).

\subsection{
Emissivity for extended mass function
}

We assume that the initial mass function has the extended form \eqref{eq:mf_pl}, i.e.\ with $ \mathrm d\bar n/\mathrm dm \propto m^\nu $ up to some mass $ M_\mathrm f $ exceeding $ M_* $\,.
We first consider the contribution from the holes with $ m > M_* $ which do not produce a low mass tail or secondary emission.
In this case, Eq.~\eqref{eq:em_avg} with $ \nu > -3 $ implies that the emissivity is given by
\begin{equation}
\bar{\mathcal E}^\mathrm P(E)
\approx
  \int_{M_*}^{M_\mathrm f}\!\mathrm dm\,
  \frac{\bar n_*}{M_*}\,
  \left(\frac{m}{M_*}\right)^\nu\,
  \frac{\mathrm d\dot N^\mathrm P}{\mathrm dE}(m,E)
\propto
  \bar n_*
  \times
\begin{cases}
E^3\,M_\mathrm f^{4+\nu}\,M_*^{-\nu-1}
& (E < M_\mathrm f^{-1})\\
E^{-1-\nu}\,M_*^{-\nu-1}
& (M_*^{-1} > E > M_\mathrm f^{-1}) \\
E\,M_*\,\mathrm e^{-E M_*}
& (E > M_*^{-1})\,,
\end{cases}
\label{eq:slope}
\end{equation}
where we have used Eq.~\eqref{eq:rate_pri} and $ \bar n_* \equiv M_*\,(\mathrm d\bar n/\mathrm dM)_* $\,.
For $ E < M_\mathrm f^{-1} < m^{-1} $\,, one is in the Rayleigh--Jeans region for the entire mass integral, so $ M_\mathrm f $ gives the upper limit; for $ M_*^{-1} > E > M_\mathrm f^{-1} $\,, one splits the mass integral into a Rayleigh--Jeans part with an upper limit at $ m \sim E^{-1} $ and a Wien part with a lower limit at $ m \sim E^{-1} $\,, both contributions scaling as $ E^{-\nu-1} $; for $ E > M_*^{-1} $\,, $ E^{-1} $ is below the lower mass limit, so one has the exponential cut-off.
For $ \nu > -1 $, the emissivity peaks at $ E \sim M_\mathrm f^{-1} $ with a value scaling as $ M_\mathrm f^{1+\nu} $\,, so the peak shifts to lower energies but gets higher as $ M_\mathrm f $ increases.
For $ \nu < -1 $ (including the favored case $ \nu = -5/2 $), the peak occurs at $ E \sim M_*^{-1} $ with a constant value.
For $ \nu < -3 $, the lower integral limit in Eq.~\eqref{eq:slope} dominates, so one replaces $ M_\mathrm f $ by $ M_* $ in the $ E < M_\mathrm f^{-1} $ case, giving $ E^3\,M_*^3 $\,.

We now consider the primary emission from the narrow mass range $ M \in [M_\mathrm c, 1.25 M_*] $\,, or equivalently $ m \in [M_\mathrm q, M_*] $\,, which produces a low mass tail with $ m < M_* $ and $ \mathrm d\bar n/\mathrm dm \propto m^2 $ but no secondary emission.
We then have
\begin{equation}
\begin{aligned}
\bar{\mathcal E}^\mathrm P(E)
\approx
  \int_{M_\mathrm q}^{M_*}\!\mathrm dm\,
  \frac{\bar n_*}{M_*}\,
  \left(\frac{m}{M_*}\right)^2\,
  \frac{\mathrm d\dot N^\mathrm P}{\mathrm dE}(m,E)
\propto
  \bar n_*
  \times
\begin{cases}
E^3\,M_*^{3}
& (E < M_*^{-1}) \\
E^{-3}\,M_*^{-3}
& (M_\mathrm q^{-1} > E > M_*^{-1}) \\
q^4\,E\,M_*\,\mathrm e^{-E M_\mathrm q}
& (E > M_\mathrm q^{-1})\,,
\end{cases}
\end{aligned}
\end{equation}
the first two cases being equivalent to Eq.~\eqref{eq:slope} with $ \nu = 2 $ and $ M_\mathrm f = M_* $\,.
This dominates the primary contributions from the $ M_\mathrm f $ holes providing $ \nu > - 1 $.
For the mass range $ M < M_\mathrm c $\,, or equivalently $ m < M_\mathrm q $\,, a high-energy tail $ \propto E^{-3} $ appears for $ E > M_\mathrm q^{-1} $ but with a suppression factor $ \alpha^{-1} $ from the mass function.
Numerical investigations show that the secondary emissivity always dominates in the energy range $ E > M_\mathrm q^{-1} $\,.
Unless there is a fine-tuned upper limit, we must integrate over the mass range $ m \in [0,M_\mathrm q] $\,.
For $ E < M_\mathrm q^{-1} $\,, the emission is determined by the jet fragmentation function \eqref{eq:frag}.
For $ E > M_\mathrm q^{-1} $\,, there is an effective upper cut-off at $ m \sim E^{-1} $\,, this necessarily being below $ M_\mathrm q $\,.
We therefore have
\begin{equation}
\begin{aligned}
\bar{\mathcal E}^\mathrm S(E)
\approx
  \int_0^{M_\mathrm q}\!\mathrm dm\,
  \frac{\bar n_*}{\alpha\,M_*}\,
  \left(\frac{m}{M_*}\right)^2\,
  \frac{\mathrm d\dot N^\mathrm S}{\mathrm dE}(m,E)
\propto
  \frac{\bar n_*}{\alpha}
  \times
\begin{cases}
q^4\,E\,M_*
& (E < M_\mathrm q^{-1}) \\
E^{-3}\,M_*^{-3}
& (E > M_\mathrm q^{-1})\,.
\end{cases}
\end{aligned}
\label{eq:em_pl}
\end{equation}

\begin{figure}[htb]
\includegraphics[scale=0.65]{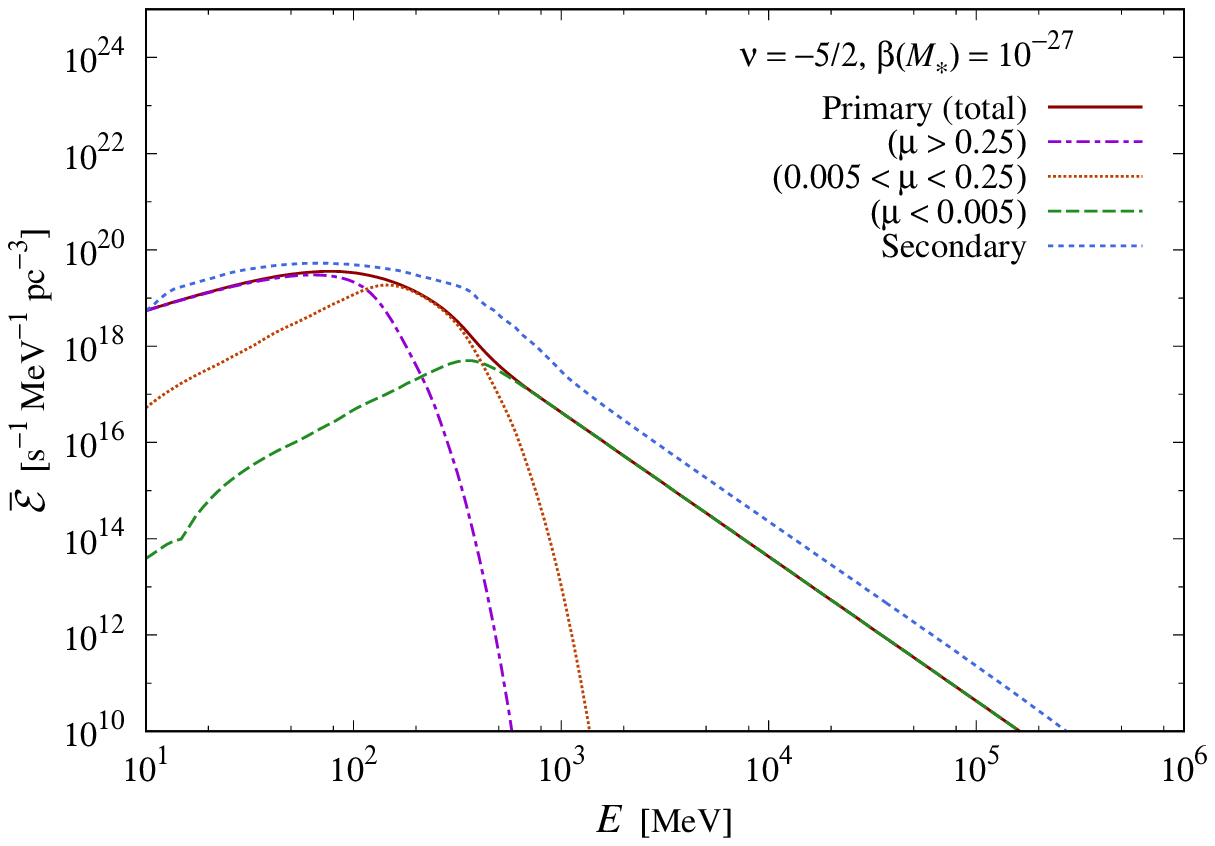}
\includegraphics[scale=0.65]{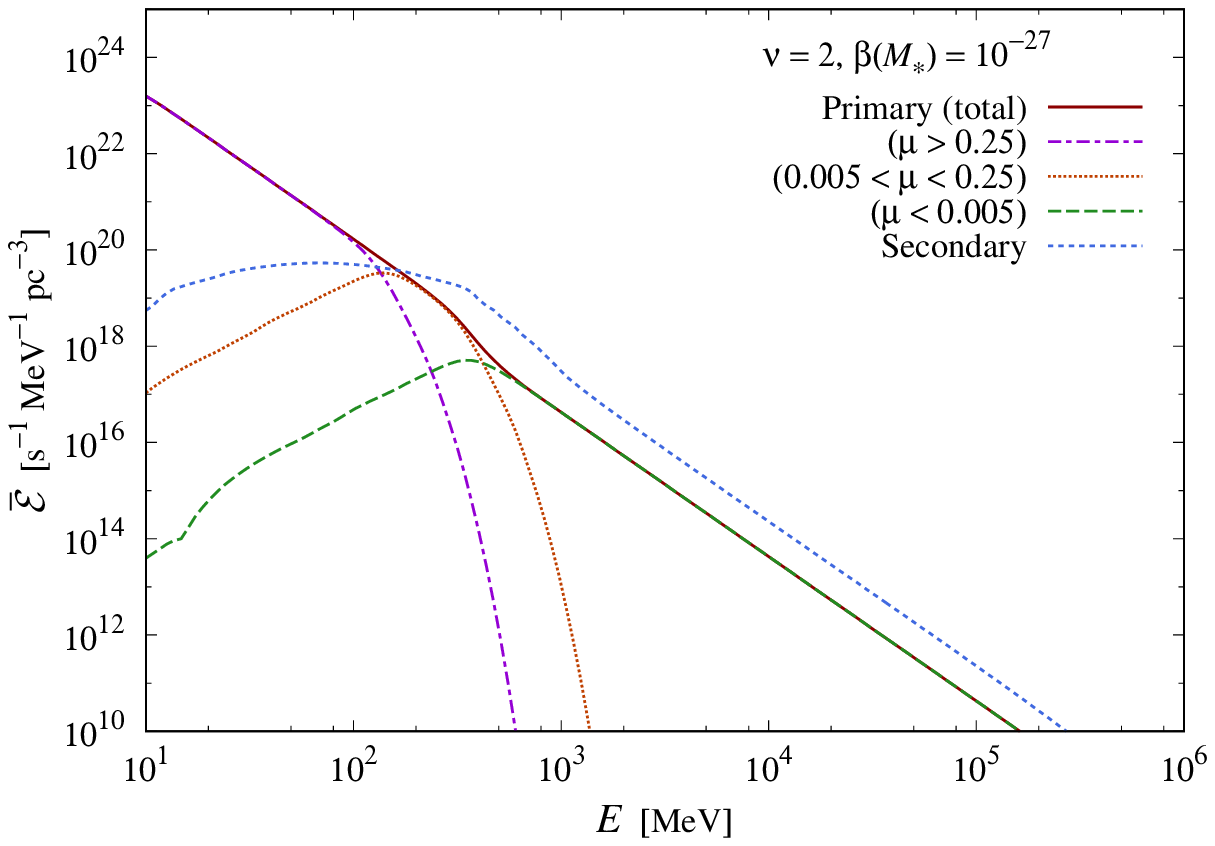} \\
\includegraphics[scale=0.65]{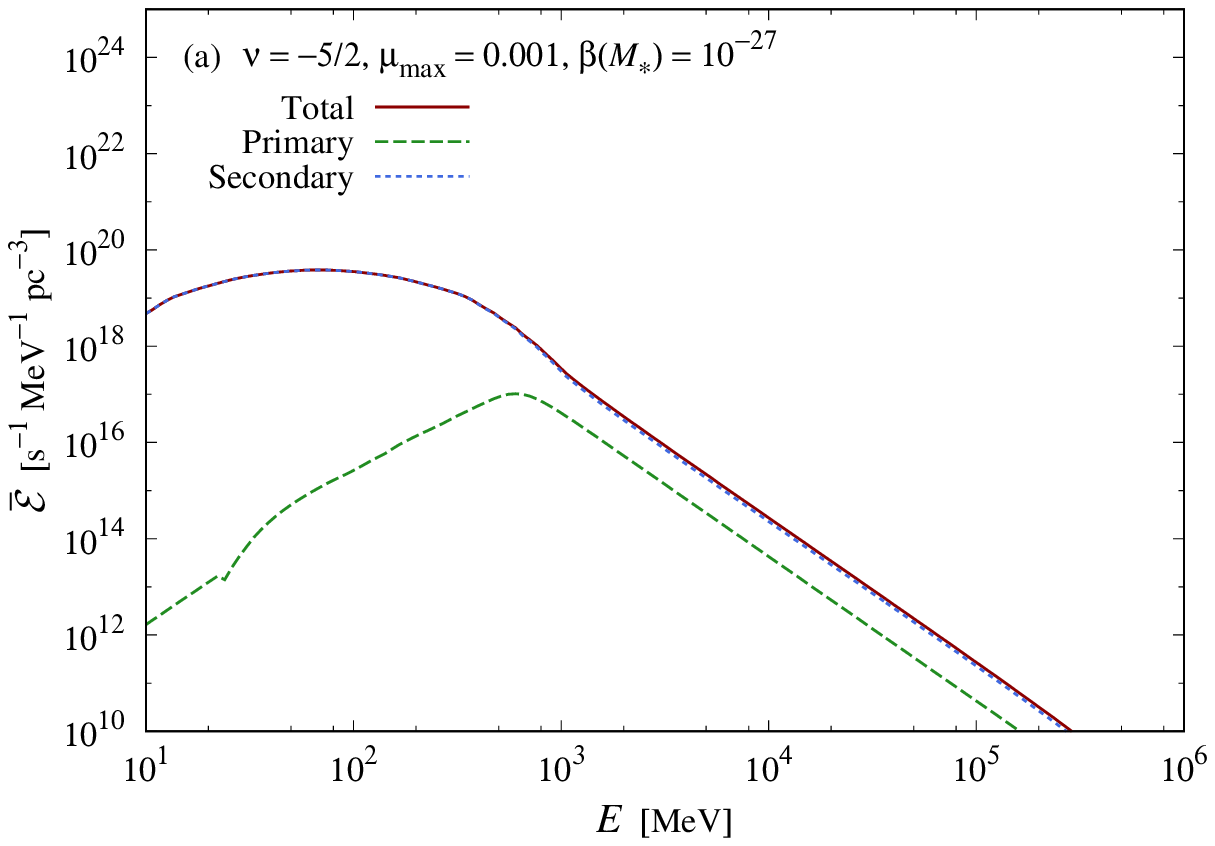}
\includegraphics[scale=0.65]{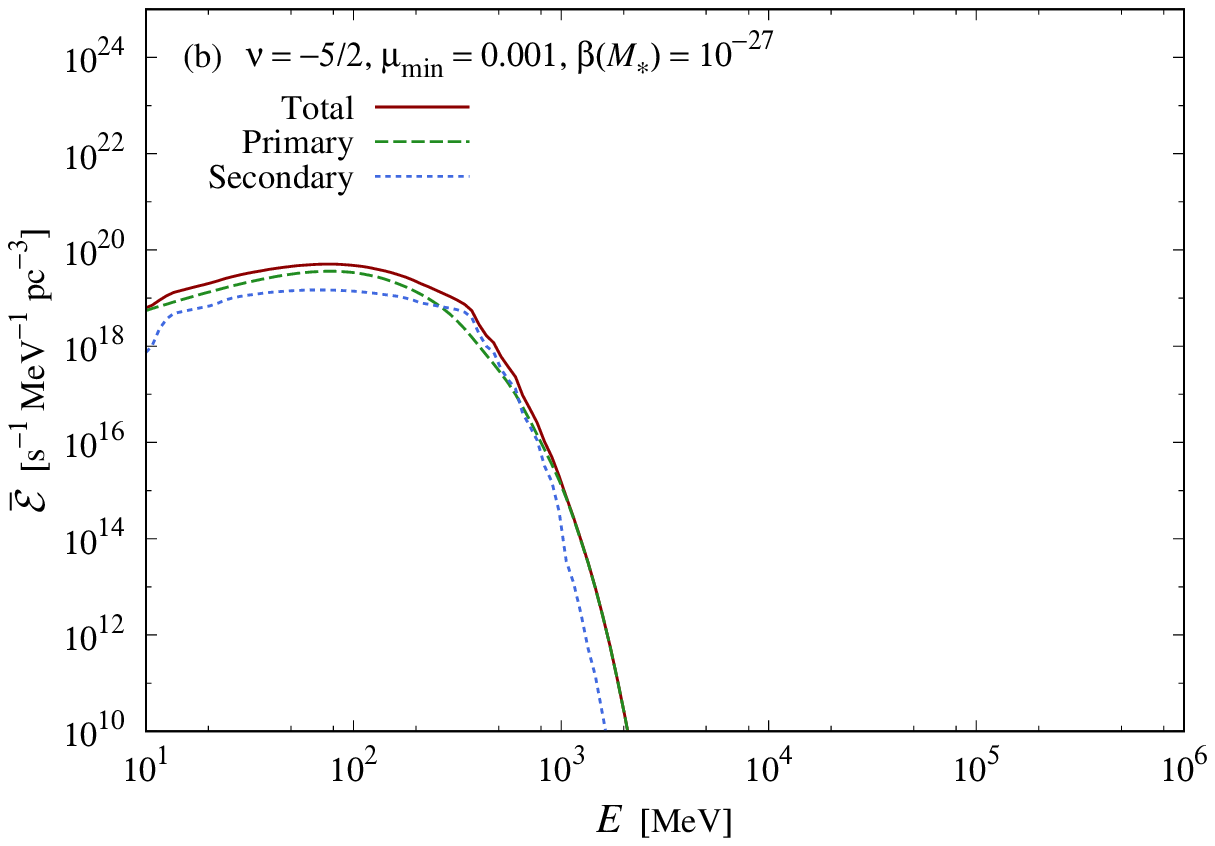}
\caption{\label{fig:em_pl}
Top: Averaged emissivity $ \bar{\mathcal E}(E) $ for extended mass functions with $ \nu = -5/2 $ (left) and $ 2 $ (right).
Bottom: Averaged emissivity $ \bar{\mathcal E}(E) $ for extended mass functions with $ \nu = -5/2 $ and fine-tuned upper (a) or lower (b) cutoffs.
}
\end{figure}

Figure~\ref{fig:em_pl} shows the averaged emissivity for extended mass functions with $ \nu = -5/2 $ and $ 2 $.
The primary emissivity is divided into the three contributions mentioned above: $ M > 1.25 M_* $\,, $ M_\mathrm c < M < 1.25 M_* $\,, and $ M < M_\mathrm c $\,.

For a fine-tuned upper cut-off at $ M_\mathrm{max} < M_\mathrm c $\,, or $ m_\mathrm{max} < M_\mathrm q $\,, one has
\begin{equation}
\begin{aligned}
\bar{\mathcal E}^\mathrm S(E)
\approx
  \int_0^{m_{max}}\!\mathrm dm\,
  \frac{\bar n_*}{\alpha\,M_*}\,
  \left(\frac{m}{M_*}\right)^2\,
  \frac{\mathrm d\dot N^\mathrm S}{\mathrm dE}(m,E)
\propto
  \frac{\bar n_*}{\alpha}
\times
\begin{cases}
q^2\,E\,m_\mathrm{max}^2\,M_*^{-1}
& (E < M_\mathrm q^{-1}) \\
E^{-1}\,m_\mathrm{max}^2\,M_*^{-3}
& (M_\mathrm q^{-1} < E < m_\mathrm{max}^{-1}) \\
E^{-3}\,M_*^{-3}
& (E > m_\mathrm{max}^{-1})\,.
\end{cases}
\end{aligned}
\end{equation}
For a fine-tuned lower cut-off at $ M_\mathrm{min} < M_\mathrm c $ or $ m_\mathrm{min} < M_\mathrm q $\,, one has
\begin{equation}
\begin{aligned}
\bar{\mathcal E}^\mathrm S(E)
\approx
  \int_{m_\mathrm{min}}^{M_\mathrm q}\!\mathrm dm\,
  \frac{\bar n_*}{\alpha\,M_*}\,
  \left(\frac{m}{M_*}\right)^2\,
  \frac{\mathrm d\dot N^\mathrm S}{\mathrm dE}(m,E)
\propto
  \frac{\bar n_*}{\alpha}
  \times
\begin{cases}
q^4\,E\,M_*
& (E < M_\mathrm q^{-1}) \\
E^{-1}\,m_\mathrm{min}^2\,M_*^{-3}
& (M_\mathrm q^{-1} < E < m_\mathrm{min}^{-1}) \\
E\,m_\mathrm{min}^4\,M_*^{-3}\,\mathrm e^{-Em_\mathrm{min}}
& (E > m_\mathrm{min}^{-1})\,.
\end{cases}
\end{aligned}
\end{equation}
The emissivity in these two cases is shown in Fig.~\ref{fig:em_pl}(a) for $ \mu_\mathrm{max} = 0.01 $ and Fig.~\ref{fig:em_pl}(b) for $ \mu_\mathrm{min} = 0.01 $.
In both cases we assume $ \nu = -5/2 $.
As expected, only in the first case is there a power-law high-energy tail.
The sum of these effectively gives the emissivity \eqref{eq:em_pl} for an extended mass function without fine-tuned cut-offs.

\subsection{
Emissivity for critical collapse mass function
}

\begin{figure}[htb]
\includegraphics[scale=0.65]{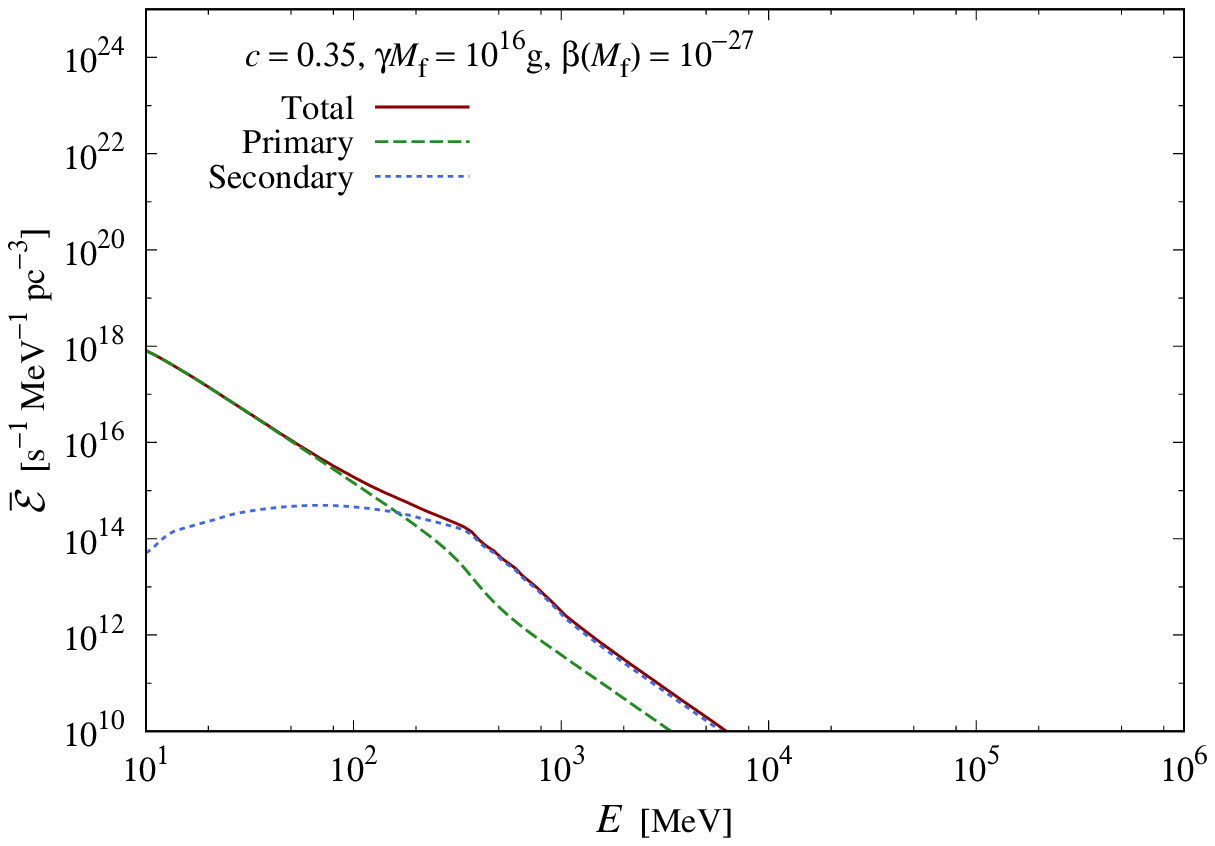}
\includegraphics[scale=0.65]{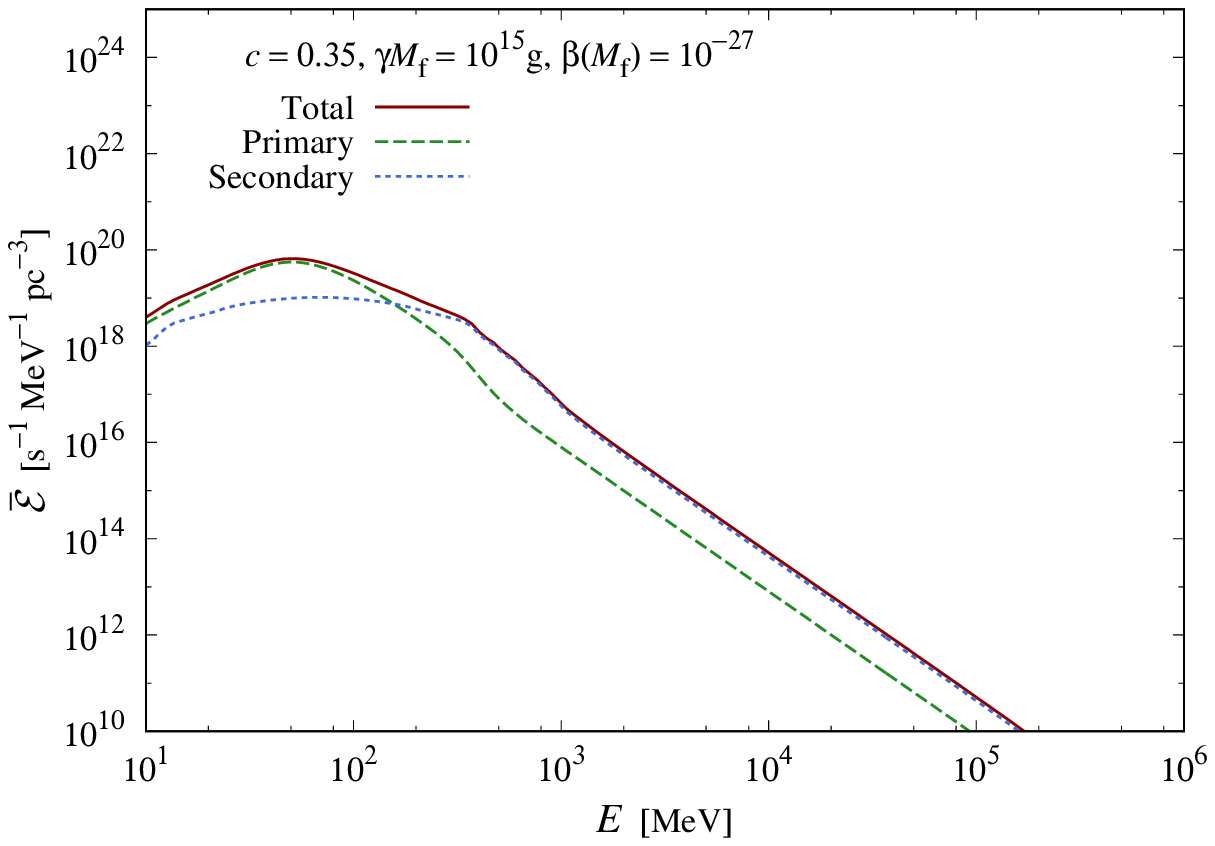} \\
\includegraphics[scale=0.65]{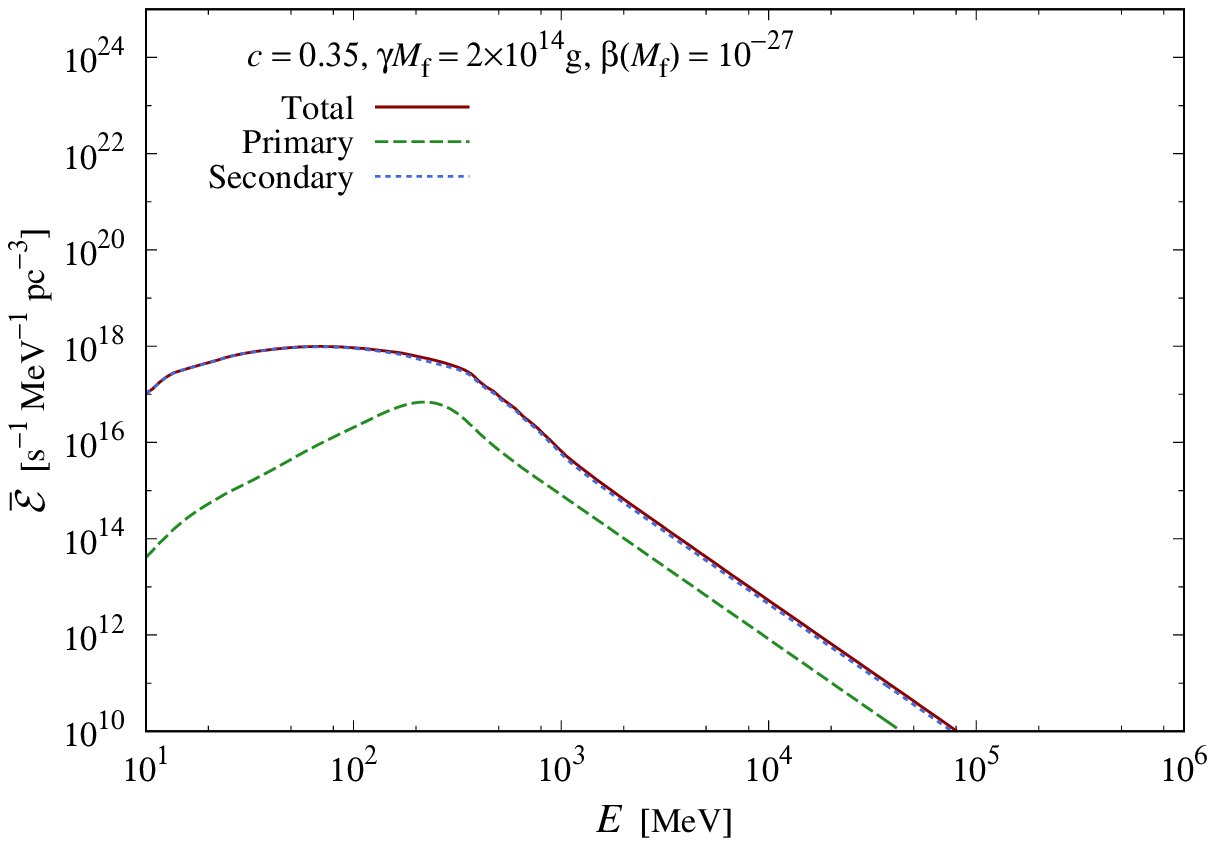}
\caption{\label{fig:em_cc}
The average emissivity $ \bar{\mathcal E}(E) $ for critical collapse with the peak mass scale $ \gamma\,M_\mathrm f = 10^{16}\,\mathrm g $ (top-left), $10^{15}\,\mathrm g $ (top-right), and $ 2 \times 10^{14}\,\mathrm g $ (bottom), respectively.
}
\end{figure}

As a specific example of an extended mass function with $ M_\mathrm f \gg M_* $\,, we assume that the PBHs form from critical collapse and have the mass function given by Eq.~\eqref{eq:currmf_cc} and shown in Fig.~\ref{fig:mfs_cc}.
We write this as
\begin{equation}
\frac{\mathrm d\bar n}{\mathrm dm}
= \beta(M_\mathrm f)\,A\,m^2\,M(m)^{1/c-3}\,\exp[-B\,M(m)^{1/c}]
\propto
\begin{cases}
m^{1/c -1}\,\exp[-B\,m^{1/c}]
& (m \gg M_*) \\
m^2
& (m < M_*)\,,
\end{cases}
\end{equation}
where $ c \approx 0.35 $, $ A $ and $ B $ are constants, and the exponential term gives an effective upper cut-off at $ M_\mathrm f \approx B^{-c} $\,.
In the $ m < M_* $ regime, although $ M $ itself depends weakly on $ m $\,, the power of $ M $ before the exponential term is only $ -0.15 $, so the main dependence comes from the $ m^2 $ term.
The emissivity can now be modelled analytically by using Eq.~\eqref{eq:slope} with $ \nu = 1.85 $.
This gives the primary emissivity
\begin{equation}
\bar{\mathcal E}^\mathrm P(E)
\propto
  \bar n_*
  \times
\begin{cases}
E^3\,M_\mathrm f^{5.85}\,M_*^{-2.85}
& (E < M_\mathrm f^{-1}) \\
E^{-2.85}\,M_*^{-2.85}
& (M_*^{-1} > E > M_\mathrm f^{-1}) \\
E\,M_*\,\mathrm e^{-E M_*}
& (E > M_*^{-1})\,.
\end{cases}
\end{equation}
In our numerical analysis, we compute the emissivity for $ \gamma\,M_\mathrm f = 10^{16}\,\mathrm g $\,, $ 10^{15}\,\mathrm g $\,, and $ 2 \times 10^{14}\,\mathrm g $ and the results are shown in Fig.~\ref{fig:em_cc} for $ \beta = 10^{-27} $\,, the maximum value allowed by the extragalactic $ \gamma $-ray background.
The primary components all show the expected $ E^3 $ rise below $ M_\mathrm f^{-1} $ and $ E^{-2.85} $ fall above $ M_\mathrm f^{-1} $\,.
The secondary components are still given by Eq.~\eqref{eq:em_pl} and show the usual $ E^{-3} $ fall above $ M_\mathrm q^{-1} $\,.

\section{\label{sec:constraint}
Constraints
}

\subsection{
Observations
}

\begin{figure}[htb]
\includegraphics[scale=0.65]{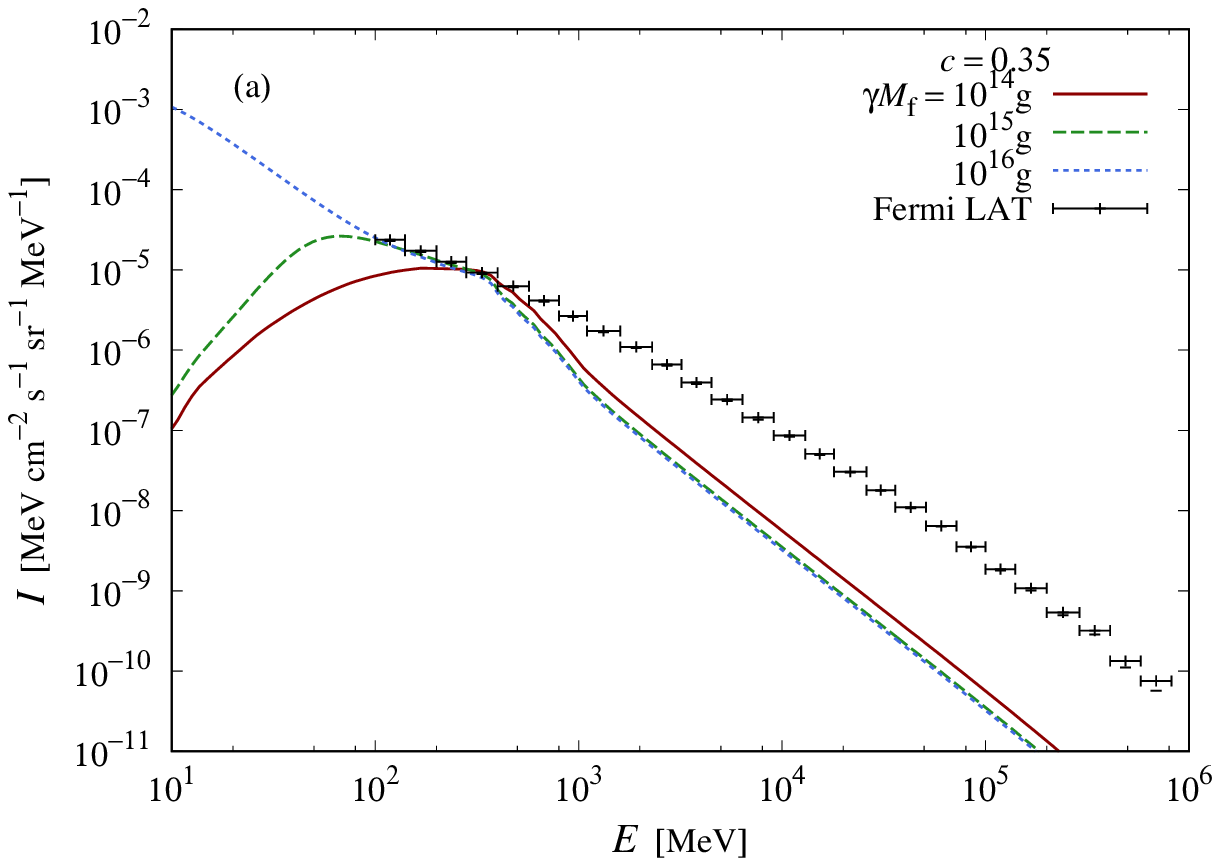}
\includegraphics[scale=0.65]{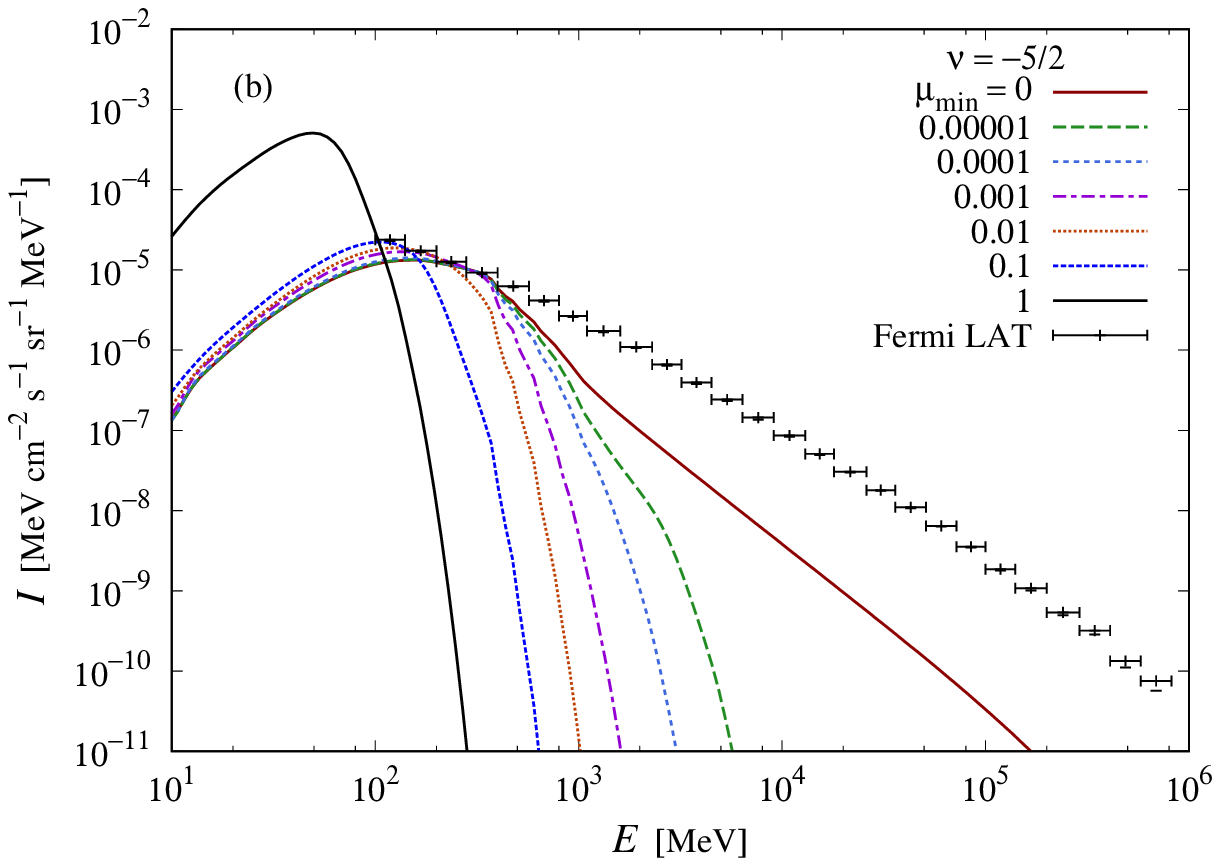}
\caption{\label{fig:flux_ext}
Comparison of the $ \gamma $-ray fluxes from PBHs with extended mass functions averaged over the latitudes $ |b| \geq 20^\circ $ with the Fermi LAT observation \cite{Ackermann:2014usa} for (a) a critical mass function with different upper cut-offs and (b) an extended mass function with different lower cut-offs.
}
\end{figure}

First, we summarize the data on the Galactic $ \gamma $-ray background from both EGRET and Fermi.
An important difference is that the Fermi data extend well above $ 250\,\mathrm{MeV} $\,.
If such high energy photons are generated by PBHs, it could only be as a result of their secondary emission.
However, in this regime most of the energy is degraded into lower energy particles, so the strongest limit may still come from observations at $ 250\,\mathrm{MeV} $\,.
The observed $ \gamma $-ray sky is regarded as the sum of Galactic and extragalactic components.
An analysis of the Fermi LAT data \cite{Ackermann:2014usa} has determined the spectrum of the Diffuse Galactic Emission (DGE) averaged over Galactic latitudes $ |b| \geq 20^\circ $\,.
In this analysis, the DGE is modeled as the sum of interstellar photons inverse-Compton scattered by cosmic-ray electrons, pion decays, bremsstrahlung due to the interactions of hydrogen atoms with cosmic rays, and other processes.
We shall require the $ \gamma $-ray background from Galactic PBHs to be below the astrophysical DGE, as illustrated in Fig.~\ref{fig:flux_ext}.

\subsection{
Wright and Lehoucq \textit{et al.} analyses
}

Constraints on the PBH scenario from EGRET observations of the Galactic $ \gamma $-ray background were first studied by Wright \cite{1996ApJ...459..487W} and Lehoucq \textit{et al.} \cite{Lehoucq:2009ge}.
Wright derived limits on the PBH clustering factor and explosion rate, while Lehoucq \textit{et al.} limited the PBH collapse fraction and density parameter.
However, both these analyses omitted one important feature.
Whereas the strongest constraint from the extragalactic background comes from the time-integrated contribution of PBHs with \textit{initial} mass $ M_* $\,, the Galactic background does not constrain these PBHs at all, since they no longer exist.
Rather it constrains PBHs with \textit{current} mass of around $ M_* $\,, since these are the ones which contribute to this background.
For example, putting $ m = M_* $ in Eq.~\eqref{eq:mmu} gives
\begin{equation}
\mu
= (2-0.05)^{1/3} - 1
\approx
  0.25\,,
\end{equation}
so the initial mass is $ 1.25\,M_* $\,.
More precisely, since the emission from PBHs with initial mass $ (1+\mu)\,M_* $ currently peaks at an energy $ E \approx 100\,(3 \mu)^{-1/3}\,\mathrm{MeV} $ for $ 1 > \mu > 0.02 $\,, this is in the EGRET range of $ 70\,\mathrm{MeV}\textnormal{--}150\,\mathrm{GeV} $ for $ 0.7 > \mu > 0.08 $.
Secondary emission can be neglected in this regime, as assumed by the earlier analyses, since this is important only for $ \mu < 0.005 $.
Although Fermi observations extend well past the QCD threshold, so we take this into account in the analysis below, we note that the high-energy tail of secondary emission from low-mass holes at $ E \gtrsim 1/M_\mathrm q $ is \emph{irrelevant} for current purposes.
This is because the high-energy tail of the $ \gamma $-ray flux obeys a power-law $ I \propto E\,\mathcal E(E) \propto E^{-2} $\,, while the spectrum of the observed Galactic $ \gamma $-rays is shallower than $ E^{-2} $\,, so the flux just below the start of the tail is most crucial for constraining PBHs.

\subsection{
Constraints for extended and critical collapse PBH mass spectrum
}

Constraints are conveniently expressed in terms of the fraction of the Universe collapsing into PBHs of mass $ M $ at the formation epoch $ t_\mathrm f $\,.
This is denoted by $ \beta(M) $ but the definition of this quantity requires care if the mass function is extended.
Ignoring evaporations, the current number density of PBHs with mass around $ M $ is roughly related to the comoving cosmologically averaged mass function by
\begin{equation}
\bar n 
= M\,\frac{\mathrm d\bar n}{\mathrm dM}(M)\,.
\end{equation}
This is related to the collapse fraction by
\begin{equation}
\beta(M)
\approx
  1.05 \times 10^{-29}\,
  \gamma^{-1/2}\,\left(\frac{g_{*\mathrm f}}{106.75}\right)^{1/4}\,
  \left(\frac{M}{M_*}\right)^{3/2}\,
  \left(\frac{\bar n}{1\,\mathrm{pc}^{-3}}\right)\,,
\end{equation}
where $ \gamma $ is the mass of the black hole relative to the particle horizon mass at formation and the factor $ g_{*\mathrm f} $ is the number of relativistic degrees of freedom then.
The latter is normalized to the value at $ 10^{-5}\,\mathrm s $\,, since it does not increase much before that in the standard model, and this is the period when PBHs are expected to form.
The current density parameter for PBHs which have not yet evaporated is then 
\begin{equation}
\Omega_\mathrm{PBH}
\approx
  \left(\frac{\beta(M)}{1.15 \times 10^{-8}}\right)\,
  \gamma^{1/2}\,
  \left(\frac{h}{0.72}\right)^{-2}\,
  \left(\frac{g_{*\mathrm f}}{106.75}\right)^{-1/4}\,
  \left(\frac{M}{M_\odot}\right)^{-1/2}\,.
\end{equation}
Note that $ \beta(M) $ refers to PBHs within a mass-band $ \Delta M \sim M $ around $ M $\,.
This is distinct from the fraction of mass in PBHs smaller than $ M $ (for $ \nu > -2 $) or larger than $ M $ (for $ \nu < -2 $), although these quantities are the same up to a numerical factor.
For a power-law mass function \eqref{eq:mf_pl}, we have
\begin{equation}
\beta(M)
= \beta(M_*)\,\left(\frac{M}{M_*}\right)^{5/2+\nu}
\approx
  10^{-27}\,
  \gamma^{-1/2}\,
  \left(\frac{g_{*\mathrm f}}{106.75}\right)^{1/4}
  \left(\frac{M}{M_*}\right)^{5/2+\nu}\,
  \left(\frac{\bar n_*}{96.5\,\mathrm{pc}^{-3}}\right)\,,
\end{equation}
where the value of $ \beta(M_*) $ is relevant to the Galactic $ \gamma $-ray background.
In calculating this background, it is sometimes interesting to assume the mass function has a lower cutoff at $ M_\mathrm{min} = (1+\mu_\mathrm{min})\,M_* $ or an upper cutoff at $ M_\mathrm{max} = (1+\mu_\mathrm{max})\,M_* $\,.

\begin{figure}[htb]
\includegraphics[scale=0.65]{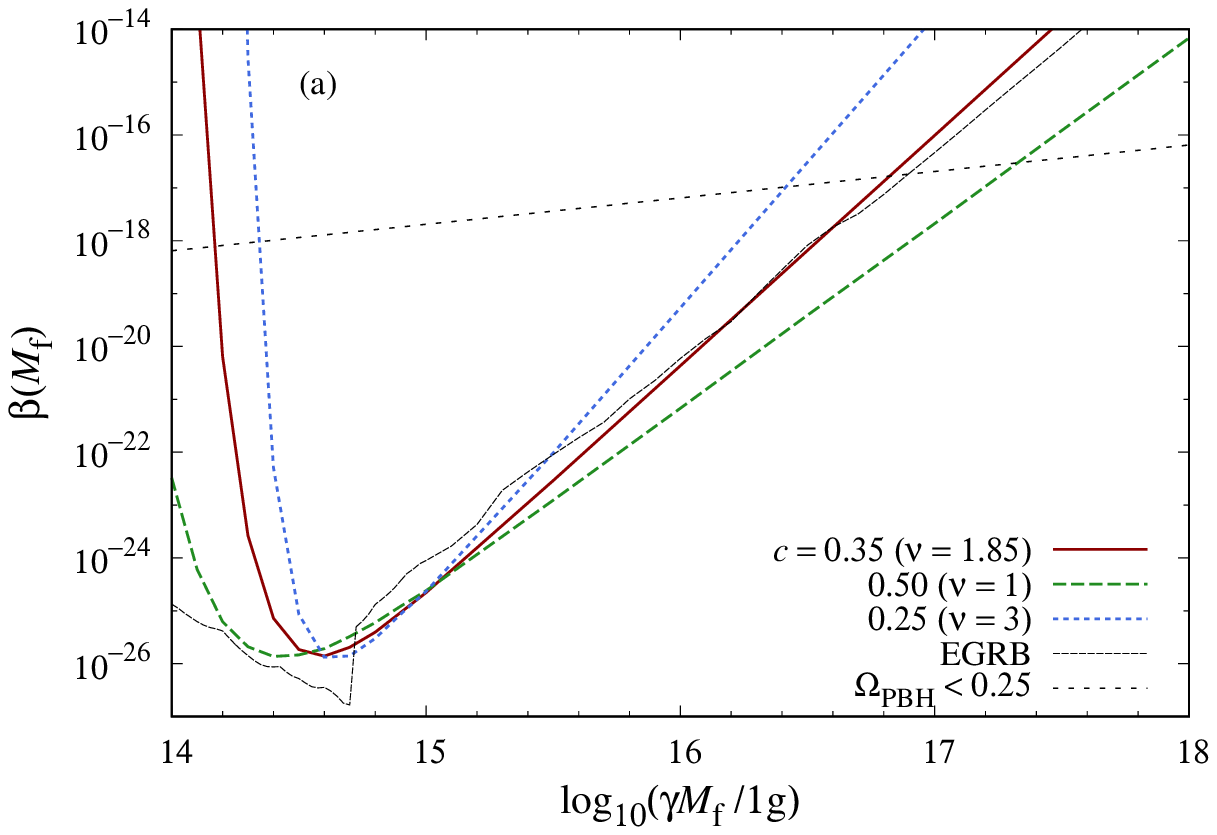}
\includegraphics[scale=0.65]{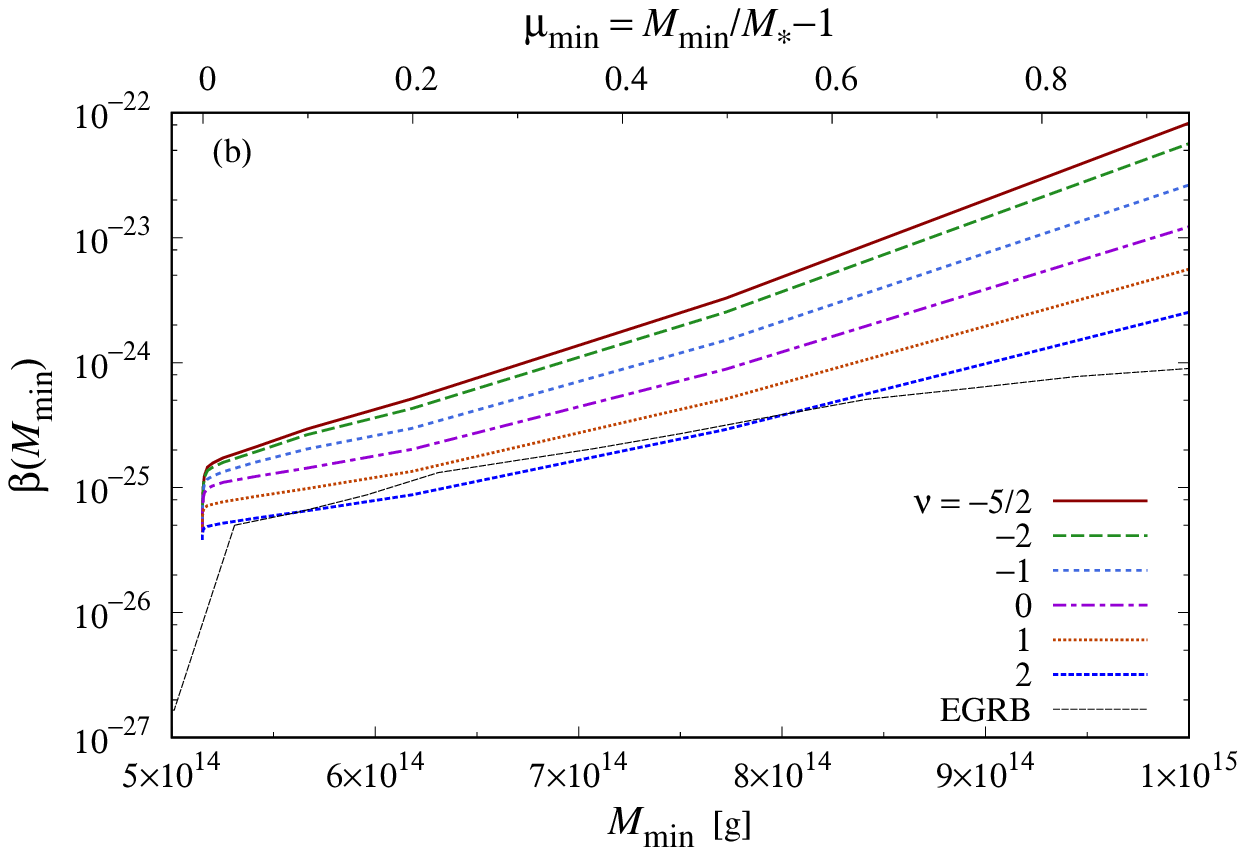}
\caption{\label{fig:ul_ext}
Upper limits from Galactic $ \gamma $-ray background on (a) $ \beta(M_\mathrm{min}) $ and (b) $ \beta(M_\mathrm f) $ for different values of $ \nu $ (including the critical case with $ \nu = 1.85 $).
The (generally weaker) limit from the extragalactic background is also shown for comparison.
}
\end{figure}

\begin{figure}[htb]
\includegraphics[scale=0.65]{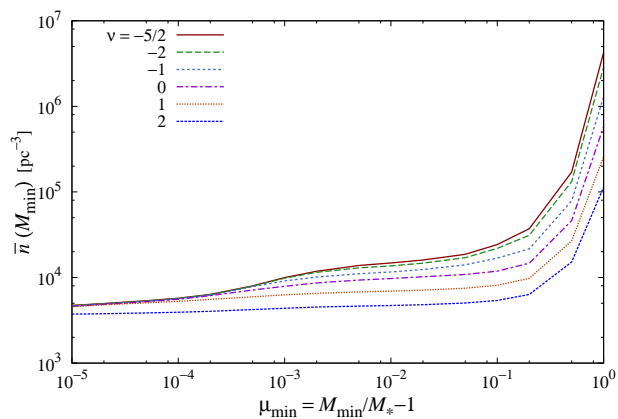}
\caption{\label{fig:uln_ext}
Upper limits on number density of PBHs for an extended initial mass function with a lower cut-off at a mass $ M_\mathrm{min} $ just above $ M_* $ as a function of $ \mu_\mathrm{min} $\,.
If there is no lower cut-off, the effective value of $ \mu_\mathrm{min} $ is $ 0 $.
}
\end{figure}

The intensity $ I \equiv E\,\Phi(E) $ for various extended mass function scenarios are shown in Fig.~\ref{fig:flux_ext}.
(a) applies for the critical collapse scenario with $ c = 0.35 $ (i.e.\ $ \nu = 1.85 $) and different values of $ M_\mathrm f $\,, (b) for mass functions with $ \nu = -5/2 $ and different values of $ \mu_\mathrm{min} $\,, a completely extended initial mass function corresponding to $ \mu_\mathrm{min} = 0 $.
Each of the curves is normalised to have the maximum flux compatible with the Fermi LAT observations.
As indicated in Fig.~\ref{fig:ul_ext}(a), for each value of $ \nu $\,, we can constrain $ \beta(M_\mathrm f) $ as a function of $ M_\mathrm f $ and this also shows the value of $ M_\mathrm f $ required for the PBHs to comprise the dark matter.
Figure~\ref{fig:ul_ext}(b) shows the limits as a function of $ M_\mathrm{min} $\,.
Figure~\ref{fig:uln_ext} shows the constraint on the number of $ M \approx M_* $ black holes as a function of $ \mu_\mathrm{min} $ for different values of $ \nu $\,.
The limit from the Milagro search for high-energy bursts \cite{Abdo:2014apa,MacGibbon:2015mya} depends on the timescale of the burst but it is always above $ 5 \times 10^4\,\mathrm{pc}^{-3}\,\mathrm{yr}^{-1} $ and weaker than the $ \gamma $-ray limit.

\subsection{
Constraints on nearly monochromatic PBH mass spectrum from Galactic background
}

\begin{figure}[htb]
\includegraphics[scale=0.65]{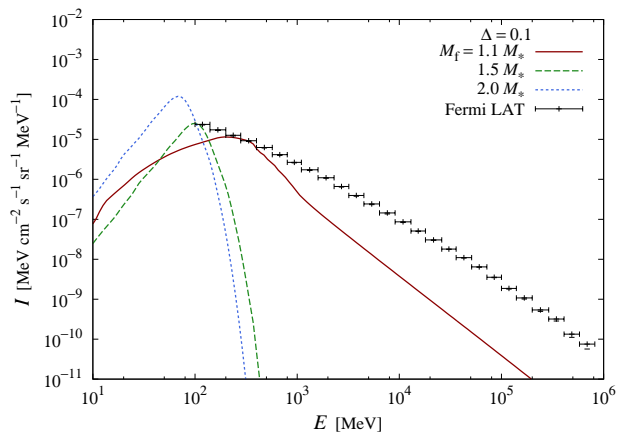}
\caption{\label{fig:flux_mono}
Comparison of the $ \gamma $-ray fluxes from PBHs, averaged over latitudes $ |b| \geq 20^\circ $, with the Fermi LAT observations \cite{Ackermann:2014usa} for nearly monochromatic initial mass functions with different values of $ M_\mathrm f $ and $ \Delta = 0.1 $.
}
\end{figure}

\begin{figure}[htb]
\includegraphics[scale=0.65]{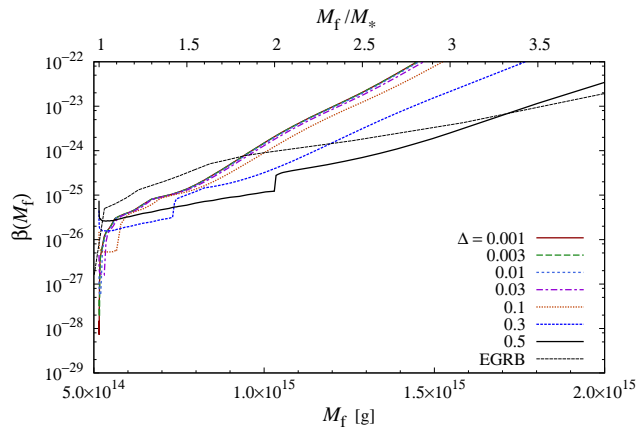}
\caption{\label{fig:ul_mono}
Upper limits on $ \beta $ with respect to the mass scale $ M_\mathrm f $ for a nearly monochromatic initial mass function with different values of $ \Delta $\,.
Upper limits only exist for $ M_\mathrm f > M_* $\,.
Also plotted is the constraint from the extragalactic $ \gamma $-ray background (EGRB) obtained in this case.
For each $ \Delta $\,, the limit has a kink at $ M_\mathrm f = M_\mathrm c/(1-\Delta) $\,, where the secondary emission stops.
The limits are $ \propto (M_\mathrm f/M_*)^{5/2} $ below the critical mass but become exponentially weaker above that.
}
\end{figure}

The intensity $ I \equiv E\,\Phi(E) $ can be calculated using the emissivities shown in Fig.~\ref{fig:em_mono} and this leads to the curves shown in Fig.~\ref{fig:flux_mono} for $ \Delta = 0.1 $ and different values of $ M_\mathrm f $\,.
Each of them is normalised to have the maximum flux compatible with the Fermi LAT observations.
In Fig.~\ref{fig:ul_mono}, upper limits on $ \beta(M_\mathrm f) $ are shown for various values of $ \Delta $\,.
For $ \Delta > 0.005 $, over almost the entire mass range $ M_\mathrm c < M_\mathrm f < M_\mathrm c/(1-\Delta) $\,, an extensive low-mass tail is formed and the secondary emission from the holes at $ m \approx M_\mathrm q $ is significant.
The upper limit on $ \beta(M_\mathrm f) $ can then be written as
\begin{equation}
\beta(M_\mathrm f)
\lesssim
  5 \times 10^{-26}\,
  \left(\frac{\Delta}{0.1}\right)\,
  \left(\frac{M_\mathrm f}{M_*}\right)^{5/2}
\quad
(M_\mathrm c < M_\mathrm f < M_\mathrm c/(1-\Delta))\,.
\end{equation}
As $ (1-\Delta)\,M_\mathrm f $ increases beyond $ M_\mathrm c $\,, the primary emission, peaking at $ E^\mathrm P = 6/[(1-\Delta)\,M_\mathrm f] $\,, becomes responsible for the constraint instead of the secondary emission.
The weakening of the constraint at that mass explains the kinks in Fig.~\ref{fig:ul_mono}.
Since the Fermi LAT observation of the Galactic background extend only down to $ 100\,\mathrm{MeV} $\,, the constraint on $ \beta $ due to the primary emission derives from the Wien tail and therefore weakens exponentially.
It roughly fits
\begin{equation}
\beta(M_\mathrm f)
\lesssim
  10^{-27}\,\exp\left(\frac{\tilde\chi\,(1-\Delta)\,M_\mathrm f}{M_*}\right)
\quad
(M_\mathrm f > M_\mathrm c/(1-\Delta))
\end{equation}
with $ \tilde\chi \approx 2 $.
The extragalactic constraint \cite{Carr:2009jm} is also shown in Fig.~\ref{fig:ul_mono}.
It is clear that this is weaker than the Galactic one below a value of $ M_\mathrm f $ which depends on $ \Delta $ and increases as $ \Delta $ increases.

Finally, we briefly discuss the transition to monochromaticity.
As regards the mass range where primary emission dominates, the constraint clearly converges as $ \Delta \to 0 $ to the value obtained in the monochromatic case.
On the other hand, the secondary emissivity increases in proportion to $ \Delta^{-1} $ as long as an extensive low-mass tail is formed.
This implies that the constraint becomes unboundedly stronger as $ \Delta \to 0 $ in the mass range $ M_* < M_\mathrm f < (1+\Delta)\,M_* $ in contrast to the monochromatic case where no mass tail emerges.

\section{\label{sec:conclusion}
Conclusions
}

In this paper we have placed constraints on the number of PBHs with mass around $ M_* $ from observations of the Galactic $ \gamma $-ray background.
There are other sources of such a background, including decaying or annihilating WIMPs, so PBHs are not definitely the explanation.
On the other hand, the observed background certainly provides the strongest constraint on the PBH density in this mass range.
In particular, it is stronger than the limit from the extragalactic gamma-ray background.

Whereas the extragalactic background comes from the time-integrated emission of all the PBHs within the particle horizon and is dominated by those with initial mass $ M_* $\,, the Galactic background comes from the instantaneous emission of all the PBHs within the Galactic halo and with initial mass slightly larger than $ M_* $\,.
However, their \textit{current} mass is smaller than $ M_* $\,, the Galactic background being dominated by the low mass tail ($ \mathrm dn/\mathrm dm \propto m^2 $) of the PBHs which have not quite completed their evaporation.
By contrast, the high-energy tail ($ \mathrm dN/\mathrm dE \propto E^{-3} $) of the extragalactic background derives from the PBHs which have just completed their evaporation, so these two components are complementary.

Our previous analysis of PBH constraints \cite{Carr:2009jm} assumed that the PBHs have a monochromatic mass function (i.e.\ a mass spread $ \Delta m $ no larger than $ m $).
However, this is inappropriate in calculating the Galactic background because the dominant contribution comes from the low-mass tail, which only represents a very narrow part of the original mass range.
Even if one expects the PBH mass spectrum to be monochromatic, there is no \textit{a priori} reason why it should encompass the particular mass $ M_* $\,.
Therefore we have focused mainly on the possibility of an extended mass spectrum, although we have also covered the nearly monochromatic scenario in order to allow comparison with the results in our previous paper \cite{Carr:2009jm}.
A particularly interesting extended mass spectrum scenario arises if the PBHs form from critical collapse.
In this case, the spectrum can be predicted very precisely and the PBH density increases with $ m $ up to some limiting mass $ M_\mathrm f $\,.
This means that the largest (unevaporated) PBHs could provide the dark matter without the $ M_* $ ones contravening the Galactic background limit.
However, this still requires fine-tuning of the mass $ M_\mathrm f $\,, since it cannot be much larger than $ M_* $\,.

Finally, it should be stressed that the analysis of this paper has shed light on a number of issues which go beyond the Galactic background problem itself.
For example, we have determined the mass $ M_\mathrm q $ at which secondary emission becomes important and the initial mass $ M_\mathrm c $ below which PBHs have evaporated to $ M_\mathrm q $ by the present epoch.
We have also calculated the ratio of the primary and secondary emission as a function of $ M $\,, and the form of the low-mass tail of currently evaporating PBHs and its connection to the high-energy tail.
This has allowed us to predict the Galactic backgrounds generated in many PBH scenarios analytically, thereby clarifying the results of our detailed numerical calculations.

\begin{acknowledgments} 
B.C.\ thanks Research Center for the Early Universe (RESCEU), University of Tokyo, for hospitality received during this work.
This work is supported in part by MEXT KAKENHI Grant Nos.~15H05889, 16H00877 (K.K.) 15H05888 (J.Y.), and by JSPS KAKENHI Grant Nos.~26105520, 26247042 (K.K.), 24111701, 26800115, 16K17675 (Y.S.), 15H02082 (J.Y.).
The work of K.K.\ is also supported by the Center for the Promotion of Integrated Science (CPIS) of Sokendai (1HB5804100).
\end{acknowledgments}

\bibliographystyle{apsrev4-1}
\bibliography{pbh}

\end{document}